\documentclass{emulateapj}
\usepackage{amssymb,amsmath}
\usepackage{verbatim}
\usepackage{color,hyperref}
\usepackage{needspace}
\definecolor{linkcolor}{rgb}{0,0,0.25}
\hypersetup{
  colorlinks=true,        
  linkcolor=linkcolor,    
  citecolor=linkcolor,    
  filecolor=linkcolor,    
  urlcolor=linkcolor      
}
\newcounter{address}
\newcommand{\ie}{i.e.}
\newcommand{\etal}{et al.}
\newcommand{\dd}{\mathrm{d}}
\newcommand{\eg}{e.g.}
\newcommand{\eqnname}{equation}
\newcommand{\Eqnname}{Equation}
\newcommand{\equationname}{\eqnname}
\newcommand{\Equationname}{\Eqnname}
\renewcommand{\tablename}{Table}
\renewcommand{\figurename}{Figure}

\newcommand{\sectionname}{$\mathsection$}

\newcommand{\feh}{\ensuremath{[\mathrm{Fe/H}]}}
\newcommand{\afe}{\ensuremath{[\alpha\mathrm{/Fe}]}}
\newcommand{\avgafe}{\ensuremath{[([\mathrm{O+Mg+Si+S+Ca]/5)\mathrm{/Fe}]}}}
\newcommand{\jk}{\ensuremath{(J-K_s)_0}}
\newcommand{\jksq}{\ensuremath{[J-K_s]_0}}

\newcommand{\dens}{\ensuremath{\nu_*}}

\newcommand{\dex}{\ensuremath{\,\mathrm{dex}}}

\newcommand{\Gyr}{\ensuremath{\,\mathrm{Gyr}}}
\newcommand{\kpc}{\ensuremath{\,\mathrm{kpc}}}
\newcommand{\pc}{\ensuremath{\,\mathrm{pc}}}

\newcommand{\inv}{\ensuremath{^{-1}}}

\newcommand{\magunit}{\,\mbox{mag}}
\newcommand{\Kunit}{\,\mbox{K}}
\newcommand{\ks}{\ensuremath{K_s}}

\newcommand{\apogee}{APOGEE}

\newcommand{\logg}{\ensuremath{\log g}}
\newcommand{\teff}{\ensuremath{T_{\mathrm{eff}}}}

\newcommand{\field}{\ensuremath{\mathrm{location}}}
\usepackage{yfonts}
\newcommand{\essf}{\ensuremath{\textswab{S}}}

\newcommand{\Rb}{\ensuremath{R_{\mathrm{peak}}}}

\newcommand{\Rf}{\ensuremath{R^{-1}_{\mathrm{flare}}}}

\newcommand{\nstars}{14,699}

\submitted{}

\begin{document}

\title{The Stellar Population Structure of the Galactic Disk}

\author{Jo~Bovy\altaffilmark{1,2,3},
  Hans-Walter~Rix\altaffilmark{4},
  Edward F. Schlafly\altaffilmark{4},
  David~L.~Nidever\altaffilmark{5},
  Jon~A.~Holtzman\altaffilmark{6},
  Matthew~Shetrone\altaffilmark{7},
  and Timothy~C.~Beers\altaffilmark{8}}
\altaffiltext{\theaddress}{\label{1}\stepcounter{address} 
  Department of Astronomy and Astrophysics, University of Toronto, 50
  St.  George Street, Toronto, ON, M5S 3H4, Canada;
  bovy@astro.utoronto.ca~}
\altaffiltext{\theaddress}{\label{2}\stepcounter{address} 
  Institute for Advanced Study, Einstein Drive, Princeton, NJ 08540,
  USA}
\altaffiltext{\theaddress}{\label{3}\stepcounter{address} 
  John Bahcall Fellow}
\altaffiltext{\theaddress}{\label{4}\stepcounter{address} 
  Max-Planck-Institut f\"ur Astronomie, K\"onigstuhl 17, D-69117
  Heidelberg, Germany}
\altaffiltext{\theaddress}{\label{5}\stepcounter{address} 
  Department of Astronomy, University of Michigan, Ann Arbor, MI
  48109, USA}
\altaffiltext{\theaddress}{\label{6}\stepcounter{address} 
  New Mexico State University, Las Cruces, NM 88003, USA}
\altaffiltext{\theaddress}{\label{7}\stepcounter{address} 
  The University of Texas at Austin, McDonald Observatory, TX 79734, USA}
\altaffiltext{\theaddress}{\label{8}\stepcounter{address} 
  Department of Physics and JINA Center for the Evolution of the Elements,
  University of Notre Dame, Notre Dame, IN 46556, USA}

\begin{abstract}  
The spatial structure of stellar populations with different chemical
abundances in the Milky Way contains a wealth of information on
Galactic evolution over cosmic time. We use data on \nstars\ red-clump
stars from the APOGEE survey, covering $4\kpc \lesssim R \lesssim
15\kpc$, to determine the structure of mono-abundance populations
(MAPs)---stars in narrow bins in \afe\ and \feh---accounting for the
complex effects of the APOGEE selection function and the
spatially-variable dust obscuration. We determine that all MAPs with
enhanced \afe\ are centrally concentrated and are well-described as
exponentials with a scale length of $2.2\pm0.2\kpc$ over the whole
radial range of the disk. We discover that the surface-density
profiles of low-\afe\ MAPs are complex: they do not monotonically
decrease outwards, but rather display a peak radius ranging from
$\approx5\kpc$ to $\approx13\kpc$ at low \feh.  The extensive radial
coverage of the data allows us to measure radial trends in the
thickness of each MAP. While high-\afe\ MAPs have constant scale
heights, low-\afe\ MAPs flare. We confirm, now with high-precision
abundances, previous results that each MAP contains only a single
vertical scale height and that low-\feh, low-\afe\ and high-\feh,
high-\afe\ MAPs have intermediate ($h_Z\approx300$--$600\pc$) scale
heights that smoothly bridge the traditional thin- and thick-disk
divide. That the high-\afe, thick disk components do not flare is
strong evidence against their thickness being caused by radial
migration. The correspondence between the radial structure and
chemical-enrichment age of stellar populations is clear confirmation
of the inside-out growth of galactic disks. The details of these
relations will constrain the variety of physical conditions under
which stars form throughout the MW disk.
\end{abstract}

\keywords{ Galaxy: abundances --- Galaxy: disk --- Galaxy: evolution
  --- Galaxy: formation --- Galaxy: fundamental parameters --- Galaxy:
  structure }

\section{Introduction}

\begin{figure*}[t!]
\includegraphics[width=0.98\textwidth,clip=]{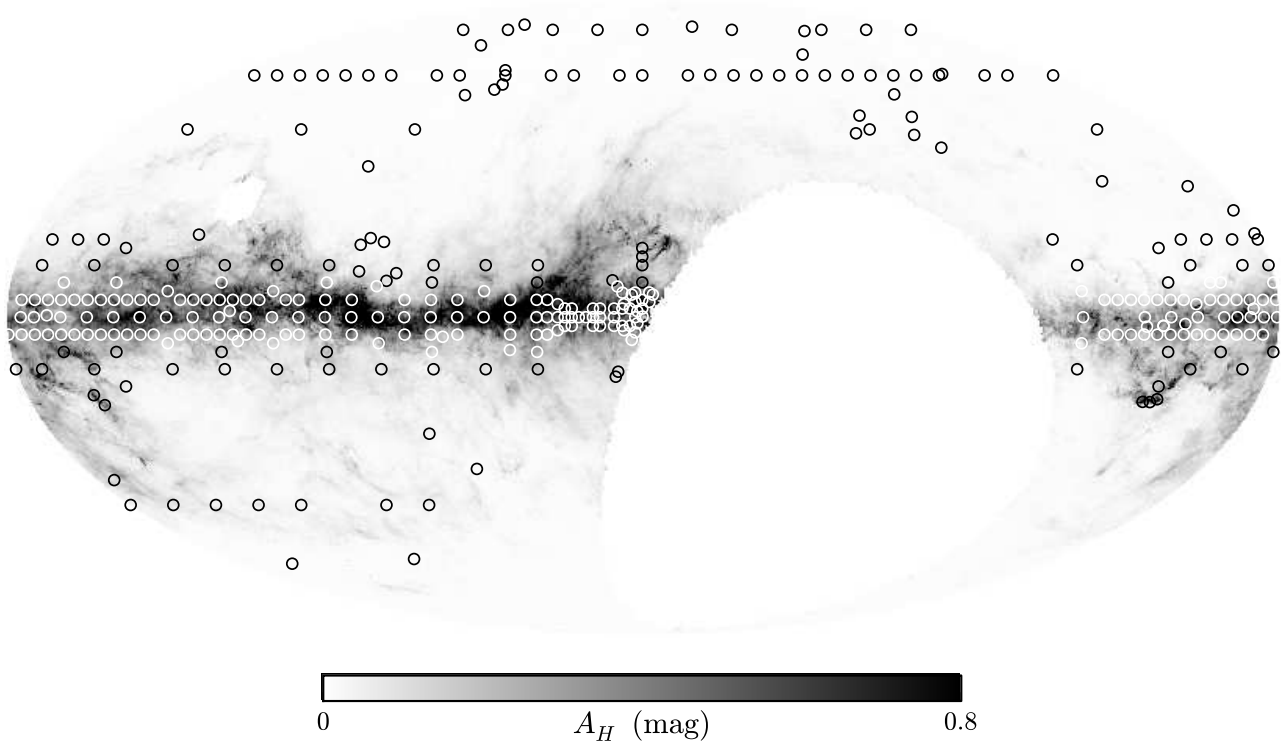}
\caption{Distribution of the APOGEE fields that contain the
  statistical APOGEE-RC sample on the sky, overlayed on the extinction
  map from \citet{Green15a} at 5\kpc. Fields at $|b| < 8^\circ$ are
  displayed in white to enhance their contrast. The upper limit of
  $A_H = 0.8\magunit$ is the highest extinction for which the RC can
  be seen to 5\kpc\ in APOGEE medium-deep fields ($H < 12.2$). The
  statistical APOGEE-RC has excellent coverage of the large portion of
  the Galactic plane that can be seen from the Northern hemisphere
  (the large white region is unobserved by both APOGEE and
  Pan-STARRS).}\label{fig:fields}
\end{figure*}

Understanding the growth and evolution of galactic disks is a central
problem in galaxy formation. Empirical studies in this area employ two
complementary approaches to investigate changes in the structure of
galactic disks over cosmic time. One of them, the ``lookback
approach'', is based on extensive samples of galaxies that span
redshifts from the time at which disks form ($z \approx 2.5$) to today
($z \approx 0$), matching corresponding samples across epochs
\citep[\eg,][]{vanDokkum13a,Wisnioski15a}.  Such studies have been
successful in establishing the evolution of the galaxy population
properties, but they are limited to characterizing individual galaxies
by quantities integrated over their stellar populations, \eg, by their
size, shape, or overall velocity dispersion. The second approach,
Galactic archaeology, aims at obtaining detailed observations of local
galaxies, either through resolving their stellar populations into
individual stars \citep[\eg,][]{Dalcanton12a} or using integral-field
spectroscopy, trying to directly reconstruct their individual
formation history.  The Milky Way (MW) is perhaps the best case for
Galactic archaeology, because it has very typical gross properties
(\eg, mass) and because we can measure the detailed
properties---six-dimensional phase--space distribution, age, elemental
abundances---of large numbers of stars \citep{Rix13a}.

The growth of the MW disk over time is encoded in the orbital
distribution of stars and their ages and elemental abundances. The
radial distribution of stars of different ages and abundances
constrains in particular the epochs and physical conditions under
which different parts of the MW formed stars. However, the large-scale
radial structure of the MW disk has been difficult to determine for
any well-defined stellar tracer, primarily due to severe dust
extinction in the midplane. Investigations have therefore been limited
to measurements of the overall radial profile, which can be
characterized as an approximately exponential disk
\citep[\eg,][]{Kent91a,Benjamin05a,Juric08a,Bovy13a}. Radial
population changes in the Galactic disk have traditionally been
characterized by a metallicity gradient, taking at each radius the
mean over the complex abundance distribution
\citep[e.g.,][]{Audouze76a,Chen03a,Anders14a,Hayden14a}. Neither of
these measurements provides a direct empirical constraint on the
growth of the MW disk over time, nor have they proven to be very
constraining for simulations of the formation of galactic disks.

\citet{BovyMAPstructure} suggested a different way to look at the
complex correlations between spatial structure and abundances of the
Galactic disk population, by separately determining disk structure of
so-called mono-abundance populations (MAPs). These are subsets of
stars selected to have very similar abundances, such as \feh\ and and
average \afe; for each star these abundances constitute life-long tags
that remain invariant even if their orbits change grossly after
birth. Beyond this simple fact, a decomposition of the disk in terms
of MAPs is essentially empirical and does not assume or imply a common
formation site or epoch; the detailed relation between MAPs and the
underlying chemical evolution of the MW requires models and
measurements of the age distributions within
MAPs. \citet{BovyMAPstructure} were the first to attempt dissecting
the radial stellar population structure of the MW disk, by measuring
the radial profiles of MAPs over a few kpc near the Sun. These
measurements revealed a complex abundance-dependence of the radial
disk structure---both on \feh\ and an average \afe---with older,
\afe-enhanced populations having centrally-concentrated profiles and
younger, solar-\afe\ populations having more extended profiles
reaching the outer disk. However, the limited radial coverage and
high-latitude survey geometry of the SDSS/SEGUE sample \citep{Yanny09}
impeded more detailed measurements of the radial surface density
profile beyond characterizing the local slope (in radius) by an
exponential-disk scale length.

The vertical profile of the MW, near the Sun and at other radii, is
also of great interest for understanding the MW, although its
interpretation in terms of stellar disk growth and evolution is still
unsettled.  In the solar neighborhood, the overall vertical stellar
density profile (that is, of all stars) can be accurately
characterized as a sum of two exponential components
\citep{Yoshii82,Gilmore83a,Juric08a,Rix13a}, with stars in the thicker
of the two components being older, more metal-poor, and \afe-enhanced
with respect to stars in the thinner components
\citep{Fuhrmann98,Prochaska00,Bensby05a,Reddy06,Haywood13a}. Dissecting
the disk into MAPs,
\citet{BovyNoThickDisk,BovyMAPkinematics,BovyMAPstructure}
demonstrated that this broader picture is the consequence of an
underlying disk structure composed of a continuum of disks, with a
smooth correlation between abundance, scale height, and velocity
dispersion. While the vertical profile near the Sun is composed of a
smooth continuum of disk thicknesses, it has become clear from
unbiased observations of element abundances of disk stars that their
distribution in $(\feh,\afe)$ is distinctly bimodal, composed of two
sequences---high- and low-\afe---that are disjoint at low-\feh\ and
that merge near solar \feh\ \citep{Adibekyan12a,Nidever14a}. The
high-\afe\ sequence becomes more prominent in the inner Galaxy, while
the low-\afe\ and in particular its metal-poor end dominate in the
outer Galaxy \citep{Bensby11a,Nidever14a,Hayden15a}, in agreement with
the scale lengths measured from the SEGUE data
\citep{BovyMAPstructure}. But a more detailed and quantitative
characterization of the radial behavior of the stellar populations
does not yet exist. Exactly how the continuity in local disk
thicknesses and the disjoint high- and low-\afe\ sequences in
abundances can be reconciled in a consistent formation scenario is
currently a mystery. The complex behavior of the radial scale lengths
with abundance found by \citet{BovyMAPstructure}, as opposed to the
smooth variation of the vertical scale height, indicates that a
thorough exploration of the radial profile of MAPs might hold the key
to resolve this tension.

\begin{figure}[t!]
\includegraphics[width=0.48\textwidth,clip=]{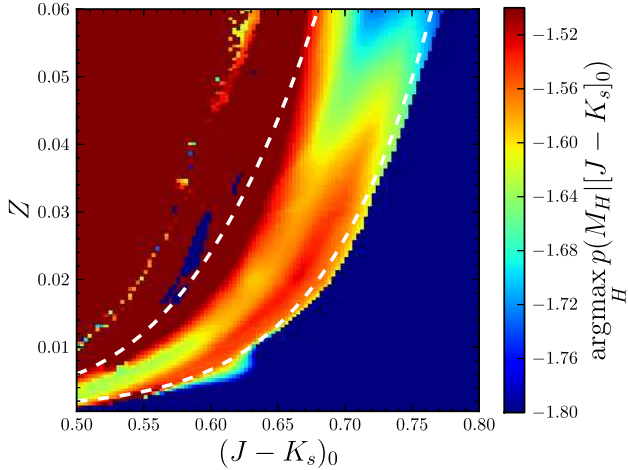}
\caption{Peak of the absolute-magnitude PDF $p(M_H |\jksq)$ as a
  function of color \jk\ and metallicity $Z$ for PARSEC isochrones in
  the RC-star region (defined using Equations [2] and [3] of
  \citealt{BovyRC}). The white dashed lines represent the region
  specified by the cuts in Equations (6) and (7) of \citet{BovyRC}
  that delineate the regions over which the distribution of absolute
  $K_s$ magnitudes is narrow; these also select regions where $M_H$ is
  narrowly distributed. The peak of the magnitude PDF does not
  strongly depend on color or metallicity over the region where the
  PDF is narrow. These $M_H(\jksq,Z)$ are combined with an overall
  offset of $\Delta M_H = 0.08$ to determine the distances for RC
  stars used in this paper.}\label{fig:rcmag}
\end{figure}

While the MW disk's overall vertical stellar density profile was long
thought to be the result of some type of heating, whether from
satellites \citep[\eg,][]{Quinn93a,Abadi03a,Brook04a} or through
radial migration
\citep[\eg,][]{Sellwood02a,Schoenrich09a,Minchev10a,Loebman11a}, the
possibility that (at least some part of) the stellar disk formed
``upside-down'' with its current thickness has recently gained favor
\citep{Bournaud09a,Stinson13a,Bird13a}. The smooth continuum of disk
thicknesses found by \citet{BovyNoThickDisk,BovyMAPstructure}
disfavors the satellite accretion and few-satellites-heating
scenarios, because those would create a discrete thick-disk component
\citep[\eg,][]{Martig14a}. However, observational evidence to prefer
either radial migration or upside-down formation is scant to date.

Radial migration of stars from well inside the MW has been proposed as
a mechanism to create the thicker, \afe-enhanced stellar disk
components in the solar neighborhood. Radial migration approximately
conserves the vertical action \citep{Solway12a,VeraCiro15a}. For
outwardly migrating stars, the decreased gravitational pull in the
outer disk leads to larger vertical excursions that for ensembles of
stars should lead to flaring. Whether radial migration causes the
whole stellar disk or its components to flare depends on which stars
participate in the migration process. If all stars migrate similarly,
then all co-eval populations should flare, but if only special
subclasses are affected (\eg, stars with low vertical excursions),
then the overall effect of migration on the vertical structure might
be small. While the current suite of simulations by no means
exhaustively covers the various migration scenarios and histories that
may have occurred in the MW, current simulations do indicate that
kinematic biases in the population of migrators are significant enough
that little or no effect on the vertical structure of the disk should
be expected \citep{Minchev12a,Roskar13a,VeraCiro14a}. Significant
radial mixing, therefore, may well happen without leading to flaring
while still having a large effect on, \eg, the present-day abundance
distribution. However, for migration to be the \emph{cause} of the
thickness of the thicker-disk components in the MW, flaring of
sub-populations must occur and determining whether individual MAPs
flare vertically towards larger radii can discriminate among formation
mechanisms for the thick-disk components.

That radial migration should and does play a large role in the
evolution of the low-\afe\ populations is becoming increasingly
clear. Radial migration provides a natural way to explain the observed
lack of correlation between age and metallicity in the solar
neighborhood \citep{Edvardsson93,Nordstrom04a} in the presence of a
strong correlation between age and \afe\ \citep{Haywood13a}. The
change in skew of the metallicity distribution from negative in the
inner MW to positive in the outer MW \citep{Hayden15a} can arise from
migration. As for the thicker components, if this migration process is
important for all low-\afe\ stars (not just a biased subset), it makes
the generic prediction of an outward flaring disk.

In this paper we set out to comprehensively determine the stellar
population structure of the Milky Way's stellar disk. We do so by
determining the radial and vertical profile of MAPs, using
SDSS-III/APOGEE \citep{Eisenstein11a,Majewski15a} data. The APOGEE
data are complementary to the SEGUE data, because they provide much
better radial coverage by using near-infrared observations of a large
number of giants close to the Galactic midplane: the sample of
red-clump giants used here spans the radial range $4\kpc \lesssim R
\lesssim 15\kpc$. This radial coverage allows us to determine the
radial profile of MAPs in detail: the data clearly require to go
beyond the simple radial-exponential profile assumed by
\citet{BovyMAPstructure} and all other previous analyses. The wide
radial range combined with good vertical coverage ($0 \leq |Z|
\lesssim2\kpc$) also makes it possible to measure any flaring of MAPs,
providing an essential test of the predictions of radial
migration. Finally, the high-resolution spectroscopic observations let
us define MAPs with high-precision $(\feh,\afe)$ abundances, such that
MAPs have no substantial contamination from neighboring MAPs. These
better APOGEE data end up confirming the smooth, continuous
distribution in scale height from the thinnest to the thickest disk
components, as found by \citet{BovyNoThickDisk}.

\begin{figure}[t!]
\includegraphics[width=0.48\textwidth,clip=]{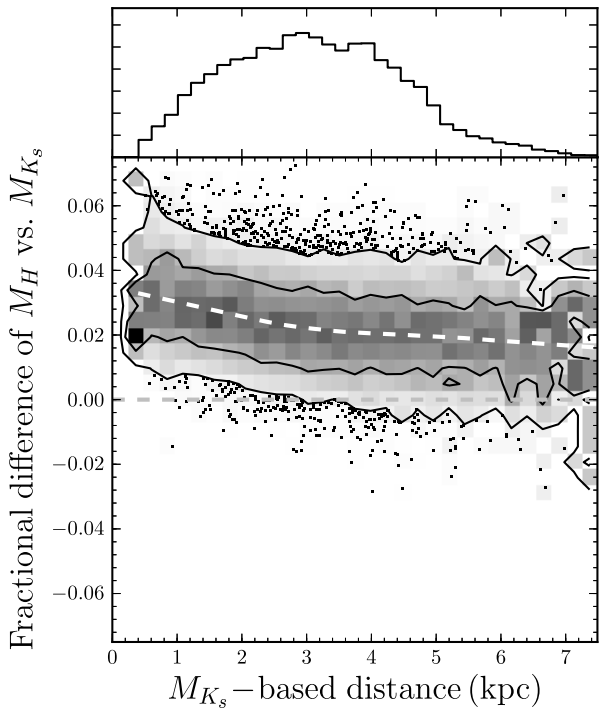}
\caption{Fractional difference between the RC distances calculated
  based on the $H$-band and $K_s$-band luminosities as a function of
  the latter distance. The conditional distribution of distance
  differences is displayed for all \nstars\ stars in the APOGEE-RC
  subsample used in this paper. The black contours contain 68\,\% and
  95\,\% of the distribution and the white dashed line shows a lowess
  trendline. The overall median difference is only $2.3\,\%$,
  consistent with the estimated $\approx2\,\%$ systematic uncertainty
  in the RC distance scale \citep{BovyRC}.}\label{fig:rcdistcomp}
\end{figure}

\begin{figure*}[t!]
\begin{center}
\includegraphics[height=0.2\textheight,clip=]{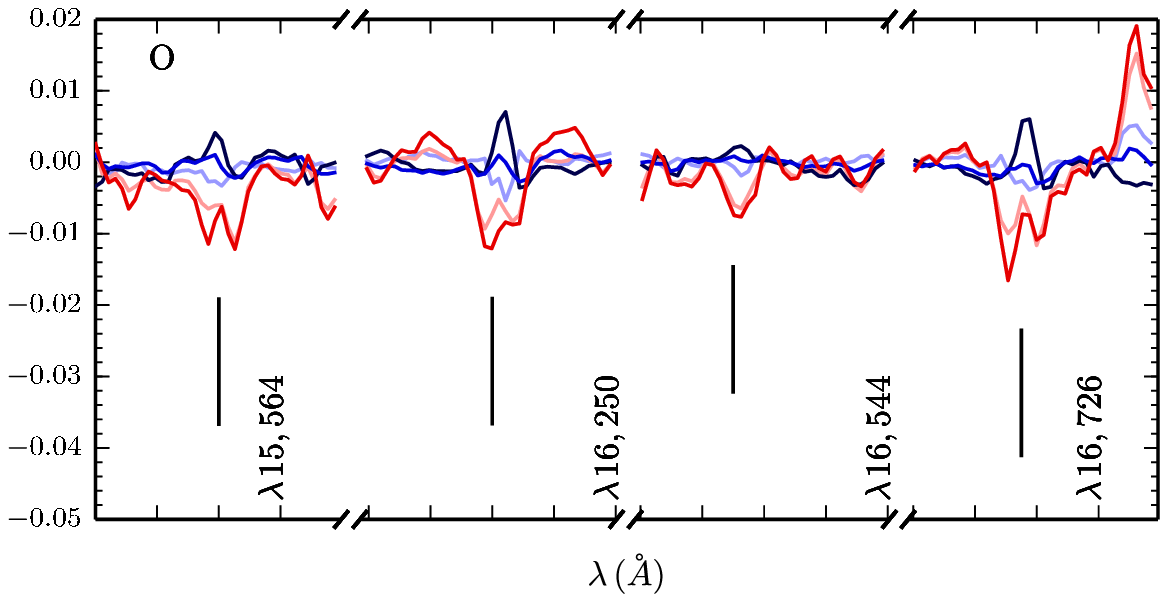}
\includegraphics[height=0.2\textheight,clip=]{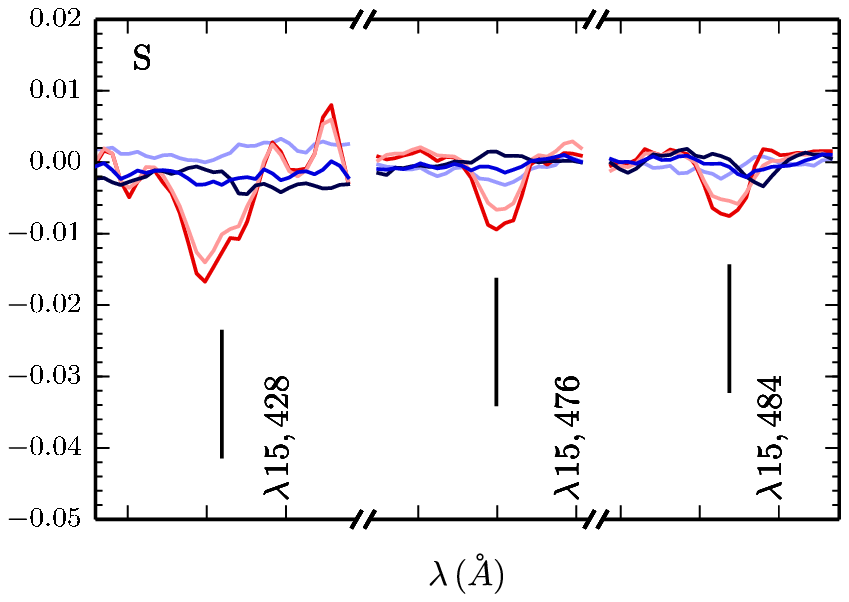}\\
\includegraphics[height=0.2\textheight,clip=]{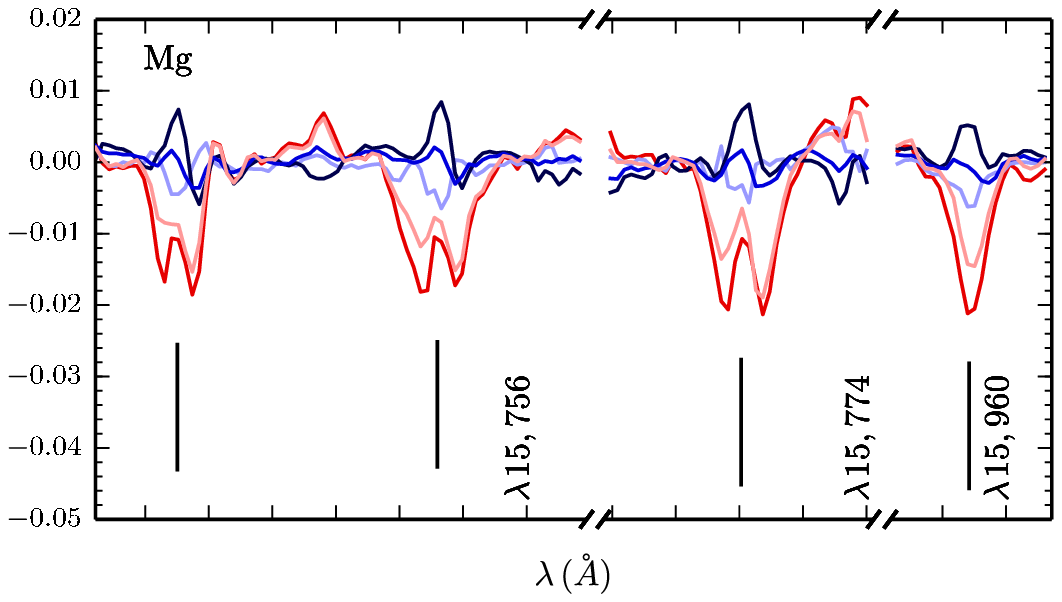}
\includegraphics[height=0.2\textheight,clip=]{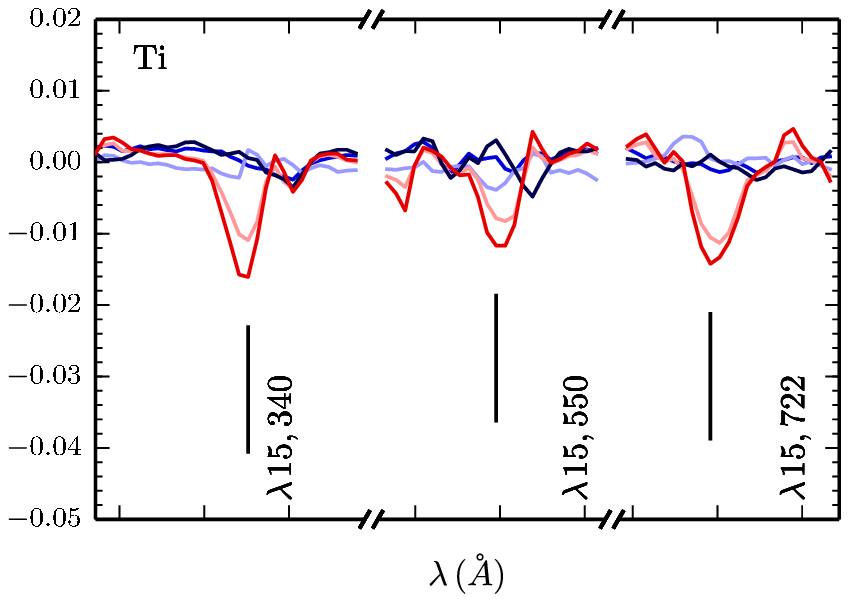}\\
\includegraphics[height=0.2\textheight,clip=]{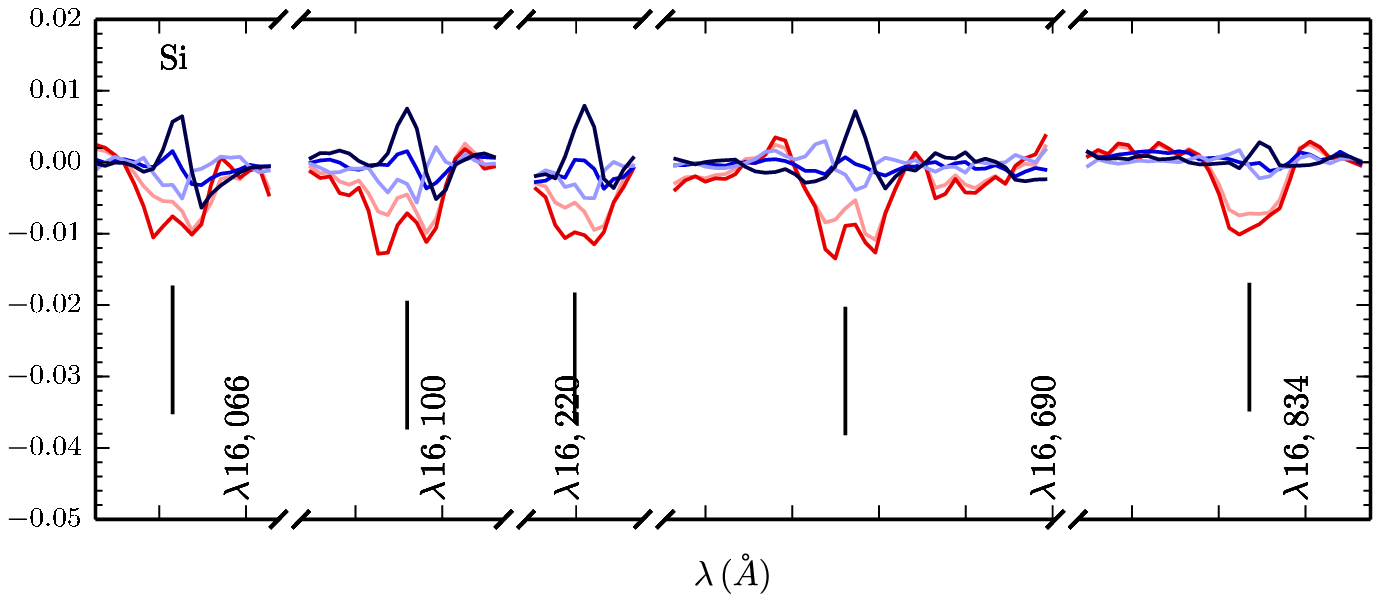}
\includegraphics[height=0.2\textheight,clip=]{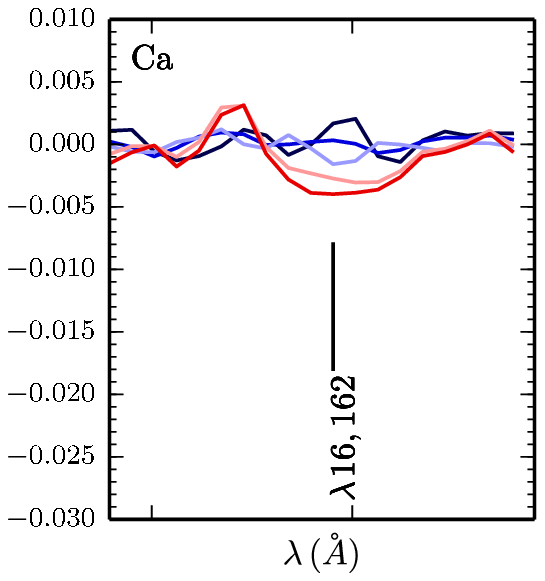}
\end{center}
\caption{Directly apparent signatures of \afe\ variations among
  APOGEE-RC stars, split by individual alpha-elements. This figure
  illustrates spectral differences in wavelength regions with strong
  features due to O, Mg, Si, S, Ca, and Ti for stars over a narrow
  range in \feh\ ($-0.45 < \feh \leq -0.35$). Each panel displays
  residuals from a mean spectrum after interpolating each spectral
  pixel to a common \teff, \logg, and \feh\ using quadratic
  interpolation. Residuals are smoothed by only including the part of
  the spectrum contained in the sub-space spanned by the eigenvectors
  of the eight largest PCA components of the residuals of all 490
  stars in this \feh\ range; this sub-space contains all of the
  variance above the measurement noise (we only include spectra with
  $S/N > 200$ in this figure). A stack of twelve random residuals each
  in five bins in $\Delta \afe = \avgafe = 0.05$---which does not
  include Ti (see text)---ranging from $0.00$ (dark blue) to $0.20$
  (dark red) is displayed. The $x$ axis only covers small parts of the
  full APOGEE spectral range in each panel and is interrupted in most
  panels to focus on features due to a specific element; the
  wavelength of the reddest tickmark in each section is indicated and
  the tickmark spacing is $2\,\AA$ everywhere. The location of the
  spectral features for each element are indicated in each panel. Note
  that the scale for the Ca panel is different due to the weakness of
  the Ca feature. This figure demonstrates the high precision in
  \afe\ and the high level of consistency between the abundances of
  different $\alpha$ elements obtained from the $H$-band APOGEE
  spectra: $\sigma_{\afe} \leq 0.02\dex$.}\label{fig:spectra}
\end{figure*}

The outline of this paper is as follows. In
\sectionname~\ref{sec:data} we present and discuss important aspects
of the data from the APOGEE-RC catalog that this analysis is based on,
paying particular attention to the abundance measurements in
\sectionname~\ref{sec:data-abu}. We describe the method for fitting
the density profiles of stellar subsamples in APOGEE in
\sectionname~\ref{sec:method}; this method is presented in more detail
in \citet{BovySF}. We apply the density-fitting methodology to broad
abundance-selected subsamples in \sectionname~\ref{sec:broad} to
explore the density profiles that represent the data well. In
\sectionname~\ref{sec:maps} we then determine the density profiles of
MAPs. We compare our results to previous work and discuss implications
for our understanding of the formation and evolution of the MW disk in
\sectionname~\ref{sec:discussion}. Finally, we present our conclusions
in \sectionname~\ref{sec:conclusion}. Readers not interested in the
details of the data and fitting methodology are encouraged to skip to
\sectionname~\ref{sec:broad}. Throughout this paper, we assume that
the Sun's displacement from the mid-plane is 25 pc toward the North
Galactic Pole \citep{Chen01a,Juric08a}, and that the Sun is located at
8 kpc from the Galactic center.

\section{Data}\label{sec:data}

\subsection{APOGEE and the APOGEE-RC catalog}

The SDSS-III/\apogee\ \citep{Majewski15a} is a high-resolution
($R\approx 22,500$) spectroscopic survey in the near-infrared (NIR;
$H$-band; 1.51 to 1.70 $\mu$m). The survey employs a 300-fiber
spectrograph (\citealt{Wilson10a}, J.~Wilson \etal\ 2016, in
preparation) to obtain signal-to-noise ratio of $100$ per
half-resolution element ($\approx141$ per resolution element) spectra
for large numbers of stars in the Milky Way, observed during bright
time on the 2.5-meter Sloan Foundation telescope \citep{Gunn06a}. The
majority of APOGEE targets are selected from the 2MASS point-source
catalog \citep{Skrutskie06a}, using a dereddened $\jk \geq 0.5$ color
cut in up to three magnitude bins in $H$ (not corrected for
extinction), with reddening corrections determined using the Rayleigh
Jeans Color Excess method \citep[RJCE;][]{Majewski11a} applied to
2MASS and mid-IR data from Spitzer-IRAC GLIMPSE-I, -II, and -3D
\citep{Churchwell09a} when available and from WISE \citep{Wright10a}
otherwise. In this paper, we employ various three-dimensional
extinction maps to take the effects of extinction on the observed
$H$-band magnitude into account, but we always use the RJCE
dereddenings for the $J-\ks$ color. A direct comparison between the 3D
extinction maps and the RJCE extinctions exhibits good agreement
between the two (see \citealt{BovySF}). Full details on the APOGEE
target selection can be found in \citet{Zasowski13a}. We use data from
the SDSS-III Data Release 12 \citep[DR12;][]{Alam15a}. The data
processing \citep{Nidever15a}, stellar-parameters and
chemical-abundances analysis (\citealt{Shetrone15a};
\citealt{Zamora15a}; \citealt{GarciaPerez15a}, and the DR12 data
analysis and calibration \citep{Holtzman15a} are performed with
automated SDSS software; we make use of standard data products as
contained in DR12.

\begin{figure}[t!]
\includegraphics[width=0.48\textwidth,clip=]{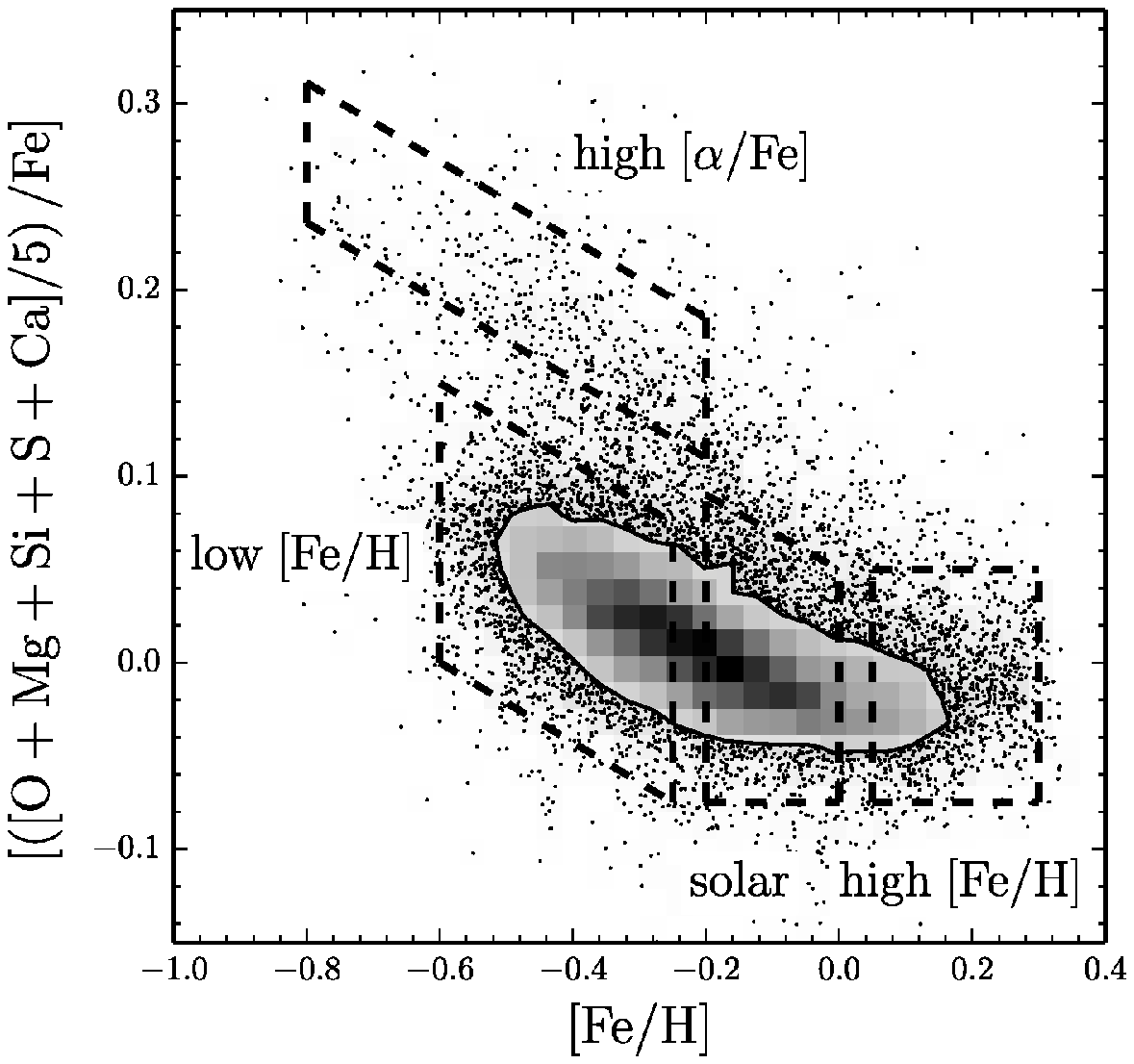}
\caption{Distribution of the \nstars\ stars in the APOGEE-RC sample
  used in this paper in the plane defined by the iron abundance, \feh,
  and the average abundance of $\alpha$ elements (see text). A linear
  binned density representation is used for the 68\,\% of the
  distribution that is contained within the shown contour. The dashed
  lines delineate the boundaries of the four broad subsamples that we
  study in \sectionname~\ref{sec:broad}, which we denote with the
  given moniker.}\label{fig:afefeh}
\end{figure}

Specifically, we base our analysis on red-clump (RC) star data from
the DR12 APOGEE-RC catalog. This catalog consists of a pure sample of
RC stars selected from the APOGEE catalog using the method described
in detail by \citet{BovyRC}, but applied to the DR12 data (see
\citealt{Alam15a}). RC stars are identified in the superset of all
APOGEE data using a combination of cuts in the surface-gravity (\logg)
/ effective-temperature (\teff) plane and the $(J-\ks)_0$ /
metallicity (\feh) plane, chosen such as to select those RC stars for
which precise luminosity distances can be determined. The catalog has
an estimated purity of $\approx 95\,\%$; the distances are precise to
$5\,\%$ and unbiased to $2\,\%$. In order to account for biases due to
the APOGEE target selection, we only use the subset of RC data that
(a) was selected as part of the ``main'' APOGEE sample (defined using
the $\jksq \geq 0.5$ color cut), and (b) are part of a line of sight
and magnitude bin combination for which the planned APOGEE
observations were complete in DR12, because these are the stars for
which we can reconstruct the sample selection function. In our
density-fitting methodology (see \sectionname~\ref{sec:method} below)
we correct for the effects of interstellar extinction using the 3D
extinction maps of \citet{Marshall06a} in the inner-disk plane and
\citet{Green15a} elsewhere (using $A_H/A_{K_s} = 1.48$ and $A_H/E(B-V)
= 0.46$; \citealt{Schlafly11a}; \citealt{Yuan13a}); the latter is
based on Pan-STARRS and 2MASS data and does not cover the entire
APOGEE footprint. We therefore remove 10 APOGEE fields that lie
outside of the \citet{Green15a} footprint. These are the fields
centered on $(l,b) = (240,-18)$, $(5.5,-14.2)$, $(5.2,-12.2)$,
$(1,-4)$, $(0,-5)$, $(0,-2)$, $(358,0)$, $(358.6,1.4)$ (difficult to
access from the Northern Hemisphere) and $(l,b) = (120,30)$,
$(123,22.4)$. This only removes 125 stars. We further remove 13 stars
with distance moduli based on their $H$-band luminosity (see below) $<
8.49$, because they could not be in our sample if we model the RC as a
standard candle with $M_H = -1.49$ (our standard assumption
below). This statistical sample contains \nstars\ RC stars. This
sample spans a range $4500\Kunit \lesssim \teff \lesssim 5200\Kunit$
and $2.25 \lesssim \log g \lesssim 3$ ($95\,\%$ ranges).

\figurename~\ref{fig:fields} displays the sky distribution of the
lines of sight included in the statistical RC sample, demonstrating
the excellent coverage of the Galactic plane at low latitude. The
fields are overlayed on the extinction map to a distance of $D =
5\kpc$ from \citet{Green15a} to illustrate the large amount of
extinction affecting the disk region. While the distribution of fields
includes a large number of fields in the region of the sky that
contains the Galactic bulge, observations in these fields were limited
to $H < 11$. Because the RC has $M_H \approx -1.49$, this implies that
the bulge is not actually included in our sample (these fields have $D
\lesssim 3\kpc$); the coverage is primarily of the disk.

\begin{figure*}[t!]
\includegraphics[height=0.195\textheight,clip=]{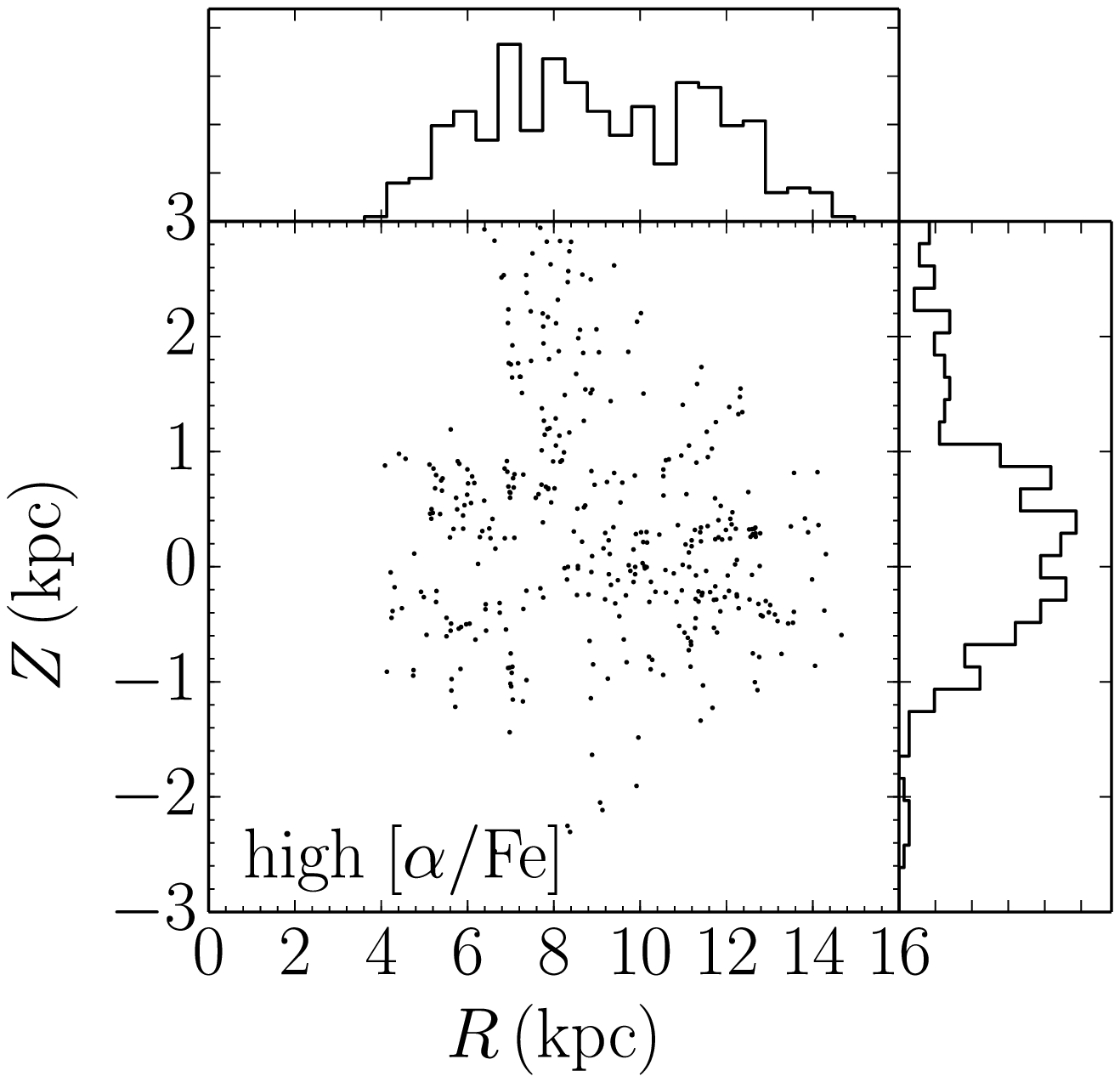}
\includegraphics[height=0.195\textheight,clip=]{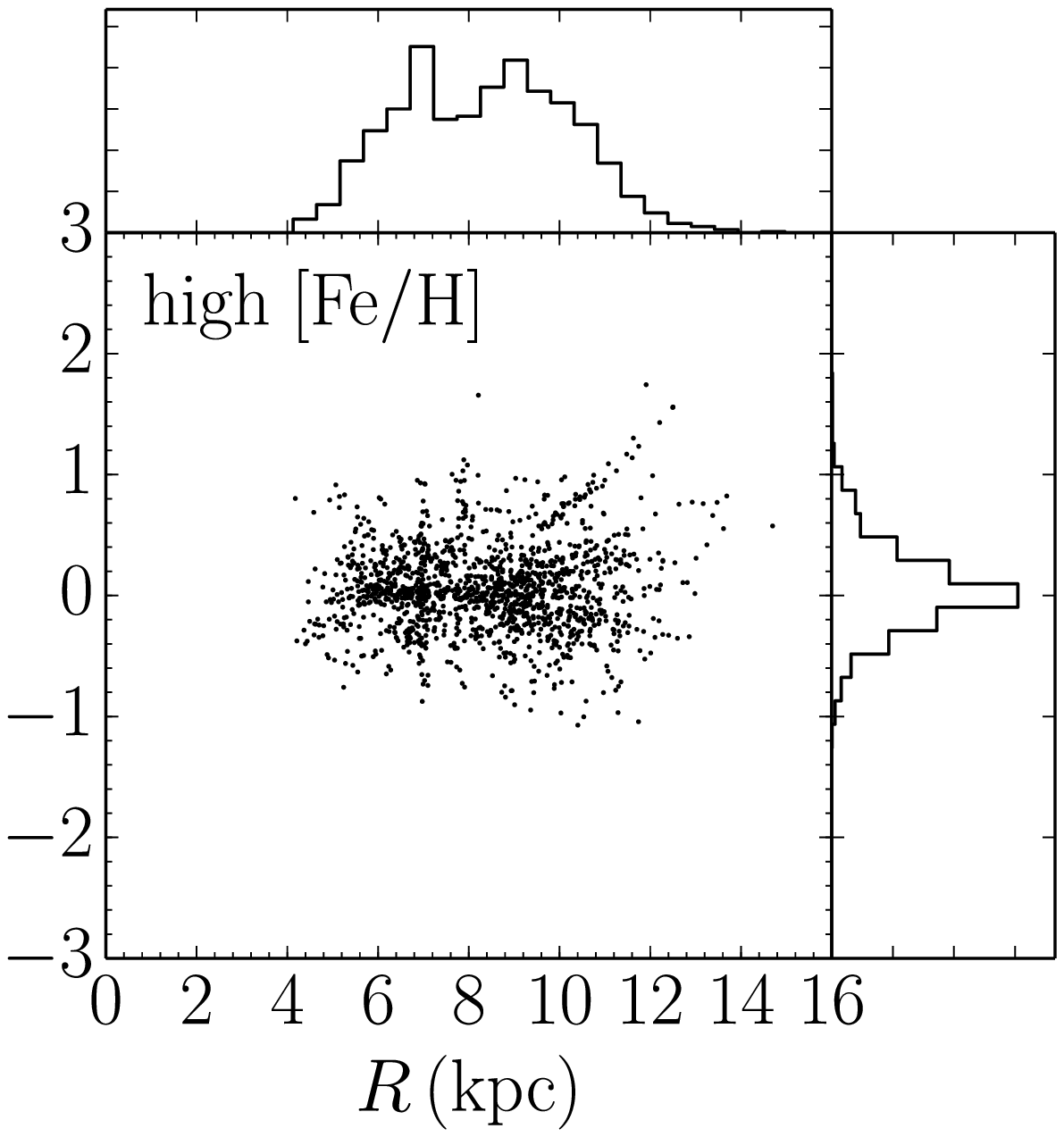}
\includegraphics[height=0.195\textheight,clip=]{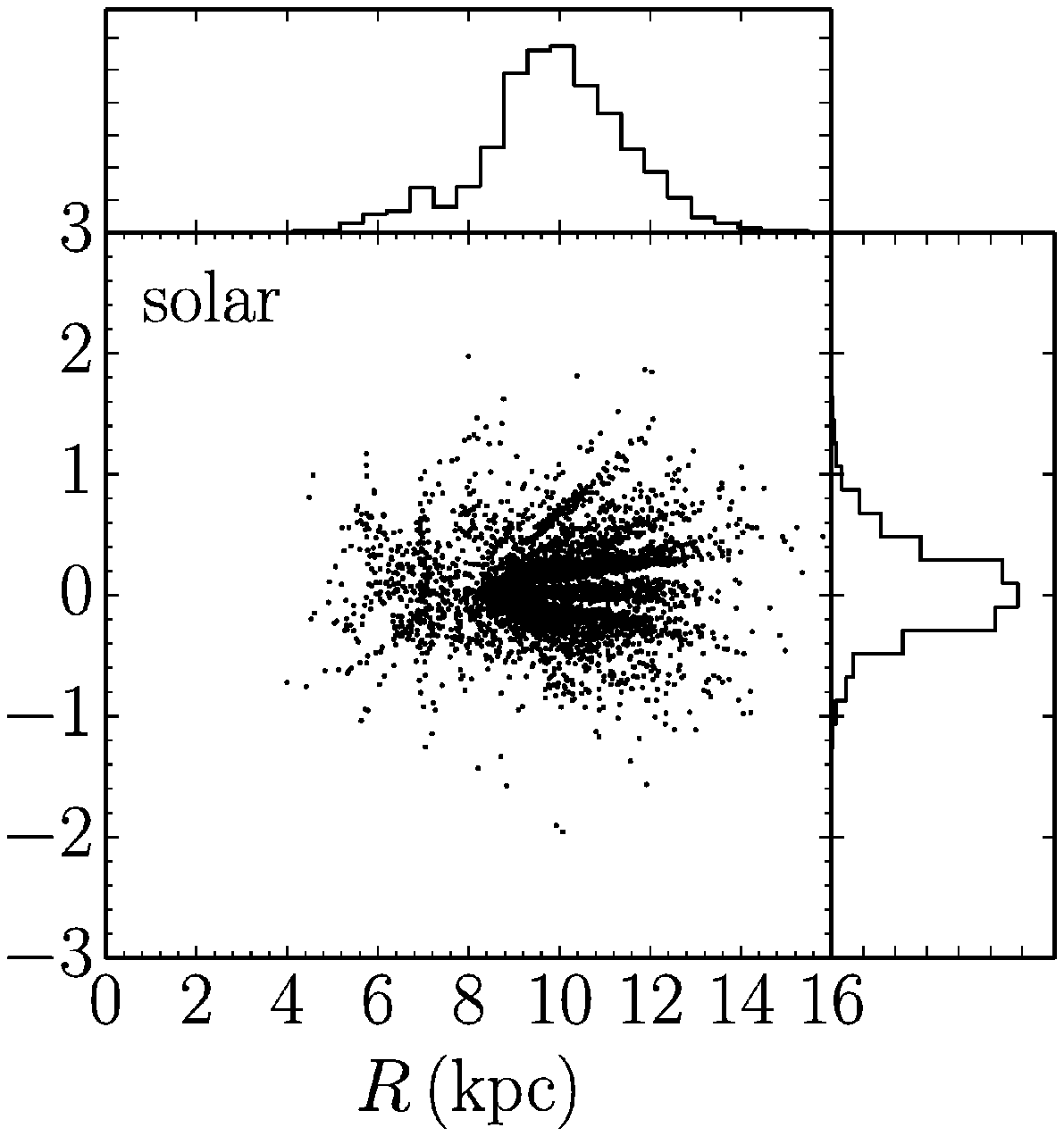}
\includegraphics[height=0.195\textheight,clip=]{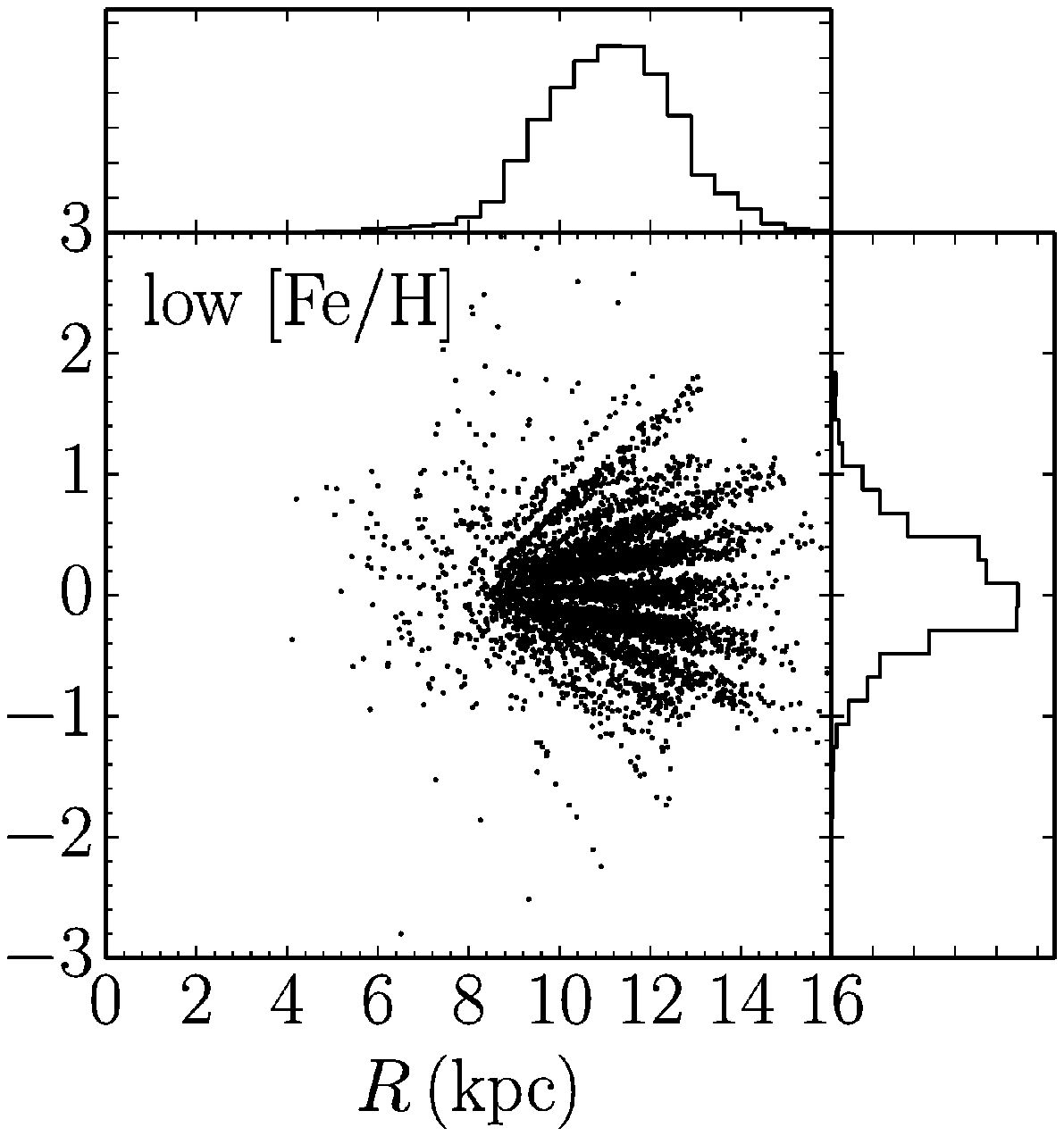}
\caption{Observed spatial distribution of stars in the four broad
  subsamples defined in \figurename~\ref{fig:afefeh}, in
  Galactocentric radius, $R$, and distance from the mid-plane,
  $Z$. The Sun is located at $(R,Z) =
  (8,0.025)\kpc$.}\label{fig:spatial-broad}
\end{figure*}

As discussed in detail in \sectionname~5 of \citet{BovyRC}, the
overall-APOGEE and RC-specific selection cuts used to define the
APOGEE-RC sample excludes stars with ages $\lesssim 800\,\mathrm{Myr}$
and metallicities $\feh \lesssim-0.9$. The youngest part of the disk,
as well as components more metal-poor than the bulk of the disk
\citep[\eg,][]{Carollo10a} are therefore not represented in the
sample; however, most of the disk mass satisfies these
constraints. Because the RC consists of evolved giants stars that
spend $\approx100\,\mathrm{Myr}$ of their lifetime in the RC phase,
the population of RC stars is a biased tracer of the underlying
stellar populations in the Milky Way. The extent of this bias is
primarily dependent on the metallicity and the (unknown) age
distribution, but the dominant effect is that the APOGEE-RC selection
overrepresents stars in the $1$ to $4\Gyr$ age range with respect to
the underlying age distribution. For example, for a relatively
constant star-formation history, the APOGEE-RC selection is expected
to be dominated by stars with ages between $1$ and $4\Gyr$, although
older ages are represented at a lower level.

Distances for stars in the APOGEE-RC catalog are determined by
combining corrections to a single absolute magnitude $M_{\ks}$, as a
function of $([J-\ks]_0,\feh)$, based on stellar-evolution models
\citep{Bressan12a}; these corrections are applied to each individual
star. The overall distance scale is calibrated against an
\emph{Hipparcos}-based determination of the RC absolute $\ks$
magnitude in the solar neighborhood \citep{Laney12a}. In this paper,
we determine distances based on the $H$-band luminosity rather than
that in the $\ks$-band, because the APOGEE sample selection is
performed in apparent $H$-band magnitude and our density-fitting
formalism is simplest in this case. We follow the same procedure as
discussed in \sectionname\sectionname~2 and 3 of \citet{BovyRC} to
compute the corrections to a single $H$-band luminosity for the RC, as
a function of $([J-\ks]_0,\feh)$\footnote{We convert the metallicity
  \feh\ to metal mass fraction $Z$ assuming $Z_\odot = 0.017$ and
  solar abundance ratios.}, and to calibrate the distance scale to the
\emph{Hipparcos} value of $M_H = -1.49$. The absolute magnitude
$M_H(\jksq,\feh)$ is displayed in
\figurename~\ref{fig:rcmag}. \figurename~\ref{fig:rcdistcomp} displays
the fractional difference between the distances for stars in our
sample determined from their \ks\ magnitude (the standard APOGEE-RC
catalog product) and from their $H$ magnitude. The fractional
differences is $\lesssim3\,\%$ for all stars, with a median of
$2.3\,\%$. This is similar to the overall accuracy of the RC distance
scale and so small that it does not impact the density fits below.

\subsection{Abundance measurements and main subsamples\label{sec:data-abu}}

We make use of the abundance measurements provided in the DR12 release
of APOGEE data \citep{Holtzman15a}. Specifically, we use the iron
content, \feh, and the average abundance of $\alpha$ elements relative
to iron. \feh\ is measured from 64 iron lines in the $H$-band. As
discussed by \citet{Holtzman15a}, \feh\ is internally corrected for
trends with \teff\ for stars in globular and open clusters observed by
APOGEE. This internal calibration only amounts to differences of
$\approx 0.04\dex$ over the $\approx700\Kunit$ spanned by the RC. We
determine the precision in \feh\ by comparing measurements of
\feh\ for different stars in the open clusters of \citet{Meszaros13a}
in the \teff\ range of the RC and find that $\sigma_{\feh} =
0.05\dex$.

We define the average $\alpha$-enhancement as an average of the
abundance of O, Mg, Si, S, and Ca. We do not include Ti, because of
issues with its measurement in the $H$ band (see \citealt{Holtzman15a}
for further discussion). Specifically, we average the abundances of
$[\mathrm{O/H}]$, $[\mathrm{Mg/H}]$, $[\mathrm{Si/H}]$,
$[\mathrm{S/H}]$, and $[\mathrm{Ca/H}]$ and subtract \feh\ to obtain
the average $\alpha$-enhancement \afe. If no measurement was obtained
for one of the five $\alpha$ elements, it is removed from the
average. $[\mathrm{O/H}]$ is measured from the abundant molecular OH
and CO features in the near-infrared that are, however, relatively
weak for the warm RC stars. Mg, Si, S, and Ca are measured from
neutral atomic lines for two (Ca) to 12 (Si) features. The abundances
of $\alpha$ elements are similarly corrected for trends with \teff\ in
clusters as \feh\ above; for the average \afe\ as defined here, this
calibration only gives differences of $0.03\dex$ over the
$\approx700\Kunit$ spanned by the RC. We determine the empirical
precision in \afe\ using the scatter in \afe\ for the calibration open
clusters described above, and find that $\sigma_{\afe} \approx
0.02\dex$, with a correlation of $-0.4$ with \feh.

We apply a simple external calibration of \feh\ and \afe\ as
follows. Using the APOGEE catalog abundances we determine the average
\feh\ and \afe\ for giants observed by APOGEE in the temperature range
of the RC ($4500\Kunit < \teff < 5200\Kunit$; see above) in M67, an
open cluster that provides an excellent chemical solar analog
\citep[\eg,][]{Onehag14a}. We find that $\feh_{\mathrm{M67}} =
0.10\dex$ and $\afe_{\mathrm{M67}} = 0.03\dex$; the large offset in
$\feh_{\mathrm{M67}}$ is at least partly due to an incorrect line list
used in DR12 \citep[see][]{Shetrone15a}. To calibrate APOGEE closer to
the solar abundance scale using M67, we apply constant offsets of
$-0.1\dex$ in \feh\ and $-0.05\dex$ in \afe. The reason for applying
an offset of $-0.05\dex$ rather than $-0.03\dex$ in \afe\ is that we
define MAPs below using bins with $\Delta \afe = 0.05\dex$; a
calibration offset of $-0.05\dex$ therefore does not change the
binning of the data, which is the same when using the catalog
abundances or the externally-calibrated abundances defined here.

\figurename~\ref{fig:spectra} demonstrates the precision of our
\afe\ abundances. This figure considers RC stars with $S/N > 200$ with
$-0.35 < \feh \leq -0.25$. We interpolate the spectra for the 490
stars in our sample with these properties onto a common $\teff =
4800\Kunit$, $\logg = 2.85$ and $\feh = -0.3$, using quadratic
interpolation. This removes the dominant effects of these stellar
properties to reveal the effect of $\afe$ on the spectra. We perform a
PCA analysis of the residuals from a mean spectrum after the
interpolation taking the measurement uncertainties into account
\citep{Bailey12a}, and only retain the components of each spectrum
corresponding to the eight PCA eigenvectors with the largest
eigenvalues. These eight eigenvectors explain all of the variance in
the residuals above the measurement
noise. \figurename~\ref{fig:spectra} displays median-stacked residuals
of twelve random stars each in five bins in \afe\ with $\Delta \afe =
0.05\dex$ ranging from $\afe = 0.00$ to $\afe = 0.20$. We perform this
median stacking of a small number of spectra to reduce the noise for
display purposes. It is clear from this figure that we measure
\afe\ at very high precision from the spectral lines of O, Mg, Si, S,
and Ca. The relative abundances of these elements with respect to each
other appear to have little variation.

\figurename~\ref{fig:afefeh} displays the distribution of the
statistical APOGEE-RC sampling in $(\feh,\afe)$. This distribution
contains the two main sequences---high- and low-\afe\---seen in other
investigations of this distribution and discussed in detail for the
APOGEE sample by \citet{Nidever14a} and \citet{Hayden15a}. Also
indicated in this figure are the four broad abundance-selected
subsamples that we consider in \sectionname~\ref{sec:broad}. The
spatial distribution of stars in these four subsamples is displayed in
\figurename~\ref{fig:spatial-broad}. With due caution about the effect
of selection biases (which we correct for in the remainder of this
paper) it is clear that the low-\afe, low-\feh\ stars are primarily
found in the outer disk, while the higher-\feh\ low-\afe\ stars are
found closer to the center. High-\afe\ stars are found throughout the
observed volume and are, in particular, more numerous at large
distances from the mid-plane.

In \sectionname~\ref{sec:maps}, we consider MAPs, defined in the same
manner as in \citet{BovyMAPstructure} as stars in bins with
$\Delta\feh = 0.1\dex$ and $\Delta \afe = 0.05\dex$. From the
discussion of the uncertainties above, it is evident that the
contamination between neighboring MAPs is slight, and that that between
non-neighboring bins is essentially non-existent.

\section{Density-fitting methodology}\label{sec:method}

\subsection{Generalities}

To fit the spatial density profile for subsamples of RC stars, we
follow the methodology of \citet{BovyMAPstructure} (see also
\citealt{Rix13a}), who model the observed rate of stars in the joint
parameter space of position, magnitude, color, and metallicity using a
Poisson process. Best-fit parameters and their uncertainties for
parameterized spatial profiles are obtained by sampling this Poisson
process' likelihood of the observed data multiplied with an
uninformative prior. As discussed by \citet{BovyRC} and above, for the
RC stars we determine the absolute magnitude $M_H$ as a function of
color \jk\ and metallicity \feh; therefore, almost the same
methodology for fitting the spatial profiles of RC stars applies here
as was used by \citet{BovyMAPstructure} to analyze G dwarfs (whose
absolute magnitude $M_r$ was similarly determined from the color
$(g-r)_0$ and \feh).

In the present application, we therefore model the rate function
$\lambda(O|\theta)$ that is a function of $O = (l,b,D,H,\jksq,\feh)$,
and is parameterized by parameters $\theta$; $\lambda(O|\theta)$ is
given by
\begin{equation}
\begin{split}
  \lambda(O|\theta) & = \dens(X,Y,Z|\theta)\times|J(X,Y,Z;l,b,D)|\\
  & \ \times\rho(H,\jksq,\feh|X,Y,Z)\times S(l,b,H)\,,
\end{split}
\end{equation}
where $\dens(\cdot|\theta)$ is the spatial density in Galactocentric
rectangular coordinates $(X,Y,Z)$ that we are ultimately most
interested in and that depends on parameters $\theta$,
$|J(X,Y,Z;l,b,D)|$ is the Jacobian of the transformation between
$(X,Y,Z)$ and $(l,b,D)$, $\rho(H,\jksq,\feh|X,Y,Z)$ is the density of
stars in magnitude--color--metallicity space, and $S(l,b,H)$ is the
survey selection function (the fraction of stars from the underlying
population of potential targets observed by the survey). In APOGEE,
the survey selection function is a function of position on the sky, is
constant with $(l,b)$ within an APOGEE field, and is a
piecewise-constant function of apparent $H$-band magnitude, because
targets were sampled in magnitude bins. The APOGEE selection function
is determined, tested, and discussed in detail in \sectionname~4 of
\citet{BovyRC}. For the present work, we have updated the selection
function to the full three-year data set presented in DR12.

As in \citet{BovyMAPstructure}, the rate has an additional amplitude
parameter. To remove this parameter from further consideration, we
marginalize the probability of the parameters of the rate function
over the amplitude of the rate. The marginalized likelihood can be
written as
\begin{equation}\label{eq:like}
  \mathcal{L}(\theta) = \sum_i\left[\ln \dens(X_i,Y_i,Z_i|\theta)-\ln \int \dd O \lambda(O|\theta)\right]\,.
\end{equation}
In this expression, we have made use of the fact that the rate
$\lambda(O_i|\theta)$ only depends on $\theta$ through
$\dens(\cdot|\theta)$ and therefore the $\dens(X_i,Y_i,Z_i|\theta)$ is
the only factor in $\lambda(O_i|\theta)$ that depends on $\theta$; the
other factors can be dropped. The integral in this equation is the
effective volume of the survey that provides the normalization of the
rate likelihood. It does not depend on the individual data point, but
is instead a property of the whole survey for a given model specified
by $\theta$.

\Equationname~(\ref{eq:like}) is similar to the equivalent likelihood
in \citet{BovyMAPstructure} (their \equationname~[8]). The only
significant difference between the two expressions is that the APOGEE
selection function depends on the apparent magnitude $H$ that is
\emph{not} corrected for extinction, while the SEGUE selection was
performed in extinction-corrected magnitude. Therefore, to calculate
the normalization integral in \equationname~(\ref{eq:like}) we require
a model for the three-dimensional distribution of extinction $A_H$ to
convert the model prediction's $H_0$ to $H$. We discuss the
methodology for efficiently calculating the effective volume for
different types of surveys in \citet{BovySF}. For a pencil-beam survey
like APOGEE, for which we can assume that the density is constant over
the area of each field, the effective volume can be efficiently
computed as
\begin{equation}
\begin{split}
  \int \dd O \, & \lambda(O|\theta) =\\
  & \sum_{\mathrm{fields}} \Omega_f\,\int \dd D\,D^2\,\dens([X,Y,Z](D,\field)|\theta)\\
  & \qquad \qquad \qquad \essf(\field,D)\,,
\end{split}
\end{equation}
where $\essf(\field,D)$ is the \emph{effective selection
  function}. For APOGEE, this is given by
\begin{equation}
\begin{split}
  & \essf(\field,D) = \sum_k S(\field,k)\\
  & \ \ \,\frac{\Omega(H_{\mathrm{min},k}-H_0(D) < A_H(l,b,D) < H_{\mathrm{max},k}-H_0(D))}{\Omega_f}\,,
\end{split}
\end{equation}
where the sum is over the different magnitude bins,
$[H_{\mathrm{min},k},H_{\mathrm{max},k}]$, that stars are selected in
along each line of sight. The numerator $\Omega(H_{\mathrm{min},k}-H_0
< A_H(D) < H_{\mathrm{max},k}-H_0)$ is the area of the APOGEE field in
question, with $A_H$ between the given boundaries, and the denominator
$\Omega_f$ is the total area of the field. $S(\field,k)$ is the APOGEE
survey selection function, \ie, the fraction of potential targets with
APOGEE spectroscopic observations in each magnitude bin. This equation
assumes that the RC is a standard candle with $M_H = -1.49$, allowing
us to compute $H_0(D)$. \citet{BovySF} demonstrate that this
assumption does not affect the density measurements here.

\subsection{Stellar Number-density Models\label{sec:models}}

\tabletypesize{\tiny}
\begin{deluxetable*}{lllccccccc}
\tablecaption{Results for broad abundance-selected samples}
\tablecolumns{10}
\tablewidth{0pt}
\tablehead{\colhead{Sample} & \colhead{Density model} &
  \colhead{Extinction map} & \colhead{$h_{R,\mathrm{in}}^{-1}$} & 
  \colhead{$h_{R,\mathrm{out}}^{-1}$} & \colhead{\Rb} & 
  \colhead{$h_Z$} & \colhead{$\Rf$ or $\beta_2$} &
  \colhead{$h_{Z,2}$} & \colhead{$\Delta \chi^2$}\\
  \colhead{} & \colhead{} & \colhead{}
  & \colhead{(kpc$^{-1}$)} & \colhead{(kpc$^{-1}$)} & \colhead{(kpc)} & 
  \colhead{(kpc)} & \colhead{(kpc$^{-1}$ or -)} & \colhead{(kpc)} & \colhead{}}
\startdata
low [Fe/H] & broken exp. w/ flare & \citet{Marshall06a} & $0.27\pm0.03$&$0.36\pm0.11$&$10.8\pm0.1$&$0.37\pm0.02$&$-0.09\pm0.01$&\ldots&0\\
(low [$\alpha$/Fe]) &  & \citet{Green15a} & 0.28&0.37&10.8&0.38&-0.08&\ldots&13\\
 &  & \citet{Sale14a} & 0.28&0.37&10.8&0.38&-0.08&\ldots&91\\
 &  & \citet{Drimmel03a} & 0.27&0.45&10.9&0.37&-0.10&\ldots&649\\
 &  & zero & 0.34&0.59&10.9&0.39&-0.10&\ldots&2162\\
 & single exp. & \citet{Marshall06a} & \ldots&$0.06\pm0.01$&\ldots&$0.45\pm0.01$&\ldots&&723\\
 & broken exp. w/ 2 $h_Z$ & \citet{Marshall06a} & $0.28\pm0.03$&$0.37\pm0.02$&$10.8\pm0.1$&$0.47\pm0.01$&$<0.01$&$0.19\pm97.44$&74\\

\\
solar & broken exp. w/ flare & \citet{Marshall06a} & $0.09\pm0.04$&$0.65\pm0.02$&$9.2\pm0.1$&$0.28\pm0.01$&$-0.09\pm0.01$&\ldots&0\\
(low [$\alpha$/Fe]) &  & \citet{Green15a} & 0.15&0.65&9.2&0.29&-0.08&\ldots&77\\
 &  & \citet{Sale14a} & 0.15&0.65&9.2&0.29&-0.08&\ldots&149\\
 &  & \citet{Drimmel03a} & 0.07&0.71&9.4&0.29&-0.09&\ldots&286\\
 &  & zero & 0.29&0.78&9.3&0.32&-0.07&\ldots&1443\\
 & single exp. & \citet{Marshall06a} & \ldots&$0.33\pm0.01$&\ldots&$0.31\pm0.01$&\ldots&&906\\
 & broken exp. w/ 2 $h_Z$ & \citet{Marshall06a} & $0.11\pm0.03$&$0.66\pm0.02$&$9.2\pm0.1$&$0.33\pm0.01$&$<0.01$&$0.11\pm29.94$&90\\

\\
high [Fe/H] & broken exp. w/ flare & \citet{Marshall06a} & $0.28\pm0.15$&$0.81\pm0.03$&$6.8\pm0.2$&$0.27\pm0.01$&$-0.14\pm0.02$&\ldots&0\\
(low [$\alpha$/Fe]) &  & \citet{Green15a} & 0.62&0.79&6.5&0.28&-0.12&\ldots&98\\
 &  & \citet{Sale14a} & 0.56&0.79&6.6&0.28&-0.12&\ldots&148\\
 &  & \citet{Drimmel03a} & 0.20&0.81&6.8&0.28&-0.14&\ldots&84\\
 &  & zero & 1.07&0.78&6.5&0.30&-0.10&\ldots&832\\
 & single exp. & \citet{Marshall06a} & \ldots&$0.60\pm0.01$&\ldots&$0.28\pm0.01$&\ldots&&430\\
 & broken exp. w/ 2 $h_Z$ & \citet{Marshall06a} & $0.43\pm0.32$&$0.80\pm0.08$&$6.6\pm1.7$&$0.28\pm0.01$&$<0.02$&$0.09\pm1.66$&95\\

\\
high [$\alpha$/Fe] & broken exp. w/ flare & \citet{Marshall06a} & $0.93\pm0.57$&$0.43\pm0.03$&$<4.4$&$0.95\pm0.05$&$0.01\pm0.03$&\ldots&0\\
 &  & \citet{Green15a} & 1.16&0.43&1.0&0.97&0.02&\ldots&6\\
 &  & \citet{Sale14a} & 0.95&0.43&1.8&0.97&0.02&\ldots&10\\
 &  & \citet{Drimmel03a} & 0.88&0.45&4.1&0.96&0.00&\ldots&14\\
 &  & zero & 0.00&0.40&1.1&1.09&0.04&\ldots&170\\
 & single exp. & \citet{Marshall06a} & \ldots&$0.43\pm0.02$&\ldots&$0.95\pm0.05$&\ldots&&0\\
 & broken exp. w/ 2 $h_Z$ & \citet{Marshall06a} & $0.90\pm0.57$&$0.43\pm0.02$&$2.2\pm1.2$&$0.95\pm0.06$&$<0.02$&$0.15\pm195.28$&0\\

\tablecomments{Lower limits are at 95\% posterior confidence. The
  seventh column is \Rf\ for the flaring model and the amplitude
  $\beta_2$ for the model with two vertical scale heights. The
  parameters $\beta_2$ and $h_{Z,2}$ are marginalized under the
  constraint that $h_{Z,2}$ is 50\,\% different from $h_Z$ (to avoid
  the massive degeneracy when $h_Z$ and $h_{Z,2}$ are allowed to be
  the same). The model consisting of a broken radial exponential with
  a flaring exponential scale height provides the best fit in all
  cases, although the high-\afe\ sample is always fit as a
  single radial exponential.}\label{table:results}
\end{deluxetable*}

\begin{figure*}[t!]
\includegraphics[width=0.2475\textwidth,clip=]{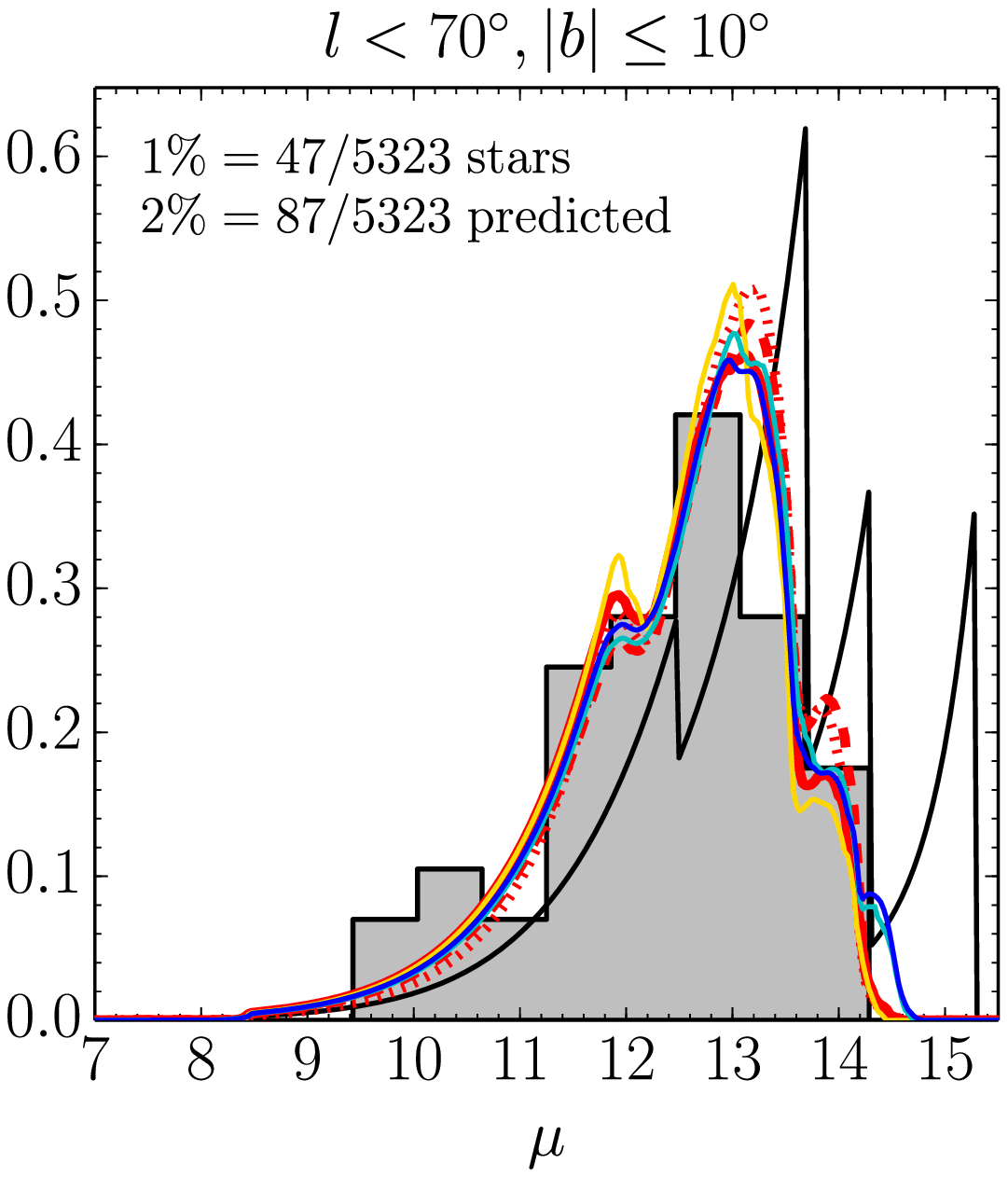}
\includegraphics[width=0.2475\textwidth,clip=]{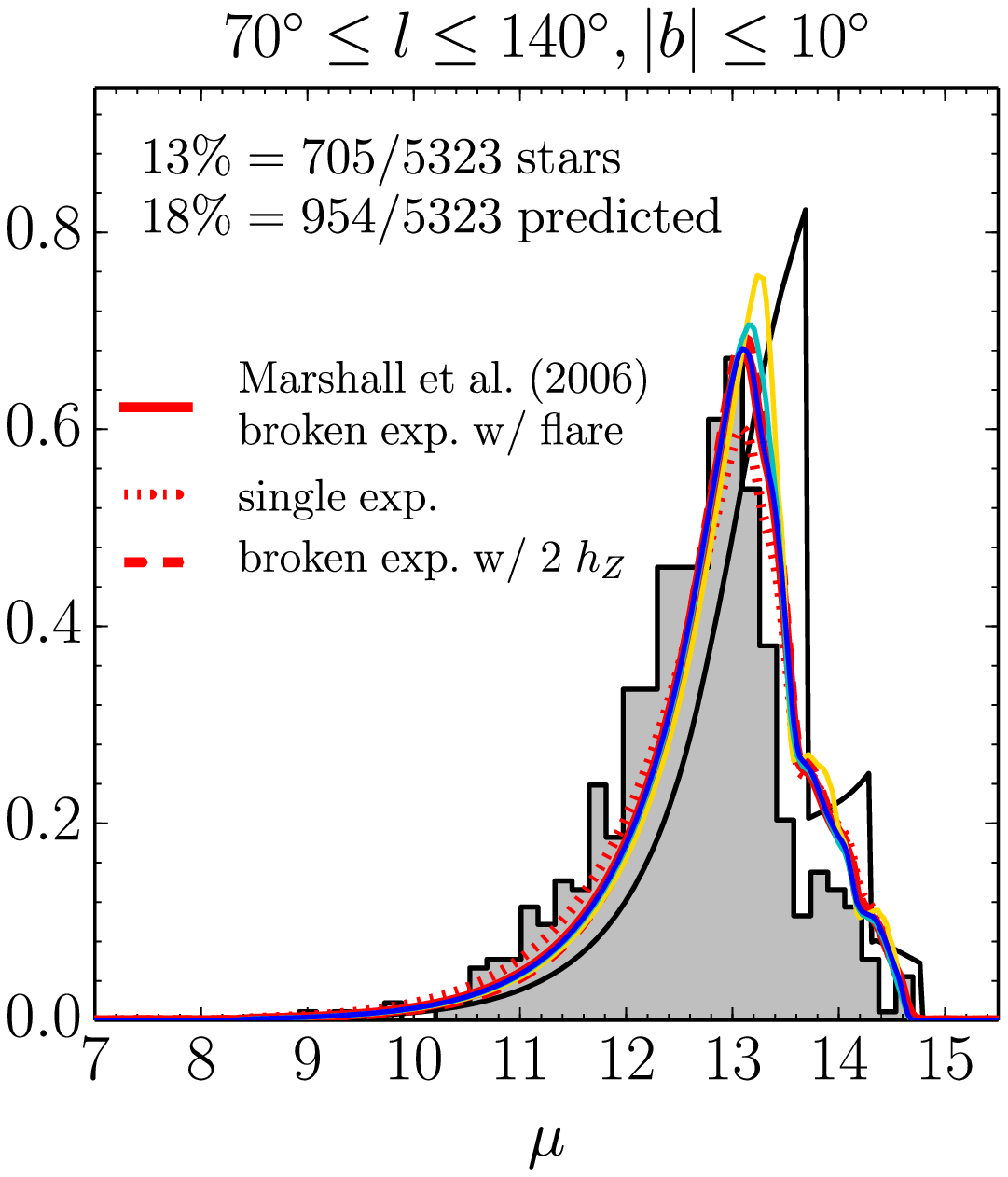}
\includegraphics[width=0.2475\textwidth,clip=]{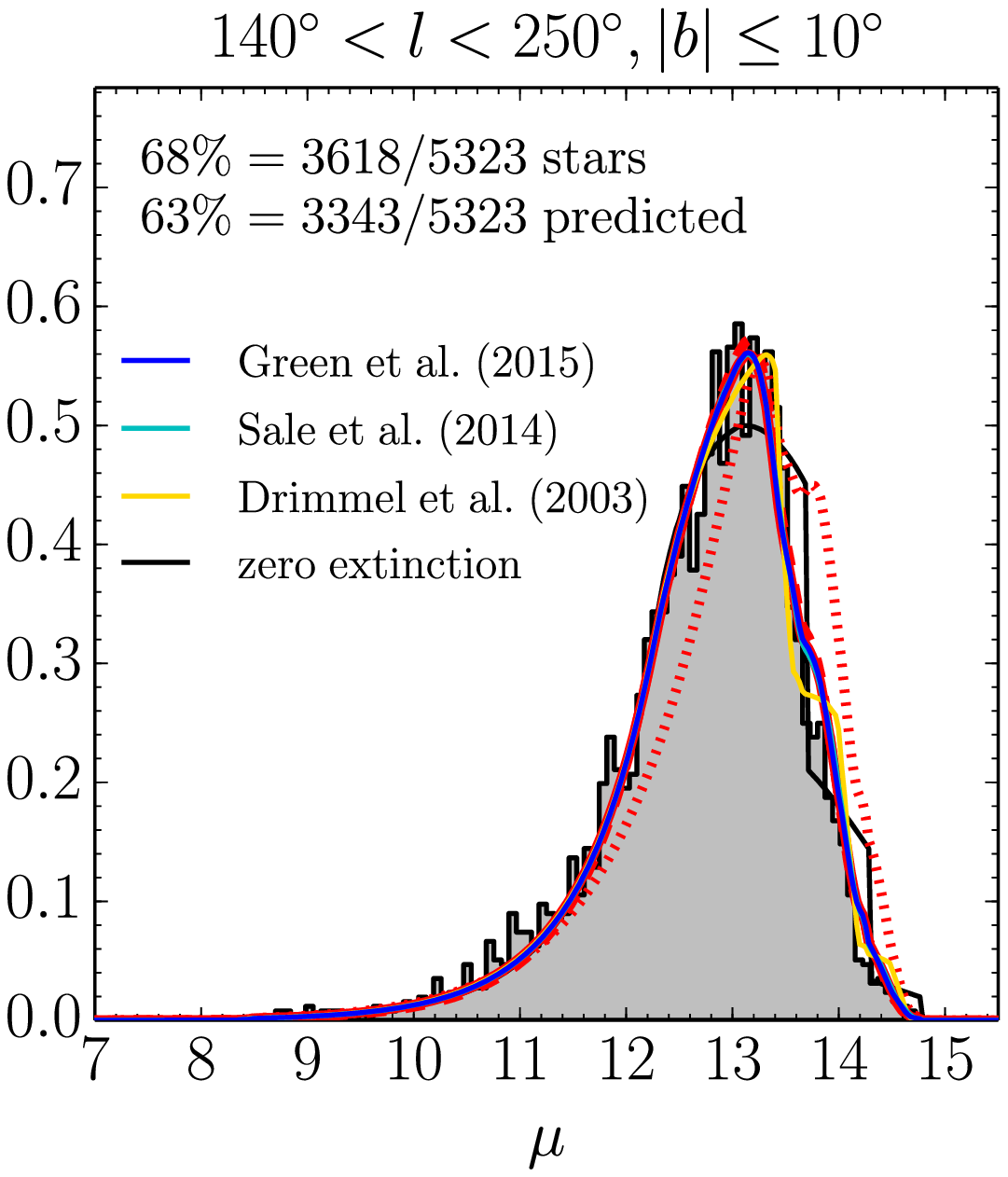}
\includegraphics[width=0.2475\textwidth,clip=]{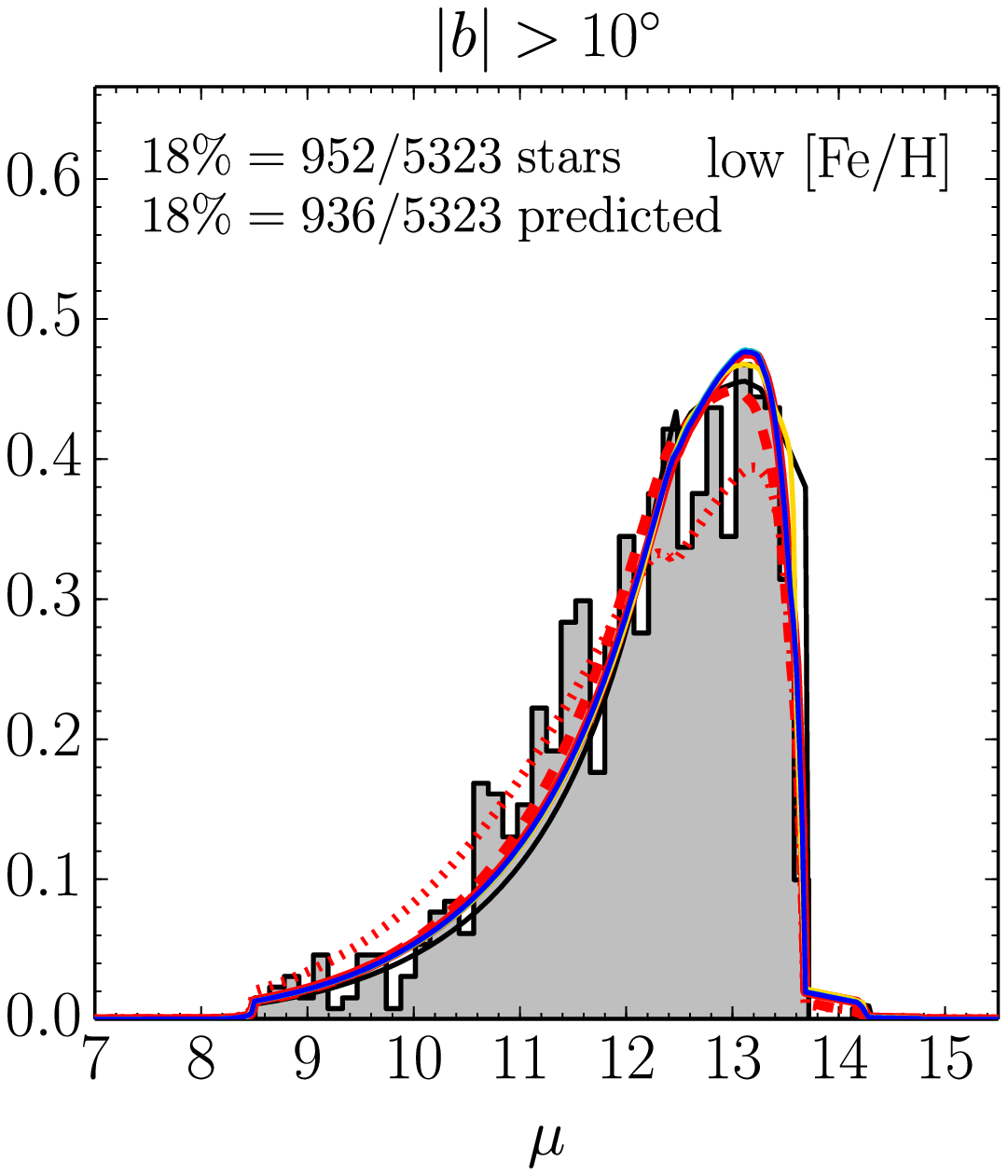}\\
\includegraphics[width=0.2475\textwidth,clip=]{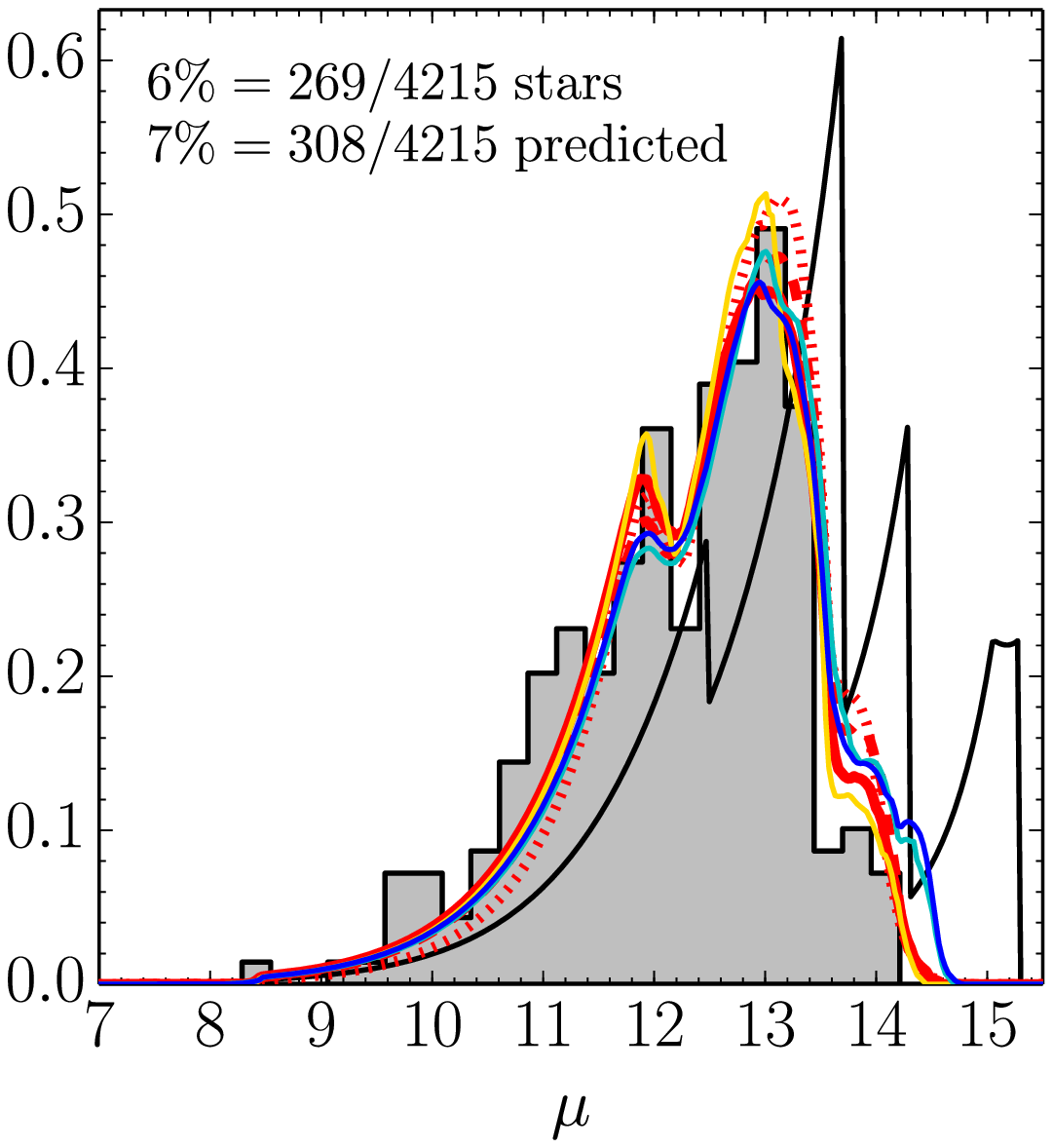}
\includegraphics[width=0.2475\textwidth,clip=]{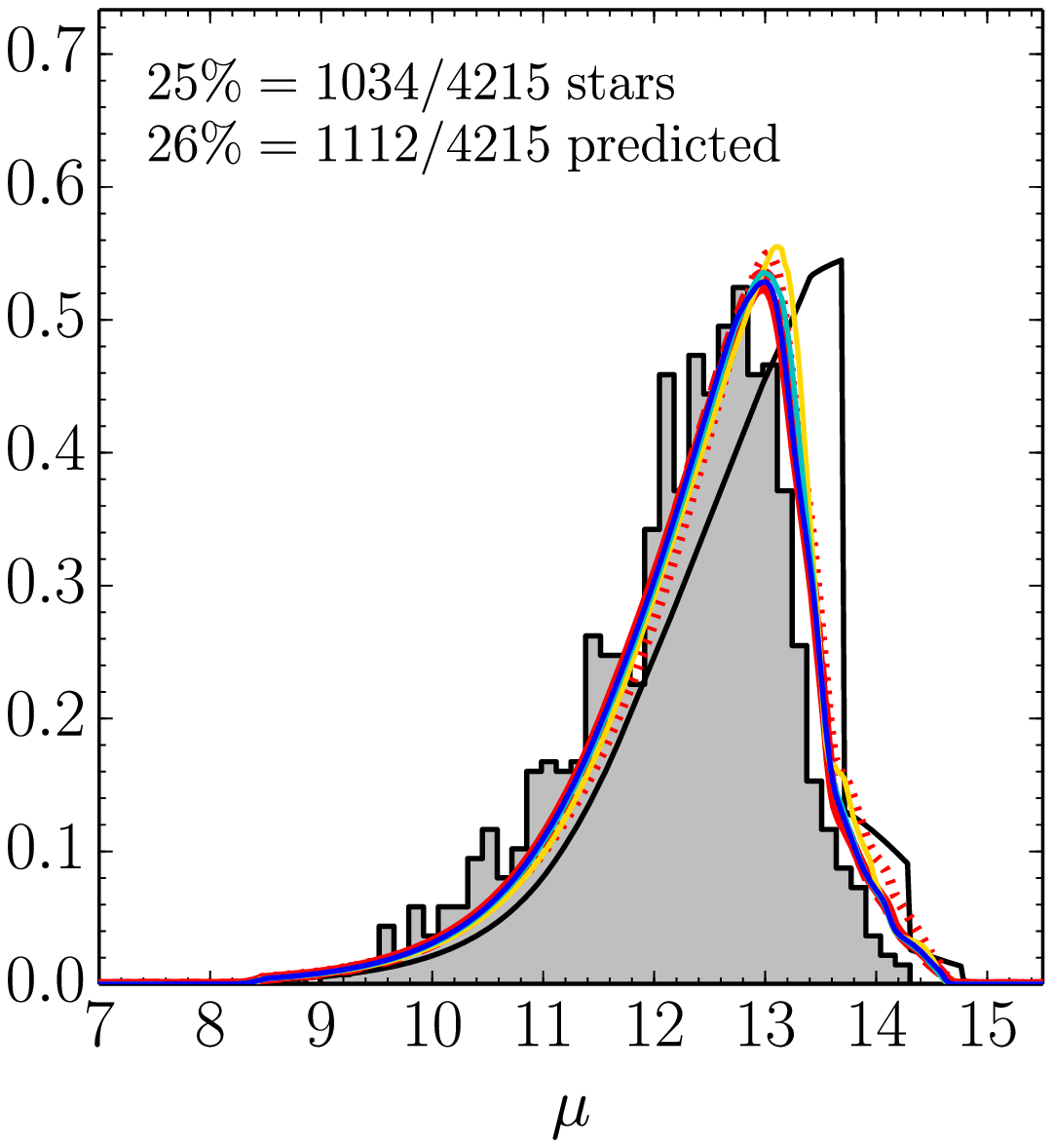}
\includegraphics[width=0.2475\textwidth,clip=]{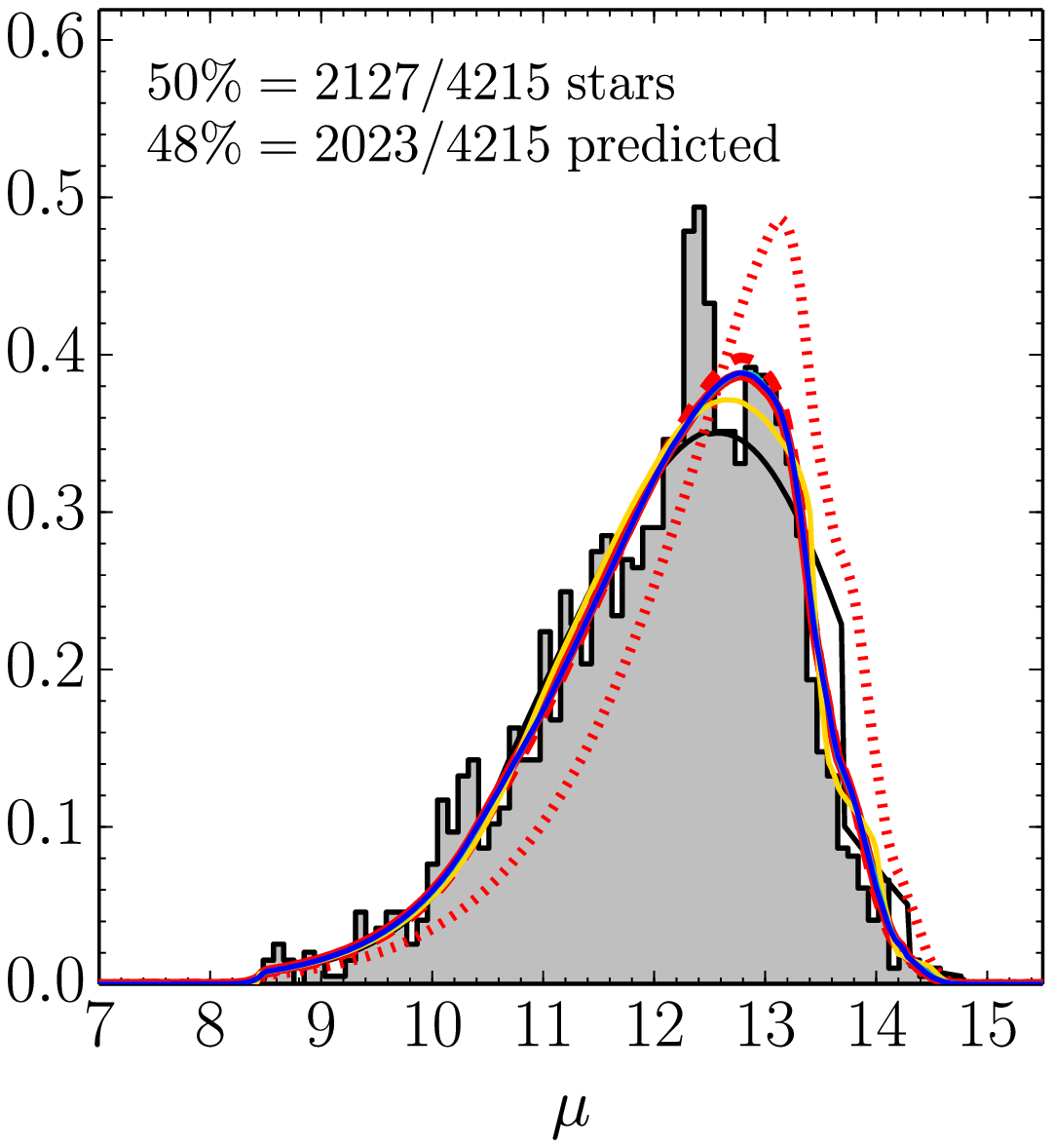}
\includegraphics[width=0.2475\textwidth,clip=]{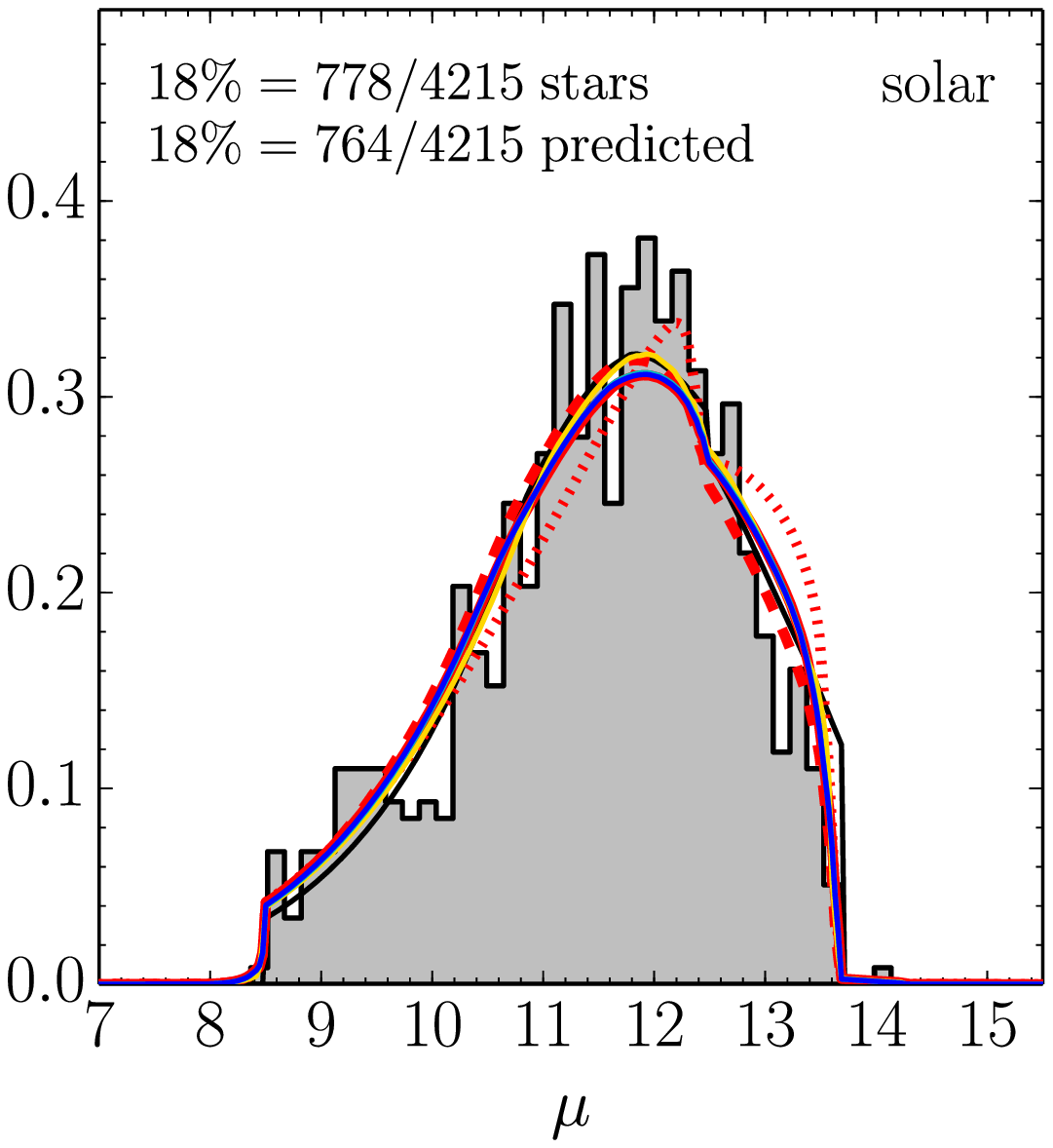}\\
\includegraphics[width=0.2475\textwidth,clip=]{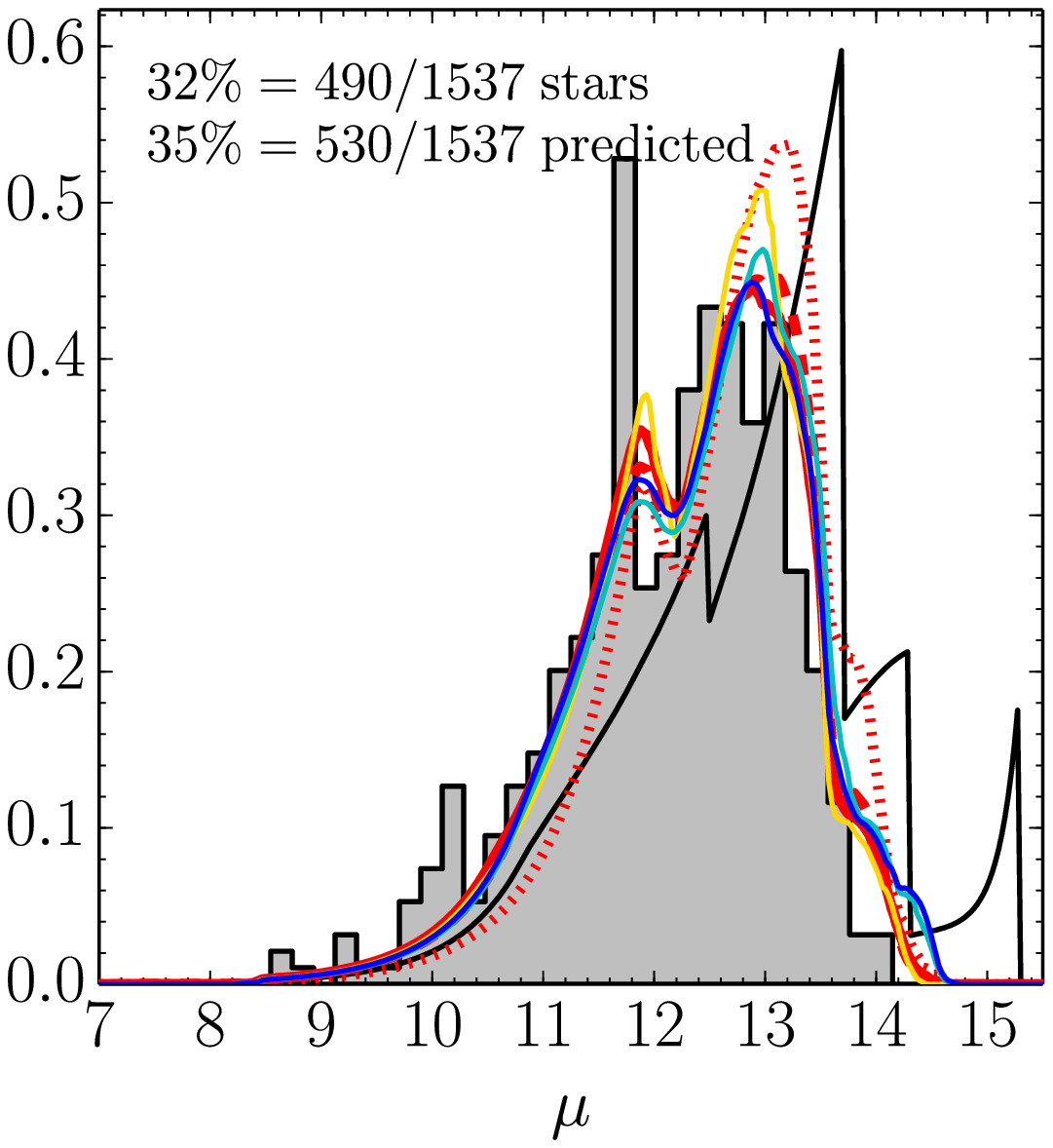}
\includegraphics[width=0.2475\textwidth,clip=]{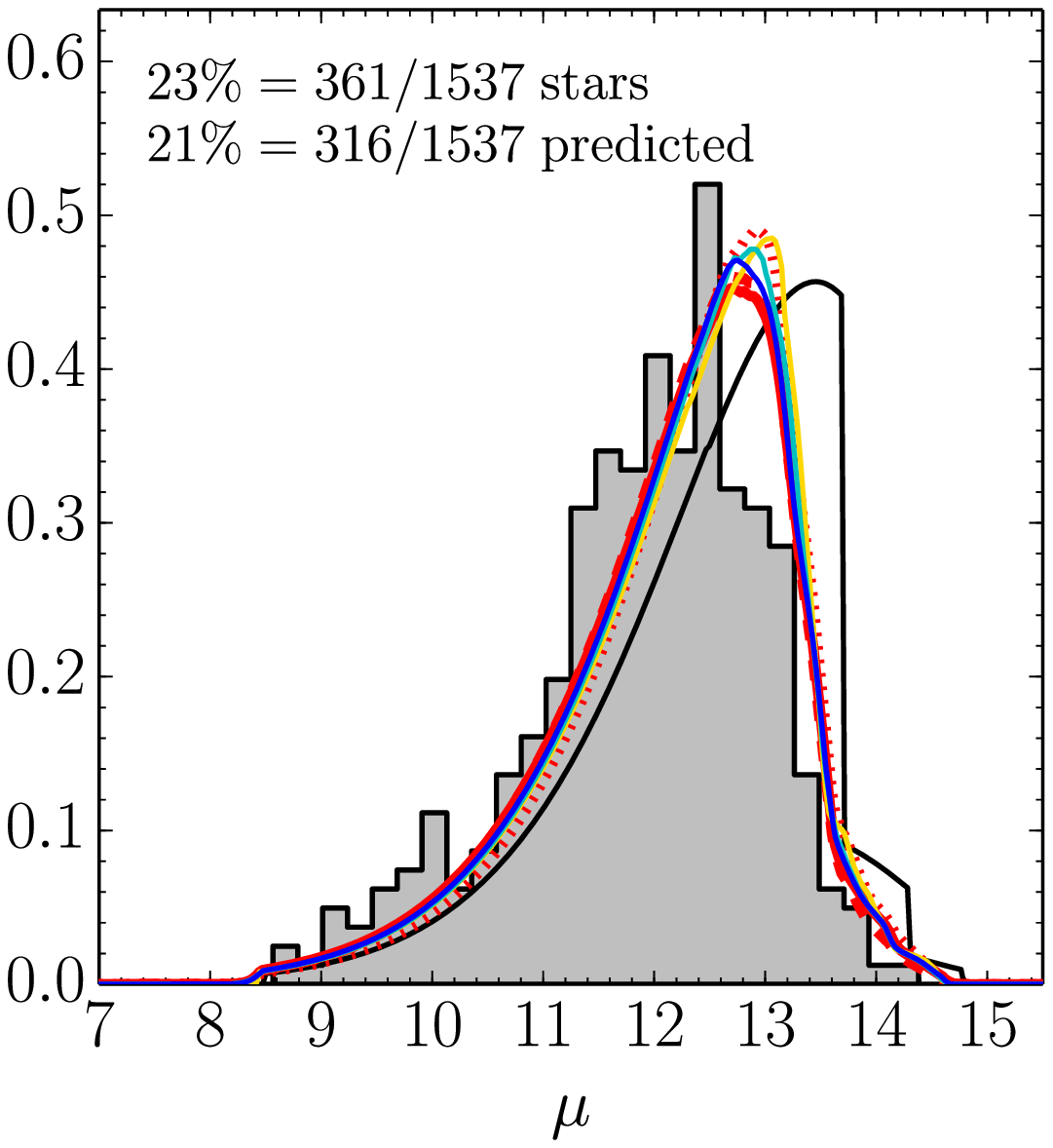}
\includegraphics[width=0.2475\textwidth,clip=]{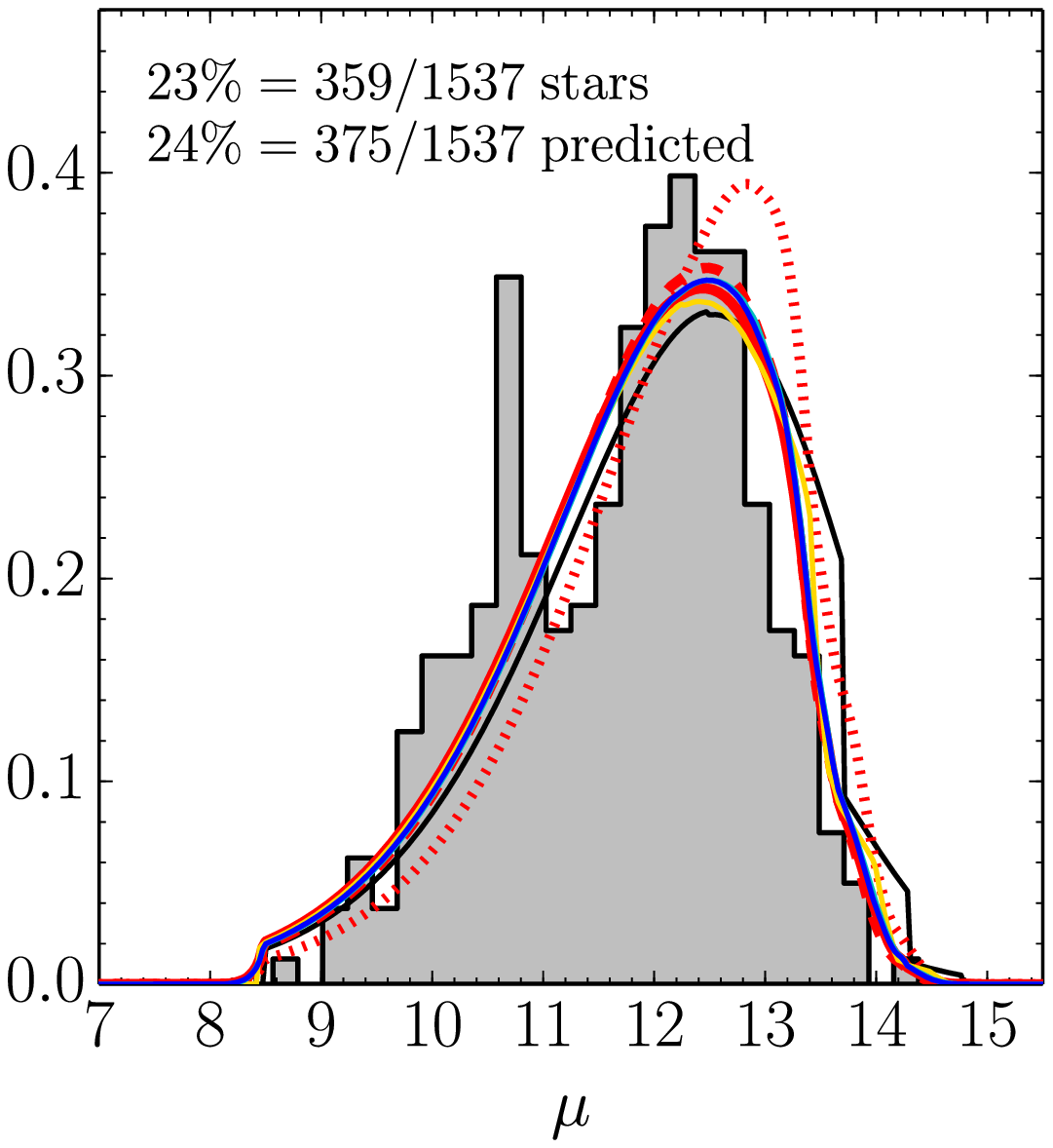}
\includegraphics[width=0.2475\textwidth,clip=]{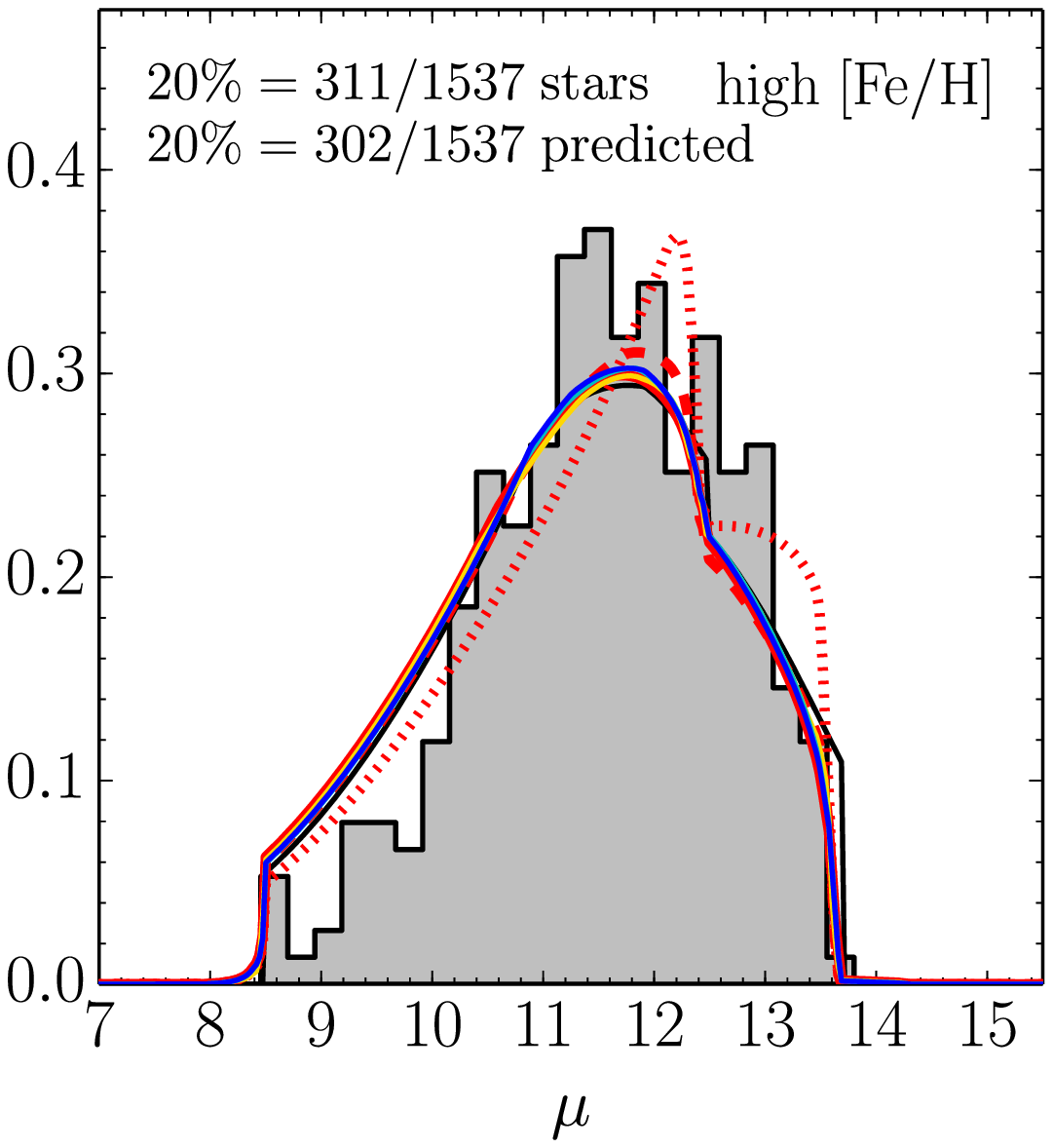}\\
\caption{Comparison between the observed (filled histogram)
  distribution of distance moduli, $\mu$, and that predicted by the
  best-fit models for the three low-\afe\ subsamples of
  \figurename~\ref{fig:afefeh} (rows) and different regions of the sky
  (columns). The red curves demonstrate the fits using a combination
  of the \citet{Marshall06a} and \citet{Green15a} 3D extinction maps
  for the different density profiles displayed in
  \tablename~\ref{table:results}. The remaining curves display the
  fits assuming different 3D extinction maps. The predicted number of
  stars in each spatial region is given for the fiducial model (broken
  exp. w/ flare with \citealt{Marshall06a} extinction). Different 3D
  extinction maps give by and large similar fits, except for in the
  inner Galaxy (left panels), where the \citet{Marshall06a} map
  performs best. A model with zero extinction (black curves) provides
  poor fits for all low Galactic latitude locations. The sharp
  features in the zero-extinction model reflect the discontinuous
  nature of the APOGEE selection function as a function of $H$; these
  are smoothed by the extinction for models with extinction (see
  \figurename~6 in \citealt{BovySF}). The density in all cases is best
  fit as a radially broken exponential with a flaring vertical scale
  height; a single radial exponential fails to explain the radial
  profile over all radii, demonstrated by the poor fit in the outer
  regions of the disk (third panels).}\label{fig:low-datacomp}
\end{figure*}

\begin{figure*}[t!]
\includegraphics[width=0.2475\textwidth,clip=]{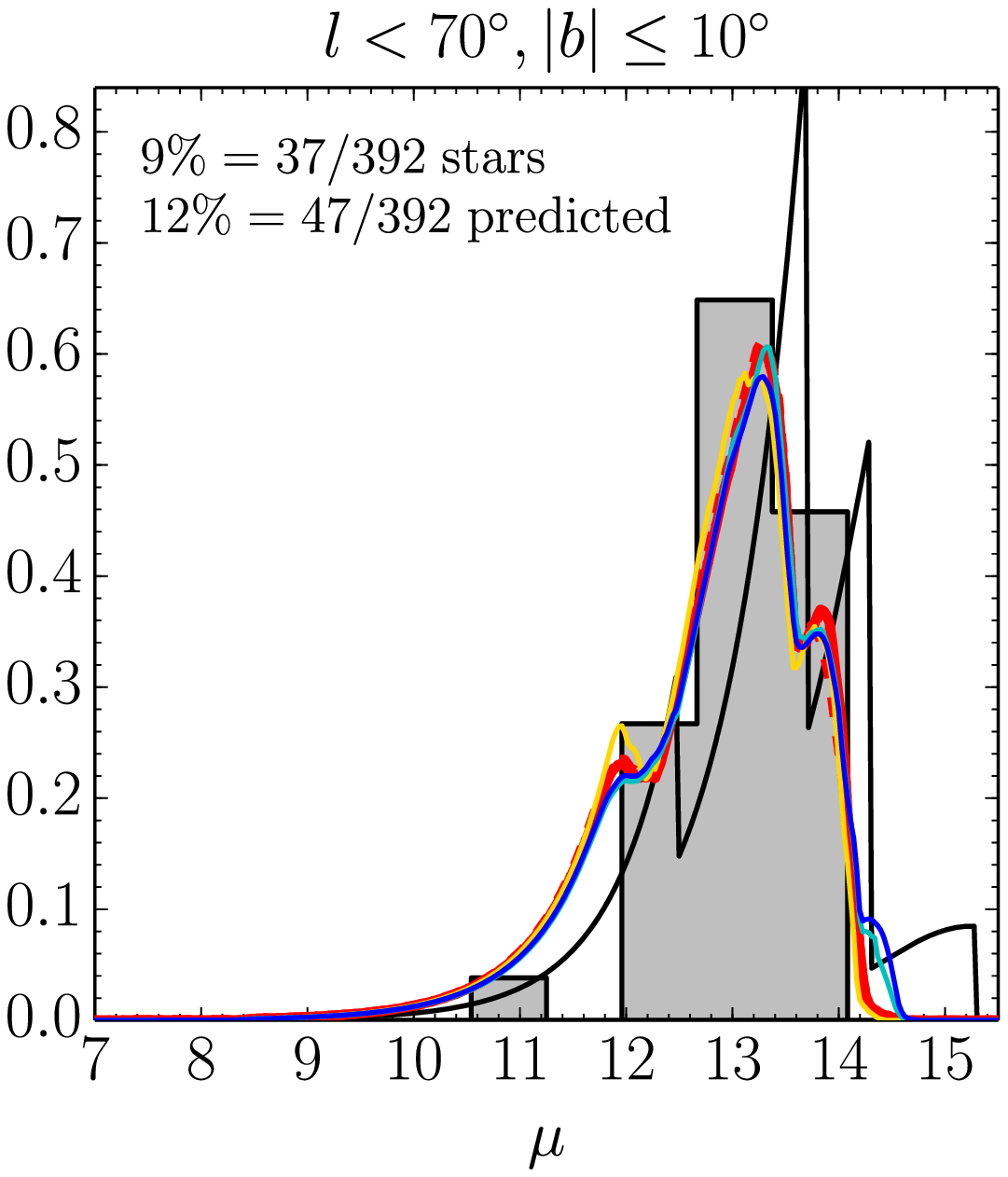}
\includegraphics[width=0.2475\textwidth,clip=]{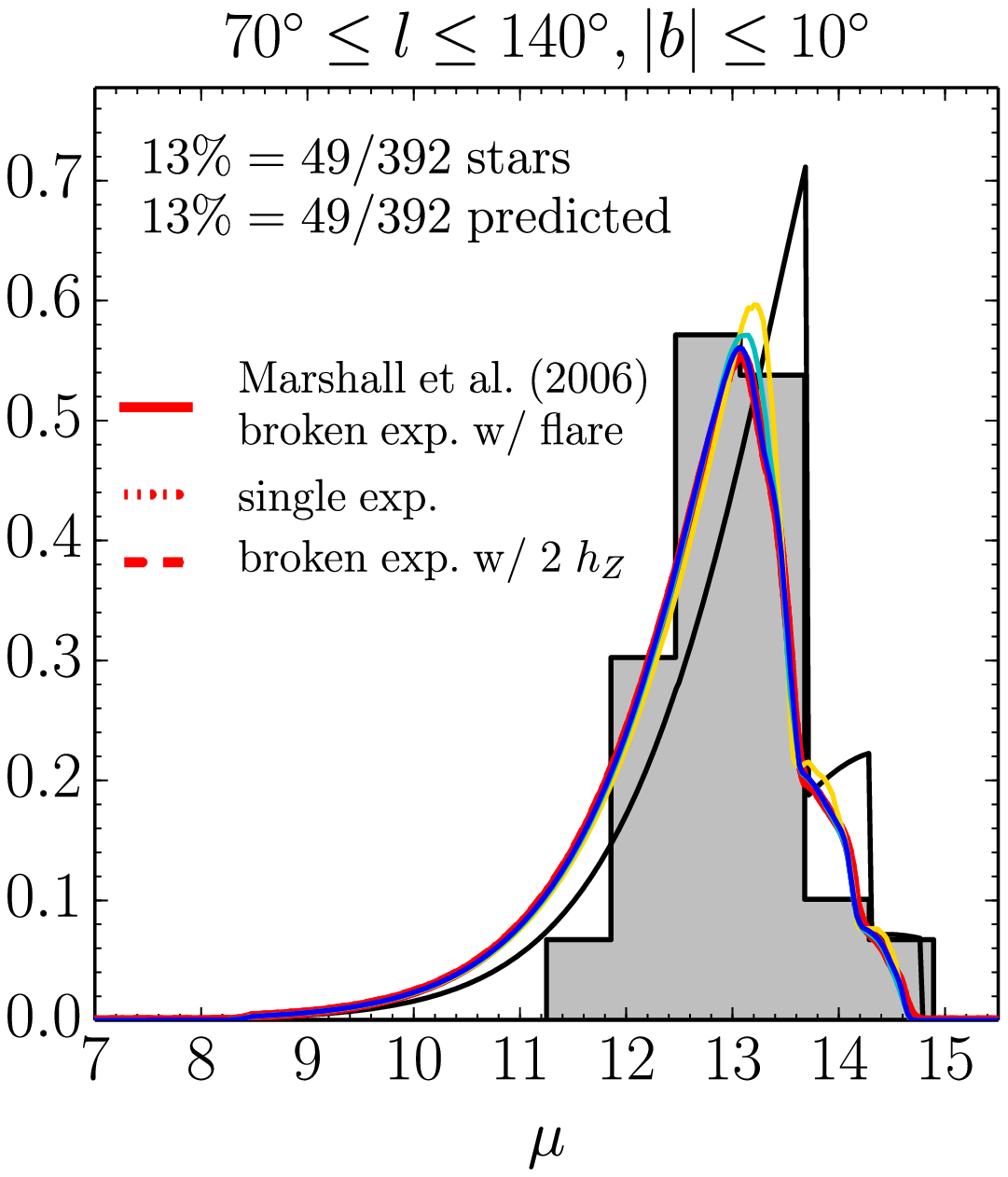}
\includegraphics[width=0.2475\textwidth,clip=]{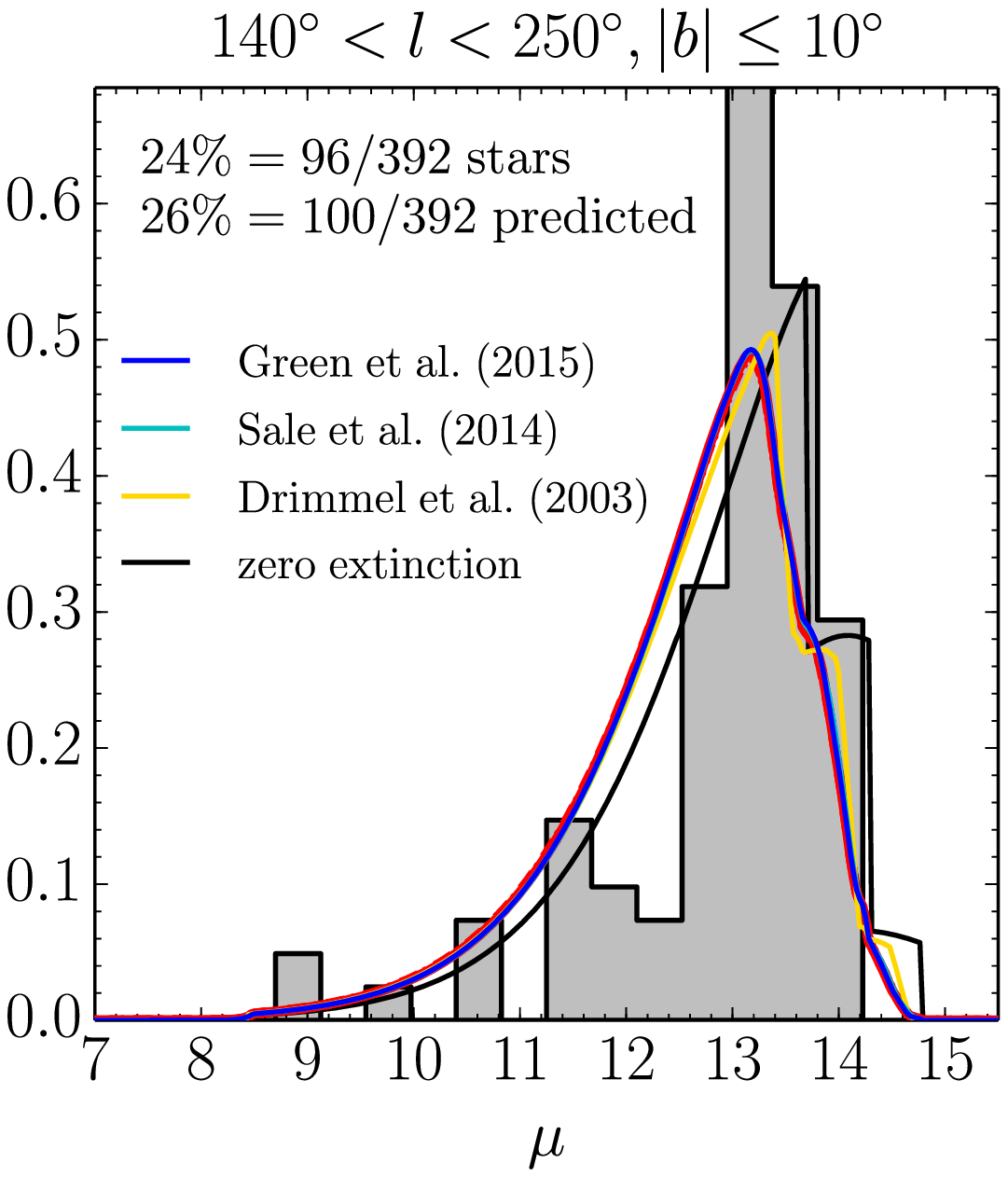}
\includegraphics[width=0.2475\textwidth,clip=]{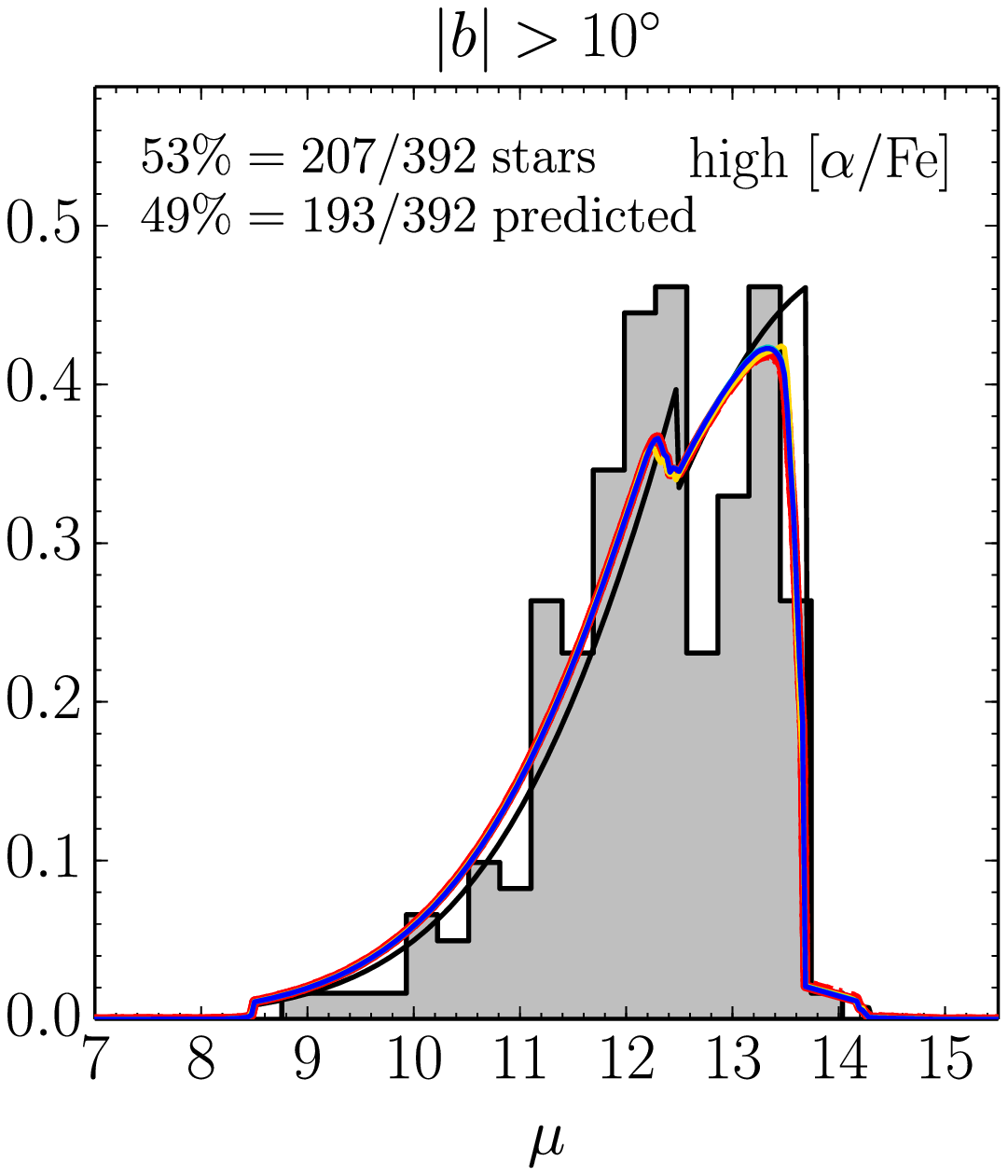}
\caption{Same as \figurename~\ref{fig:low-datacomp}, but for the
  high-\afe\ sample. All different density profiles provide equally
  good fits, demonstrating that the density is very close to a single
  radial and vertical exponential for this
  population.}\label{fig:high-datacomp}
\end{figure*}

\begin{figure*}[t!]
\includegraphics[width=\textwidth,clip=]{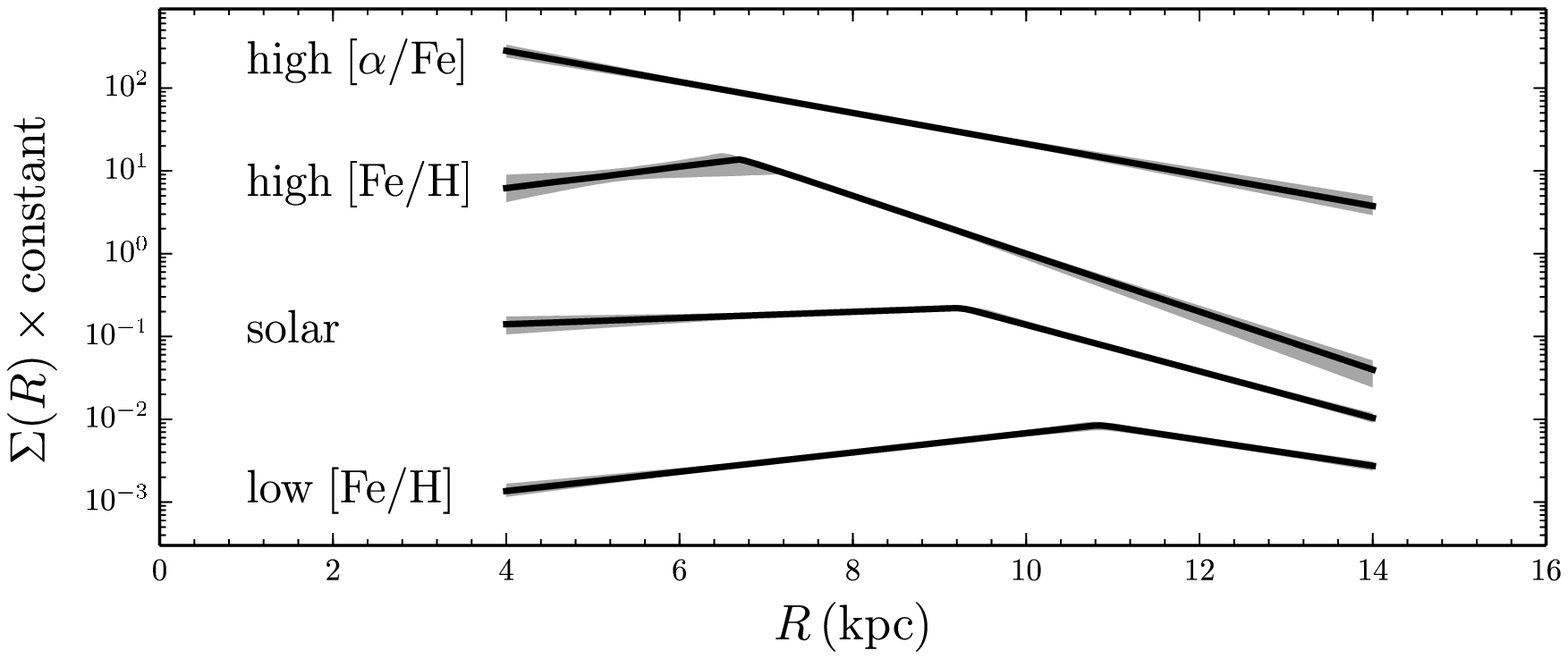}
\caption{Radial surface profile $\Sigma(R)$ of the four broad
  abundance-selected subsamples indicated in
  \figurename~\ref{fig:afefeh}. The gray region gives the 95\,\%
  uncertainty range. All profiles are relative to the density at
  $R=8\kpc$; an arbitrary offset in the vertical direction has been
  applied to separate the four profiles. The three
  low-\afe\ subsamples are best represented as a broken exponential,
  while the high-\afe\ subsample consists of a single exponential
  distribution over the full radial range that is
  observed.}\label{fig:broad-radial}
\end{figure*}

We fit a variety of models for the stellar number density of
abundance-selected populations. All of the models for which results are
given in this paper assume that the density is axisymmetric and that
the radial profile is separable from the vertical profile, such that
\begin{equation}
  \dens(R,\phi,Z) = \Sigma(R)\,\zeta(Z)\,,
\end{equation}
where we define $\zeta(Z)$ such that $\int \dd Z\, \zeta(Z) = 1$. We
have performed fits that add a constant density to capture any
outliers, but find that the contribution from outliers is negligible
in all cases; therefore we do not include them in this
description. The basic model for $\Sigma(R)$ that we consider here is
that of a broken exponential
\begin{equation}\label{eq:brokexp}
  \ln\Sigma(R) \propto \left\{ \begin{array}{ll}
    -h_{R,\mathrm{in}}^{-1} (R-R_0) & R \leq \Rb\,,\\
    -h_{R,\mathrm{out}}^{-1} (R-R_0) & R > \Rb\,.\end{array}\right.
\end{equation}
The relative normalization of these exponentials is set to produce a
continuous $\Sigma(R)$ at $\Rb$. In MCMC explorations of the parameter
constraints, the inverse scale lengths and the logarithm of $\Rb$ are
given flat priors. We also consider models where $\Sigma(R)$ is a
single exponential; such models essentially have $\Rb \equiv 0$.

We have explored additional functional forms for $\Sigma(R)$, such as
a Gaussian centered on \Rb, but find that the density is typically
best described as a broken exponential. However, because the density
drops quickly as one moves away from \Rb, and because the disk has a
large range of $\Rb$ (see below), determining the exact form of the
radial profile is difficult with the current data.

We consider four distinct functional forms for $\zeta(Z)$. The first
is that of a single exponential, with a scale height that is an
exponential function of radius $R$ (\ie, a flaring model)
\begin{equation}\label{eq:flare}
  \ln \zeta(Z) \propto
  h_Z^{-1}\,\exp\left(\Rf\,[R-R_0]\right)\,|Z|\,.
\end{equation}
To investigate the flaring profile further, we also consider
non-exponential flaring that is either linear or inverse-linear, \ie
\begin{equation}\label{eq:flarealt}
  \ln \zeta(Z) \propto
  h_Z^{-1}\,\left(1\pm\Rf\,[R-R_0]\right)^{\pm1}\,|Z|\,,
\end{equation}
where the signs are such that $\Rf$ is always negative for
outwardly-increasing $h_Z$. In MCMC explorations of the PDF, the
inverse scale height $h_Z^{-1}$ and the inverse flaring scale length
\Rf\ are given uniform priors. A non-flaring vertical profile has $\Rf
= 0$. The second form for $\zeta(Z)$ is a sum of two exponentials
\begin{equation}\label{eq:twoexp}
  \zeta(Z) = \frac{1-\beta_2}{2\,h_Z}\,\exp\left(h_Z^{-1}\,|Z|\right)
    +\frac{\beta_2}{2\,h_{Z,2}}\,\exp\left(h_{Z,2}^{-1}\,|Z|\right)\,.
\end{equation}
We also briefly consider the generalization of the models in
\equationname s~(\ref{eq:flare}) and (\ref{eq:twoexp}), \ie, the
two-exponential model where each scale height flares exponentially as
in \equationname~(\ref{eq:flare}). As demonstrated below, the
two-exponential form for the vertical profile does not fit the data as
well as the flaring single-exponential form. Our main use of the
second form is to investigate whether any MAPs show evidence of a
secondary component with a different scale height.

We optimize the likelihood in \equationname~(\ref{eq:like}) for each
density model and each data set below using a downhill simplex
algorithm. We then use this optimal solution to initiate an MCMC
sampling of the posterior PDF, obtained using an affine-invariant
ensemble MCMC sampler \citep{Goodman10a,ForemanMackey12a}. Reported
parameter estimates are based on the median and standard deviation of
one-dimensional projections of the MCMC chain.

\subsection{Tests on mock data}

We have performed a suite of mock-data tests to validate our code
(\ie, checking that we recover the correct input density profiles when
fitting a model that includes the input assumptions) and to determine
the impact of the uncertainty in the three-dimensional extinction
map. We generate mock data with profiles similar to a few of the
best-fit profiles for the broad abundance-selected subsamples
below. In particular, we produce mock data with a flat radial profile,
a single-exponential profile with a scale length of $3\kpc$, or
broken-exponential profiles with peak radii of $8$ and $11\kpc$; all
mock data have a constant thickness with a scale height of $300\pc$,
except that we also generate mock data with the exponential flaring
profile of \equationname~(\ref{eq:flare}) with a scale length of
$10\kpc$ for the broken-exponential profiles. All mock data assume
extinction according to the \citet{Green15a} map and properly take
into account the variation of the density profile within each APOGEE
field, therefore testing that this can be ignored in the calculation
of the effective volume (see above). Each mock-data sample has 20,000
stars.

When fitting the mock data with profiles that include the input
density profile, we find that our code recovers the correct profile to
within the uncertainties that we find for the real data below. In
particular, we confirm that radial profiles consisting of a single
exponential are recovered as such, even when fit with more the general
broken-exponential profile. Furthermore, we recover that no
constant-scale-height mock data set displays any flaring and that the
flaring scale length is correctly recovered in the mocks with
flaring. For the latter, we also find that the radial profile is
almost perfectly recovered even when they are fit with a constant
scale height. This confirms the expectation that the radial and
vertical profiles are measured almost independently of each other due
to the many lines of sight at $b=0^\circ$ in the APOGEE footprint (see
\figurename~\ref{fig:fields}).

To determine the impact of the uncertainty in the interstellar
extinction, especially in the inner MW, we fit the mock data assuming
extinction according to \citet{Marshall06a} near the mid-plane, rather
than the \citet{Green15a} extinction assumed in generating the mock
data; the latter is typically smaller than the former. We find only
small changes in the best-fit parameters when using the incorrect
extinction map; these differences are smaller or similar to the
uncertainties in the best-fit parameters, and do not lead to
qualitatively-different inferred density profiles. Even employing the
map of \citet{Drimmel03a} in the fit only changes the best-fit
parameters by insignificant amounts, similar to what we find for the
real data below. Therefore, we conclude that the uncertainty in the
three-dimensional extinction does not limit our understanding of the
spatial density profiles of stellar subsamples in APOGEE.

\section{The spatial structure of broad abundance-selected subsamples}\label{sec:broad}

We now discuss the results from fitting the spatial density of the
broad abundance-selected subsamples (\figurename s \ref{fig:afefeh}
and \ref{fig:spatial-broad}), using the functional forms described in
\sectionname~\ref{sec:models} above. We start by fitting broad
sub-samples first, because of the (relatively) large sample size that
allows us to precisely determine the shape of their density
profiles. In \sectionname~\ref{sec:maps} we then refine our results by
fitting the preferred models from this section to the MAPs, which have
much smaller sample sizes. The well-populated broad subsamples also
make it possible to demonstrate the goodness of the fits of different
spatial profiles by directly overlaying the best-fit models on the
observed distribution of stars.

\subsection{The surface-density profile\label{sec:broad-radial}}

Results from fitting the density profiles given in
\sectionname~\ref{sec:models} to the broad abundance-selected
subsamples are given in \tablename~\ref{table:results}. Comparisons
between different model fits discussed below and the observed distance
distribution are displayed in \figurename~\ref{fig:low-datacomp} for
the three low-\afe\ subsamples, and in
\figurename~\ref{fig:high-datacomp} for the high-\afe\ subsample. We
fit the radial and vertical profiles simultaneously, but focus on
discussing the resulting radial profiles $\Sigma(R)$ in this
subsection.

It is clear from \tablename~\ref{table:results} and
\figurename~\ref{fig:low-datacomp} that a single-exponential radial
profile $\Sigma(R)$ (dotted line in
\figurename~\ref{fig:low-datacomp}) does not provide a good fit to the
data for the low-\afe\ subsamples. A broken exponential of the type as
in \equationname~(\ref{eq:brokexp}) provides a much better
fit. Discrepancies between the observed and predicted distribution of
distances in \figurename~\ref{fig:low-datacomp} are small for this
model. In addition to the broken-exponential model, we have also
performed fits with a Gaussian radial profile. These only gave similar
or worse fits to the data, but with $\Delta \chi^2$ of only tens. The
exact shape of the radial profile can therefore not be determined at
high confidence, but it is clear that $\Sigma(R)$ for the low
\afe\ subsamples increases up to a peak radius \Rb\ and declines
beyond that. $\Sigma(R)$ is highly inconsistent with being a single
exponential.

For the high-\afe\ subsample the single-exponential model provides a
good fit, while fits with a broken-exponential or other radial profile
all end up as close to a single exponential as possible. For example,
the broken-exponential fit places \Rb\ outside of the observed volume
($\Rb < 4.4\kpc$; \tablename~\ref{table:results}) such that this fit
is equivalent to a single exponential. The Gaussian radial profile
does the same, and adjusts the width parameter such that the profile
closely approximates a single exponential. We can therefore be
confident that $\Sigma(R)$ for the high-\afe\ subsample is very close
to a single exponential.

The fits in \tablename~\ref{table:results} and \figurename
s~\ref{fig:low-datacomp} and \ref{fig:high-datacomp} are performed for
several different extinction maps. The standard extinction map used is
that of \citet{Green15a}. Used on its own it is labeled as
``\citet{Green15a}''. When we replace the part of it at $-100^\circ
\leq l \leq 100^\circ$ and $|b| \leq 10^\circ$ with the map of
\citet{Marshall06a}, we label this as ``\citet{Marshall06a}'';
Likewise, when we replace the part of it that overlaps with the map of
\citet{Sale14a}, we refer to that model as ``\citet{Sale14a}''. We
also test the performance of two other extinction maps: that of
\citet{Drimmel03a}, which is defined over the entire sky, and a model
without any extinction (labeled ``zero'').

In all cases the combination of the \citet{Marshall06a} and the
\citet{Green15a} maps provides the best fit. The map of
\citet{Sale14a} performs slightly worse than that of \citet{Green15a}
where \citet{Sale14a} overlaps the latter. The map of
\citet{Drimmel03a}, which consists of a simple model for the
three-dimensional distribution of dust and has lower angular
resolution than the other maps, gives very poor fits. The model with
zero extinction clearly provides a bad fit to the data, both from the
$\Delta \chi^2$ in \tablename~\ref{table:results}, and directly from
the comparison between the model and the data in
\figurename~\ref{fig:low-datacomp}, especially at $l < 70^\circ$. All
of the different extinction maps, however, give very similar best-fit
parameters for the basic model for the spatial density.

The radial profile and its uncertainties for the standard model of a
broken exponential for the four broad subsamples is displayed in
\figurename~\ref{fig:broad-radial}. It is clear that the uncertainty
in this (parametric) model is small, and that the peak radius \Rb\ for
the low-\afe\ subsamples is larger for decreasing \feh. The shape of
the radial profile around \Rb\ is quite similar for all three
low-\afe\ subsamples, with a shallow rise at $R < \Rb$ and a steep
decline at $R > \Rb$. The high-\afe\ subsample could be thought of as
having $\Rb \lesssim 4\kpc$, and therefore be the continuation of the
trend of the low-\afe\ subsamples, but with the current radial
coverage we cannot test that scenario.

\subsection{The vertical profile\label{sec:broad-vertical}}

Having determined that a broken-exponential $\Sigma(R)$ provides the
best fit, we fit three different vertical profiles to the APOGEE-RC
data. The simplest model is that of a single-exponential model with a
radially constant scale height $h_Z$. By fitting more complex models,
we find that this model is strongly ruled out for the
low-\afe\ subsamples. They are instead better fit with a model where
$h_Z$ is a function of $R$, and we employ the flaring model of
\equationname~(\ref{eq:flare}) (the flaring profile is explored in
more detail for the MAPs below using the alternative flaring
models). An alternative model is that each subsample consists of the
sum of two exponentials (the model of
\equationname~[\ref{eq:twoexp}]); this model does not fit as well (see
\tablename~\ref{table:results}). We have also fit a generalized model
where the vertical profile consists of the sum of two exponentials
that flare exponentially with the same scale length. In all cases, the
best fit for this general model reverts to that of the
single-exponential, flaring model. All of the low-\afe\ subsamples are
consistent with a common flaring scale length of $\Rf \approx
-0.1\kpc\inv$. We refine this measurement in
\sectionname~\ref{sec:maps-vertical} below.

Like for the surface-density profile, the high-\afe\ subsample is
consistent with the simplest model, in this case a single vertical
exponential with a constant $h_Z(R)$. That is, we see with high
confidence that the high-\afe\ subsample does not display the same
kind of flaring as the low-\afe\ subsamples, but $h_Z(R)$ is instead
constant. We refine the quantitative constraint in
\sectionname~\ref{sec:maps-vertical} below.

\section{The spatial structure of MAPs}\label{sec:maps}

In this section we repeat the density fits from the previous section,
but we perform them on abundance-selected subsamples that are narrower
in \feh\ and \afe. That is, we use MAPs, defined here as abundance
bins with widths of $\Delta \feh = 0.1\dex$ and $\Delta \afe =
0.05\dex$. We do not make use of other abundances for defining MAPs,
but stress that our empirical description does not assume chemical
homogeneity beyond $(\feh,\afe)$. As discussed at the end of
\sectionname~\ref{sec:data-abu}, these widths are about twice as large
as the uncertainties in these quantities and the contamination between
MAPs is therefore small. Because of the small number of stars in the
statistical APOGEE-RC at high \afe, we perform fits for MAPs with at
least 15 stars; the measurements for MAPs with so few stars are noisy,
but informative enough to help establish trends. We again discuss the
results for the surface-density and vertical profiles separately, but
both were measured simultaneously for all MAPs.

\subsection{The surface-density profile}

\begin{figure}[t!]
\includegraphics[width=0.48\textwidth]{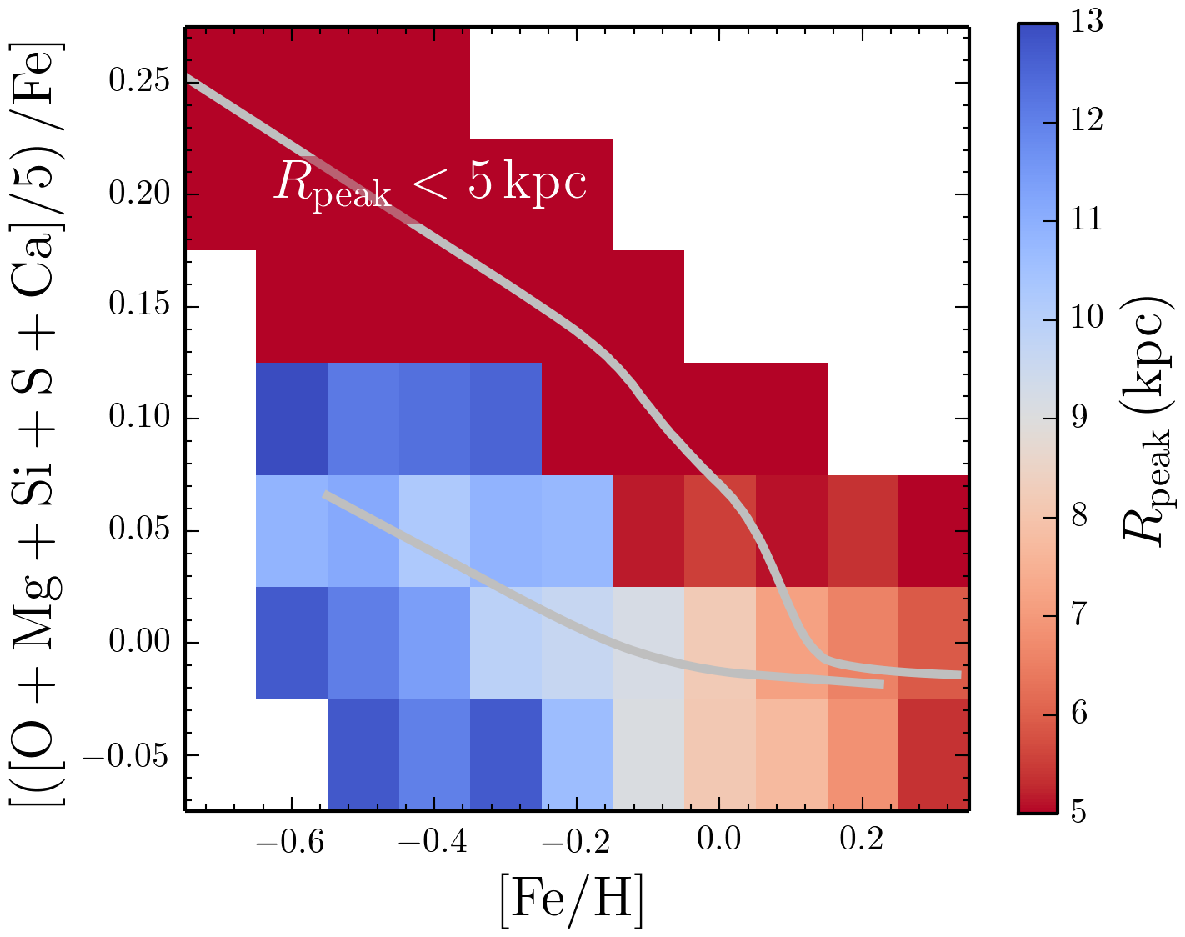}
\caption{Peak radius of the radial profiles of MAPs. This figure
  displays the Galactocentric radius at which the surface density of
  each MAP peaks when fit with a broken-exponential radial profile
  that is constrained to be increasing within \Rb. The locus of the
  high- and low-\afe\ sequences in \figurename~\ref{fig:afefeh} are
  indicated with gray curves. MAPs along the high-\afe\ sequence do
  not display a peak in their radial profile within the observed $R$
  range; they are best represented as a single radial exponential and
  are indicated as ``$R_{\mathrm{peak}} < 5\kpc$''. MAPs along the
  low-\afe\ sequence show a striking increase in \Rb\ with decreasing
  \feh.}\label{fig:rpeak}
\end{figure}

\begin{figure*}[t!]
\begin{center}
\includegraphics[width=0.9\textwidth,clip=]{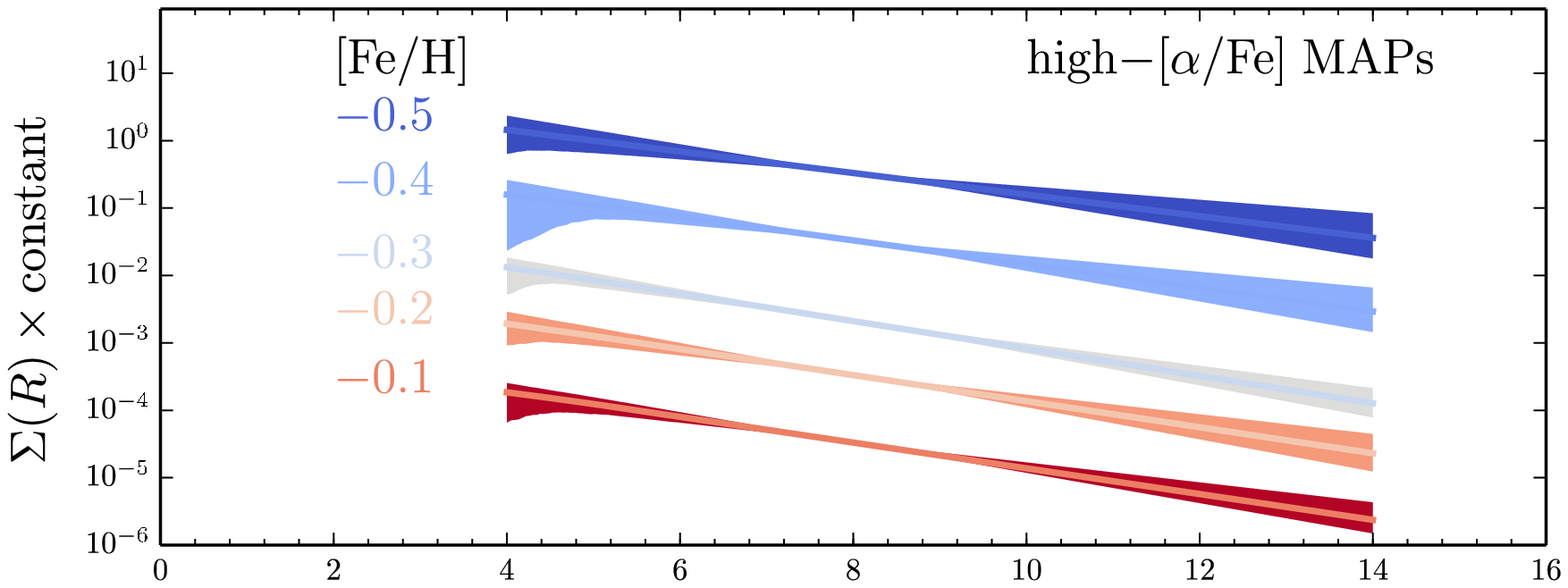}
\includegraphics[width=0.9\textwidth,clip=]{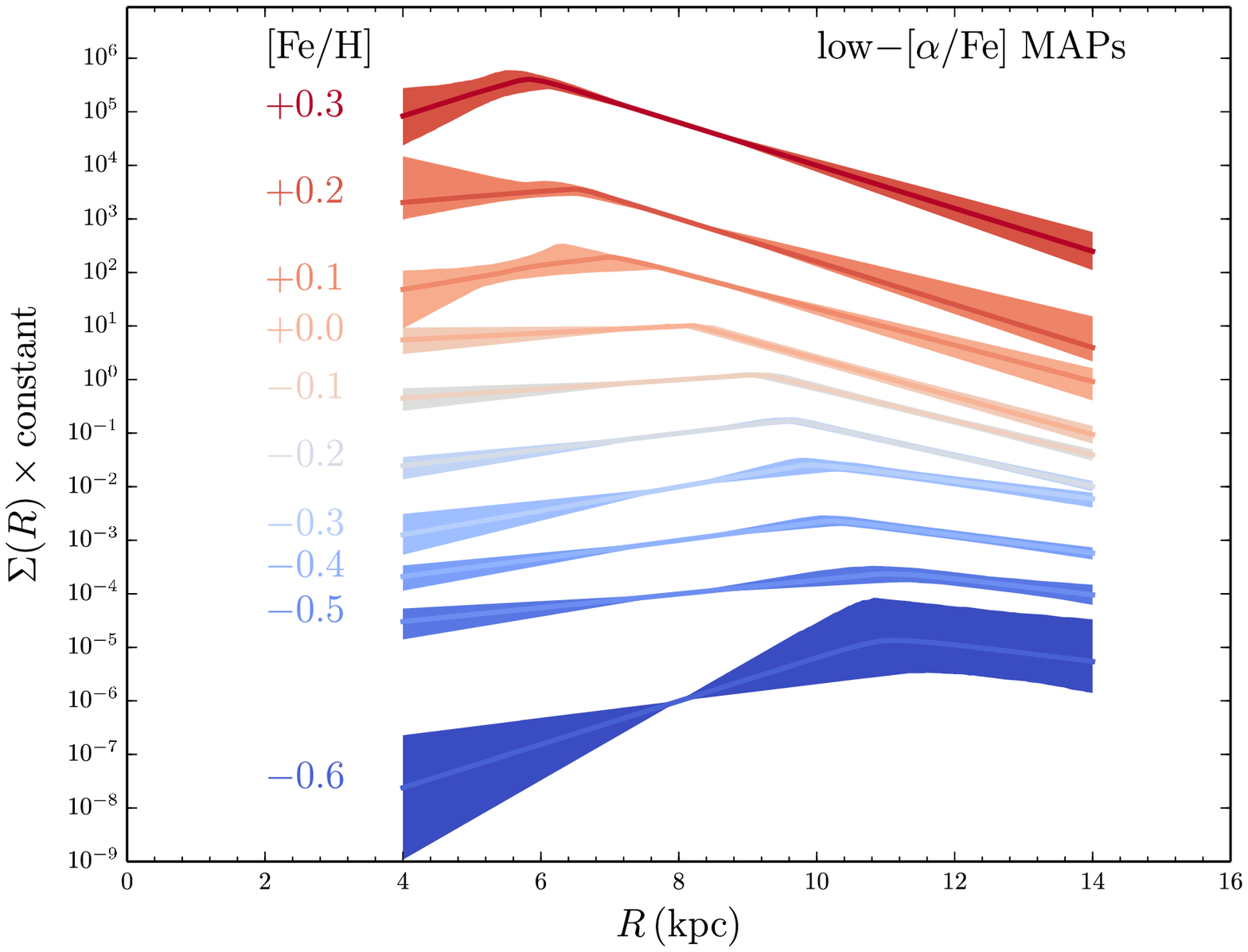}
\end{center}
\caption{Radial surface profile of MAPs. For display purposes, MAPs
  along the well-populated low- and high-\afe\ sequences are shown in
  the bottom and top panel, respectively, but the trends are the same
  for all MAPs. For the low-\afe\ sequence these are the MAPs with
  $\afe = +0.05$ up to $\feh = -0.4$ and $\afe = 0.0$ at higher
  $\feh$. For the high-\afe\ sequence these are the MAPs with $\afe =
  +0.20$ for $\feh= (-0.5,-0.4)$, $\afe = +0.15$ for $\feh =
  (-0.3,-0.2)$, and $\afe = +0.10$ for $\feh = -0.1$. The colored bands
  give the 95\,\% uncertainty region. The radial profiles of
  high-\afe\ MAPs are given by single exponentials with a common scale
  length of $2.2\pm0.2\kpc$. The metal-poor low-\afe\ MAPs peak in the
  outer disk ($\Rb \gtrsim10\kpc$) and are spread over a wide range of
  radii, with a relatively shallow outer scale length of about
  $3\kpc$. The metal-rich low-\afe\ MAPs are very
  centrally-concentrated ($\Rb\lesssim8\kpc$), with outer scale
  lengths of only $\approx1.25\kpc$. The inner, rising scale length is
  universally $\approx3\kpc$, in all MAPs where it can be
  constrained.}\label{fig:maps-radial}
\end{figure*}

\begin{figure}[t!]
\includegraphics[width=0.48\textwidth,clip=]{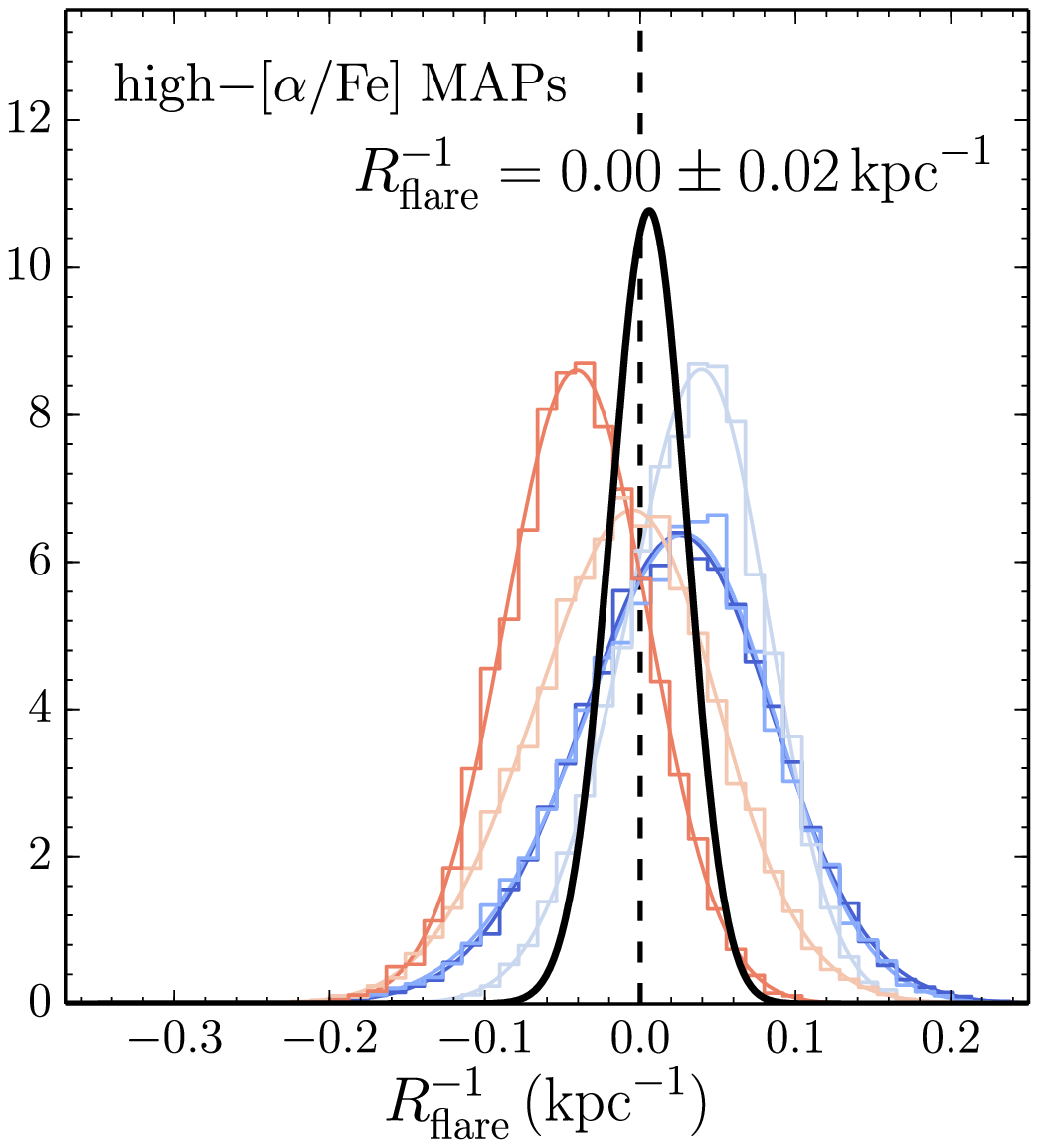}\\
\includegraphics[width=0.48\textwidth,clip=]{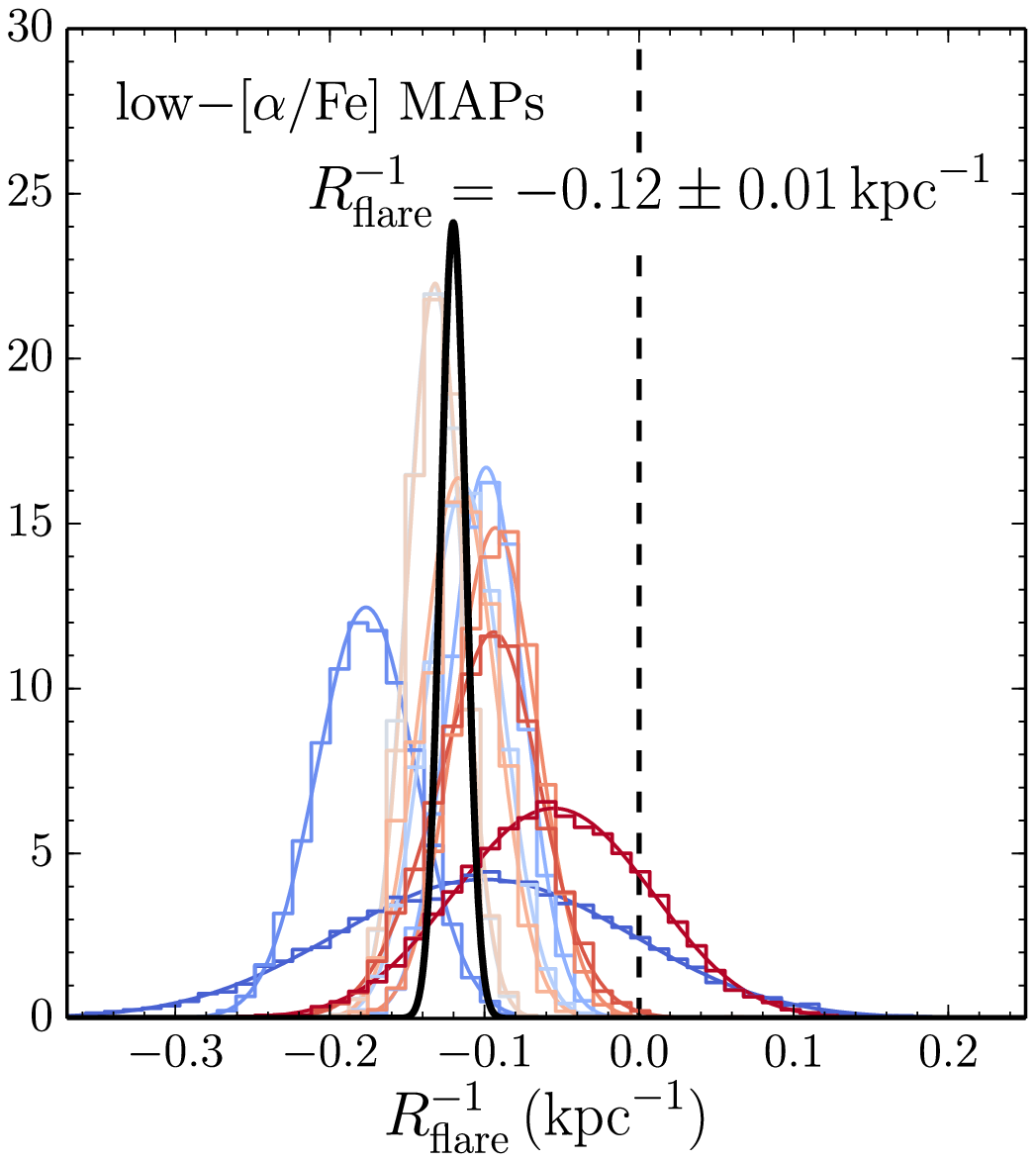}
\caption{Flaring of individual MAP disk components. The PDFs for the
  inverse flaring scale length \Rf\ for individual MAPs for
  well-populated high-\afe\ (\emph{top panel}) and
  low-\afe\ (\emph{bottom panel}) MAPs are displayed (histogram),
  together with smooth fits to these PDFs with a sum of two Gaussians
  (colored curves). The combined PDF obtained by multiplying the
  individual smooth PDFs is shown in black; its mean and standard
  deviation are given in the top right. The displayed MAPs are those
  of \figurename~\ref{fig:maps-radial}, with colors ranging from blue
  to red for low- to high-\feh\ MAPs. The dashed vertical line
  indicates the limit of flaring disks (to its left); the
  high-\afe\ MAPs display no flaring to high precision, while the
  low-\afe\ MAPs are consistent with a single flaring scale length of
  $R_{\mathrm{flare}}\approx8.5\pm0.7\kpc$.}\label{fig:mapfits-flare}
\end{figure}

\begin{figure*}[t!]
\begin{center}
\includegraphics[width=0.9\textwidth,clip=]{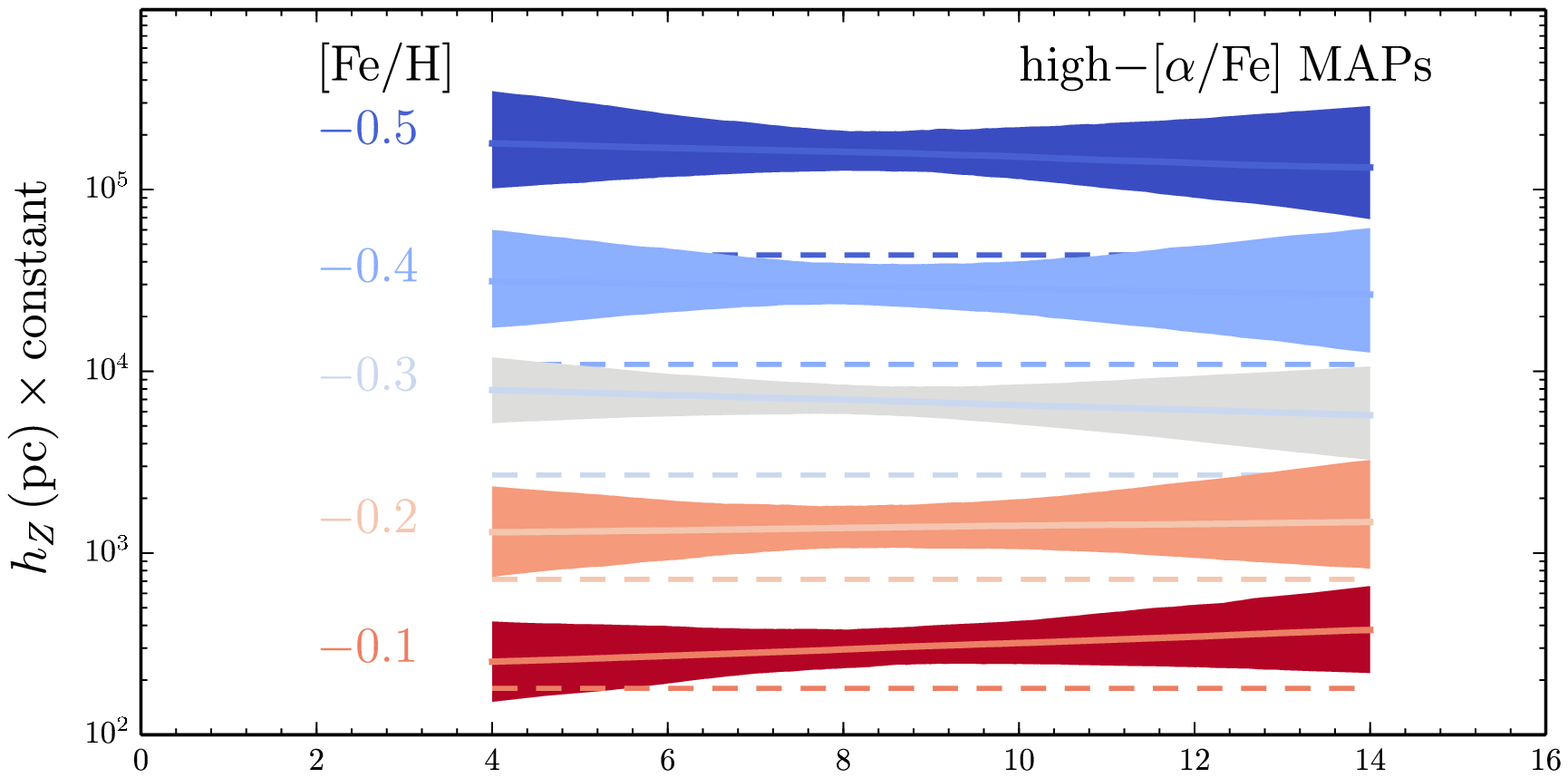}\\
\includegraphics[width=0.9\textwidth,clip=]{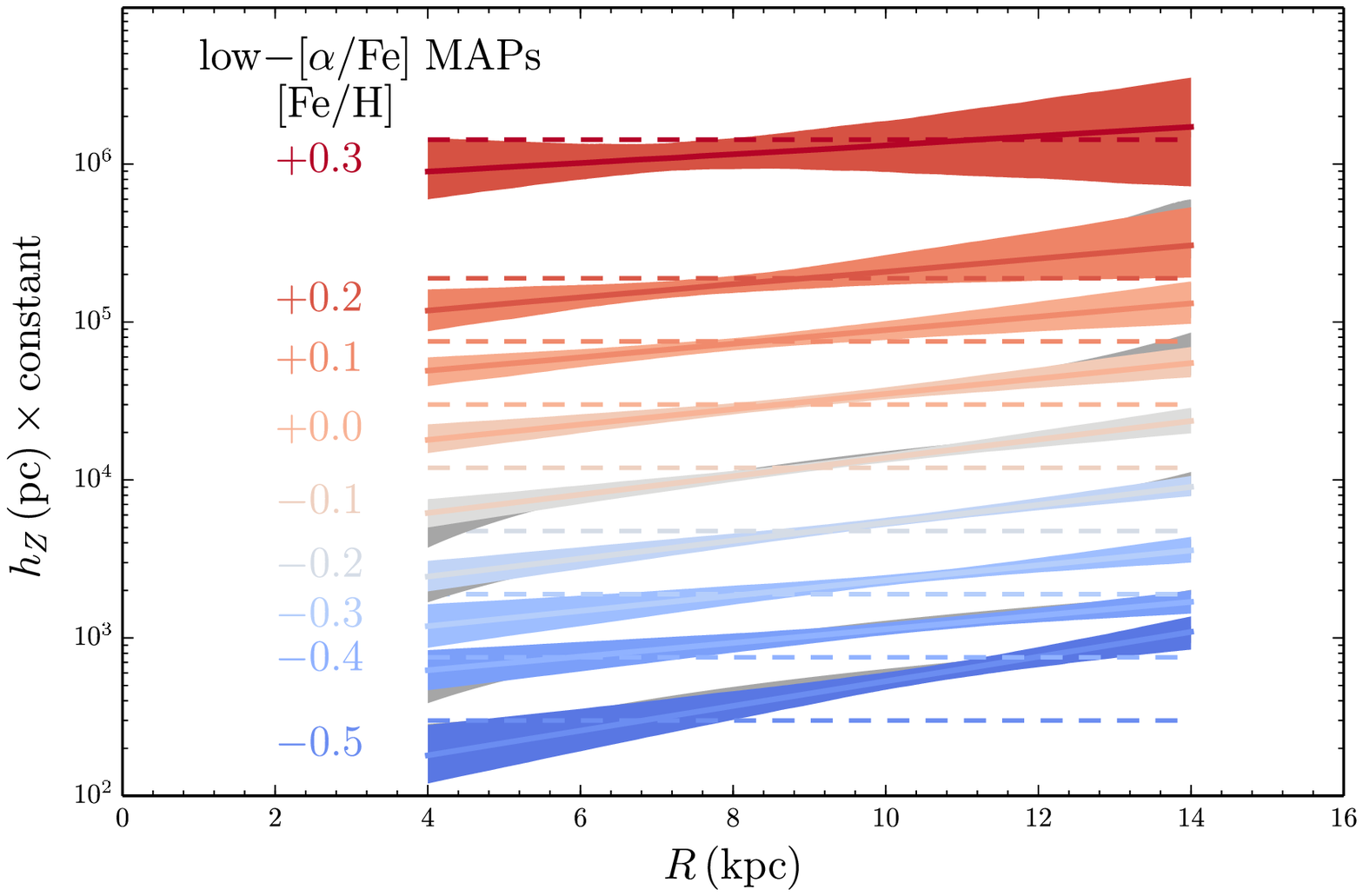}
\end{center}
\caption{Vertical profile of MAPs. This figure displays the radial
  dependence of the scale height of MAPs along the well-populated
  high-\afe\ (\emph{top panel}) and low-\afe\ sequence (\emph{bottom
    panel}; those of \figurename~\ref{fig:maps-radial}, except for the
  lowest-\feh\ MAP). The colored bands give the $95\%$ uncertainty
  region around the median displayed as a solid line for the
  exponential flaring model. The gray band (when visible) shows the
  range in median profile spanned by all three flaring models
  (exponential, linear, and inverse-linear; see \equationname
  s~[\ref{eq:flare}] and [\ref{eq:flarealt}]). A \feh-dependent offset
  has been applied to the $y$ axis to separate the different MAPs; the
  dashed horizontal line gives the position of $h_Z = 300\pc$ for each
  MAP. The high-\afe\ MAPs do not flare, while all low-\afe\ MAPs are
  consistent with an exponential flaring profile with a scale length of
  $R_{\mathrm{flare}}\approx8.5\pm0.7\kpc$ (see
  \figurename~\ref{fig:mapfits-flare} above). The scale height
  increases smoothly from high- to
  low-\feh\ MAPs.}\label{fig:maps-radialflare}
\end{figure*}

\begin{figure}[t!]
\includegraphics[width=0.48\textwidth]{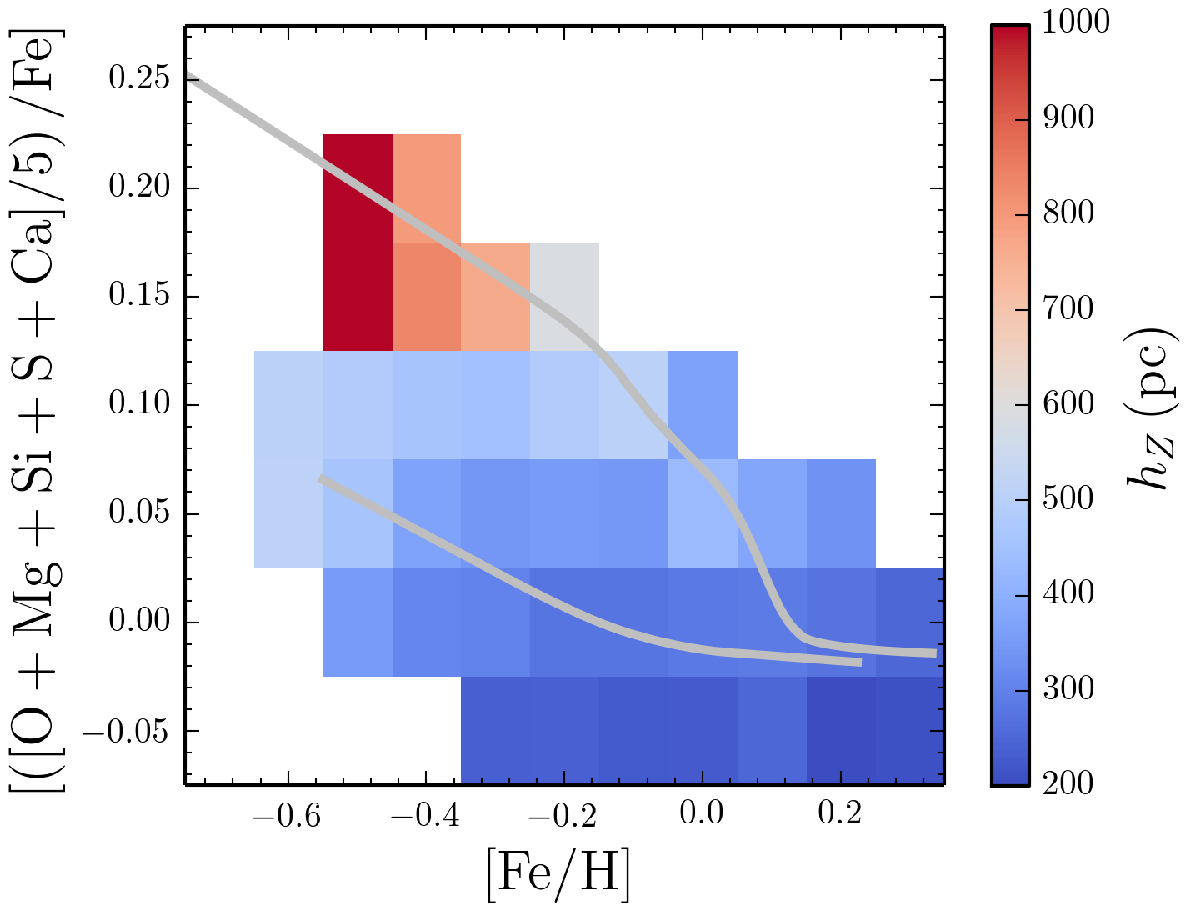}
\caption{Vertical scale height of MAPs. The scale heights, $h_Z$, for
  fits assuming a constant scale height with Galactocentric radius for
  MAPs best fit as a single radial exponential (those labeled
  ``$R_{\mathrm{peak}} < 5\kpc$'' in \figurename~\ref{fig:rpeak}) and
  for fits with a flaring vertical profile with a flare scale length
  of $10\kpc$ for all other MAPs. A few MAPs along the upper and left
  edges for which the scale height cannot be precisely determined from
  the present data are omitted.  The locus of the high- and
  low-\afe\ sequences are indicated as in
  \figurename~\ref{fig:rpeak}. MAPs displays a smooth increase in
  $h_Z$ as a function of declining \feh\ and increasing \afe, even for
  the conservative flaring model chosen here.}\label{fig:hz}
\end{figure}

\begin{figure}[t!]
\includegraphics[width=0.48\textwidth,clip=]{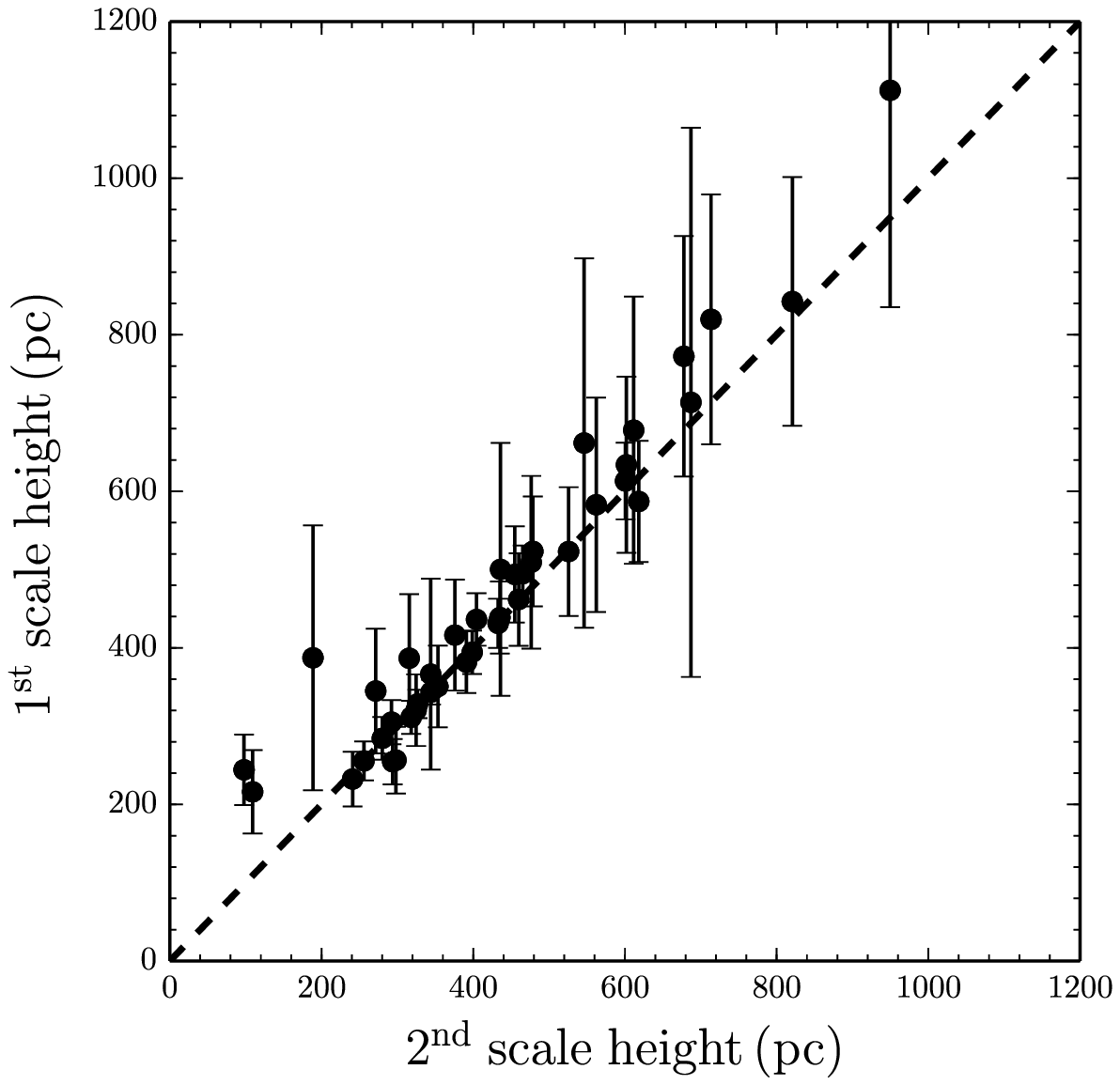}
\caption{Scale height of the dominant component versus that of the
  secondary component in two-vertical-exponential fits to the MAP
  spatial densities. The scale heights displayed here are the median
  of samples from the posterior PDF for which the secondary component
  provides significantly to the density (defined as contributing more
  than $15\,\%$ of the local column density), to avoid the massive
  degeneracy in the scale height when the amplitude is allowed to be
  close to zero. The two scale heights are the same for all MAPs,
  demonstrating that they are well represented as a single vertical
  exponential. MAPs with very low scale heights, however, show some
  evidence of a second, even lower scale height.}\label{fig:twohz}
\end{figure}

Inspired by the fits to the broad abundance-selected subsamples in
\sectionname~\ref{sec:broad-radial}, we fit a broken-exponential
$\Sigma(R)$ to each well-populated MAP. We constrain these
broken-exponential models to have an inner profile that is increasing
with $R$ and an outer profile that is decreasing, although the vast
majority of MAPs have well constrained profiles without this prior
constraint that satisfy it. For MAPs that are best fit as a single
exponential, this constraint forces \Rb\ to lie at small $R$; without
this constraint, the fit would have a degeneracy between very small
\Rb\ and very large \Rb\ (as long as \Rb\ is outside of the observed
volume). We always use the combined \citet{Marshall06a} and
\citet{Green15a} extinction map, which provided the best fit to the
broad subsamples above.

We display the dependence on $(\feh,\afe)$ of the peak of $\Sigma(R)$
in \figurename~\ref{fig:rpeak}. We determine $\Rb$ typically to
$0.3\kpc$, while the range of $\Rb$ covers about $8\kpc$. Thus, the
smooth trends seen in \figurename~\ref{fig:rpeak} are determined at
high significance. These more detailed results confirm the behavior
found for the broad subsamples in
\sectionname~\ref{sec:broad-radial}. Low-\afe\ MAPs have an increasing
$\Rb$ with decreasing \feh, ranging from $\Rb \approx 5\kpc$ at the
highest \feh, to $\Rb \approx 13\kpc$ at the lowest \feh. This trend
has a weak dependence on \afe. We have indicated the locus where the
low-\afe\ sequence is well populated (that is, where \Rb\ is best
determined) and along this sequence the correlation between \feh\ and
\Rb\ is incredibly tight.

The behavior of individual high-\afe\ MAPs also confirms that
high-\afe\ MAPs do not display a break in their surface-density
profiles, but are instead consistent with a single exponential. For
all MAPs along the high-\afe\ sequence, \Rb\ is constrained to lie
outside of the observed volume.

The radial dependence of $\Sigma(R)$ is displayed in
\figurename~\ref{fig:maps-radial}. This figure only shows MAPs along
the well-populated high- and low-\afe\ sequences for clarity, but the
behavior of other MAPs is similar, albeit noisier. It is clear that
the radial profiles for all but the lowest \feh\ MAP are well
constrained to have the broken-exponential form with an almost
universal shape around \Rb. The inner profile is typically shallower
than the outer profile, except for lower-\feh\ MAPs. However, these
MAPs are only sparsely populated at the large distance from the
Galactic center where their outer profiles are constrained, as is also
evident from the uncertainties. The top panel displays $\Sigma(R)$ for
the high-\afe\ MAPs. In addition to all being consistent with \emph{a}
single exponential, they are all consistent with \emph{the same}
single exponential, with a scale length of $h_R = 2.2\pm0.2\kpc$.

\subsection{The vertical profile}\label{sec:maps-vertical}

Results from fitting the MAPs with the standard model for the vertical
density---a single exponential with an exponentially-increasing scale
height---are displayed in \figurename s~\ref{fig:mapfits-flare},
\ref{fig:maps-radialflare}, \ref{fig:hz}, and \ref{fig:twohz}. The
vertical profile of individual MAPs is difficult to determine from the
APOGEE-RC data, because the sample is dominated by low-latitude fields
that give little leverage for measuring the vertical density drop
off. When allowing the inverse scale length of the flare (\Rf) to be
free, there is a large degeneracy between $\Rf$ and the scale height
at the solar circle $h_Z$. \figurename~\ref{fig:mapfits-flare}
displays PDFs for $\Rf$ for MAPs along the well-populated high- and
low-\afe\ sequences. For these MAPs, $\Rf$ is relatively well
constrained. We see from these PDFs that the high-\afe\ MAPs are all
consistent with having a constant $h_Z(R)$, while low-\afe\ MAPs show
strong evidence for a flaring $h_Z(R)$. The constraints on $\Rf$ for
both \afe\ groups are tight: $\Rf = 0.00\pm0.02\kpc^{-1}$ for the
high-\afe\ MAPs and $\Rf = -0.12\pm0.01\kpc^{-1}$ for the
low-\afe\ MAPs. The latter corresponds to a flaring scale length of
$8.5\pm0.7\kpc$. We illustrate the flaring (and non-flaring) of the
MAPs in \figurename~\ref{fig:maps-radialflare}. To further explore
what the radial dependence of $h_Z$ is, we have also fit each MAP with
the linear and inverse-linear flaring profiles of
\equationname~(\ref{eq:flarealt}). \figurename~\ref{fig:maps-radialflare}
contains these results: the median of the MCMC samples from these fits
are very close to that using the exponential model. The gray band for
each MAP displays the range in $h_Z$ spanned by the different
models. This gray band is only visible for a few low-\afe\ MAPs and
mainly near the edges of the radial range, indicating that the
low-\afe\ MAPs have an exponential flaring profile to within their
uncertainties. For high-\afe\ MAPs the results from the alternative
flaring profiles are almost indistinguishable from those with the
exponential model.

To determine the scale heights, $h_Z$, of all of the MAPs, we then
repeat the density fits while fixing $\Rf$. Specifically, we set $\Rf
= 0$ for all of the high-\afe\ MAPs, defined here as those with $\Rb <
5\kpc$ (see \figurename~\ref{fig:rpeak}), and $\Rf = -0.1\kpc^{-1}$
otherwise. The resulting $h_Z(\feh,\afe)$ is displayed in
\figurename~\ref{fig:hz}. In addition to the requirement that each MAP
needs to have at least 15 stars, we have further removed MAPs for
which the uncertainty in $h_Z$ is larger than $20\,\%$; this removes a
few MAPs at low \feh\ (both at high and low \afe). It is clear from
this figure that the dependence of $h_Z$ on $(\feh,\afe)$ is very
smooth, and similar to that found by \citet{BovyMAPstructure}. In
particular, intermediate components with $h_Z \approx 500\pc$ are
prevalent at the low-\feh\ end of the low-\afe\ sequence and at the
high-\feh\ end of the high-\afe\ sequence.

In \sectionname~\ref{sec:broad-vertical} we determined that a vertical
profile consisting of the sum of two exponentials does not provide a
good fit to the broad abundance subsamples. We repeat this exercise
here, fitting a model with two vertical exponentials with a constant
scale height. In \figurename~\ref{fig:twohz} we compare the scale
height of the dominant component with that of the secondary
component. We constrain the secondary component to contribute at least
$15\,\%$ of the local column density, to remove the large degeneracies
at small amplitudes of the secondary component. We find that whenever
the secondary component makes a substantial contribution to the column
density, its scale height is the same as that of the dominant
component. That is, the vertical profile is consistent with being a
single exponential.

\section{Discussion}\label{sec:discussion}

This paper is the first to dissect in detail---and over a wide range
of Galactocentric radii---the radial structure of the MW's stellar
disk in terms of MAPs, abundance-selected stellar sub-populations. As
some of the results of this analysis may appear unorthodox, we first
show the extent to which these results are compatible with earlier
analyses, before proceeding to interpret them in a galaxy formation
context.

\needspace{8ex}
\subsection{Comparison to Bovy \etal\ (2012c)}

For the high-\afe\ MAPs our results on the radial profile are in
perfect agreement with those of \citet{BovyMAPstructure}. They find
for a single-exponential fit to $\Sigma(R)$ for high-\afe\ MAPs that
$h_R = 2.01\pm0.05\kpc$, consistent with our measurement here of $h_R
= 2.2\pm0.2\kpc$. Their result is more precise because they have
$\approx14,000$ high-\afe\ stars compared to only $\approx500$
here. However, the RC distance scale is more accurately known than
that of the G dwarfs used by \citet{BovyMAPstructure}, so larger
samples of RC stars should eventually lead to a more accurate and
precise measurement of $h_R$. The improved radial coverage of the
APOGEE-RC sample over the G-dwarf sample also allows us to ascertain
that $\Sigma(R)$ is indeed a single exponential; this was an
assumption in \citet{BovyMAPstructure}.

The present analysis has shown that the radial surface-density profile
of low-\afe\ MAPs is \emph{not} a single exponential, but is much
better fit as a broken-exponential profile, rising to a peak radius
\Rb, before falling off.  This makes it difficult to meaningfully
compare our measurements of $\Sigma(R)$ to measurements in the
literature, which typically consist of single-exponential fits to a
mix of MAPs.

\citet{BovyMAPstructure} were the first to dissect the MW disk into
narrow abundance bins and while they also fit single-exponential
models to $\Sigma(R)$, their results can be more easily compared to
the current measurements.  For comparison, we have carried out
single-exponential fits to the APOGEE data $\Sigma(R)$ (see
\tablename~\ref{table:results}), but the different radial coverage
makes this a qualitative exercise rather than a quantitative one. On
this level, our results are consistent with those of
\citet{BovyMAPstructure} for the low-\afe\ MAPs. When fit with a
single-exponential $\Sigma(R)$, low-$\feh$ MAPs have a flat profile,
solar-$\feh$ MAPs have a scale length of $h_R = 3.0\pm0.1\kpc$, and
high-$\feh$ MAPs have $h_R = 1.67\pm0.03\kpc$, similar to the results
in Figure 5 of \citet{BovyMAPstructure}. It is clear, however, that we
significantly refine the results of \citet{BovyMAPstructure} for the
low-\afe\ MAPs here by determining the shape of $\Sigma(R)$.

We find that the vertical profile of the low-\afe\ MAPs consists of a
single exponential, but with a scale height that is flaring outward
with an approximately exponential profile.  \citet{BovyMAPstructure}
assumed a constant $h_Z(R)$, because (unpublished) investigations
using the SEGUE G-dwarf data set demonstrated that all but the most
extreme flaring models were consistent with the data, and $h_Z(R) = $
constant was the simplest possible assumption. The inability to
determine the flaring of low-\afe\ MAPs by \citet{BovyMAPstructure}
was due to the limited radial coverage (spanning only about $\pm 2$
kpc) and the lack of low latitude lines of sight in SEGUE. We have not
attempted to refit the SEGUE G-dwarf data using the best-fit flaring
model from this paper, but the slow exponential flaring of $\Rf =
-0.12\pm0.01\kpc^{-1}$ is such that it is most likely consistent with
the SEGUE G-dwarf data.

The measurements of the scale heights in this paper are much more
uncertain than those of \citet{BovyMAPstructure}, because most of the
APOGEE lines of sight are concentrated near the plane. Because we do
not currently possess the relative calibration of $(\feh,\afe)$ in
APOGEE and SEGUE we cannot directly compare measurements of $h_Z$ in
SEGUE and in APOGEE. However, it is clear that the main trends are the
same: (a) $h_Z(\feh,\afe)$ spans the range from $200\pc$ to $1000\pc$,
with a smooth transition between these two extremes, (b) the thinnest
components have low \afe\ and high \feh, (c) the thickest components
have high \afe\ and low \feh, and (d) intermediate ($h_Z
\approx500\pc$) components lie at the high-\feh\ end of the
high-\afe\ sequence and the low-\feh\ end of the
low-\afe\ sequence. Point (a) here is particularly remarkable, because
we measure $h_Z$ much closer to the plane and for isothermal MAPs
\citep{BovyMAPkinematics} in dynamical equilibrium, we expect $h_Z$ to
become larger near the plane. We also confirm that each MAP consists
of a single vertical exponential component.

\subsection{Comparison to star-count measurements}

Any quantitative comparison with previous measurements of $\Sigma(R)$
is problematic, because these measurements typically fit
single-exponential $\Sigma(R)$, while we find that the radial profile
of MAPs is better fit as a broken exponential with a peak radius. We
are therefore forced to qualitatively interpret older measurements
based on their radial and vertical coverage. The most reliable
measurements of the radial structure are based on IR data. Among
these, GLIMPSE measurements based on $|b| \leq 1^\circ$ at $10^\circ
\leq l \leq 65^\circ$ give $h_R = 3.9\pm0.6\kpc$ \citep{Benjamin05a},
while COBE near-IR and far-IR disk data at $|b| \leq 30^\circ$ lead to
$h_R = 2.3\kpc$ \citep{Drimmel01a}, and near-IR 2MASS data in the
outer disk are best fit with $h_R = 2.2\kpc$ \citep{Reyle09a}. We can
explain these differences qualitatively as follows: the GLIMPSE star
counts at $|b| \leq 1^\circ$ are dominated by the low-\afe\ MAPs,
which in the inner MW are approximately a combination of the ``solar''
and ``high-\feh'' populations in \figurename~\ref{fig:broad-radial}
and therefore have quite flat $\Sigma(R)$. The COBE measurements that
extend to higher $|b|$ contain a much higher contribution from the
short $h_R$ high-\afe\ MAPs, and therefore lead to $h_R$ closer to
$2\kpc$. The outer-disk 2MASS sample from \citet{Reyle09a} is
dominated by the steep outer scale length of low-\afe\ ``solar''
populations ($h_R \approx 1.5\kpc$), and the somewhat steep outer
profile of ``low-\feh'' populations ($h_R \approx 2.7\kpc$; see
\tablename~\ref{table:results}); that an overall profile with $h_R =
2.2\kpc$ would result appears somewhat likely. While these qualitative
comparisons are interesting, they mostly point toward the necessity to
perform proper comparisons between the radial profile obtained from
different types of tracers and from different parts of the MW disk.

Star counts in the outer disk have found evidence of a break in the
stellar surface density (\ie, an abrupt steepening of the already
decreasing surface density), with the majority of studies converging
on a break around $R \approx 13.5\kpc$ (\eg,
\citealt{Reyle09a,Sale10a,Minniti11a}; R.~Benjamin \etal, 2016, in
preparation), with a steep density decline beyond the break ($h_R =
1.2\pm0.3\kpc$; \citealt{Sale10a}). We have found that each MAP has a
peak radius, similar in kind to the break radius found in the outer
disk, with outer scale lengths $h_{R,\mathrm{out}} \lesssim 2\kpc$
(see \figurename~\ref{fig:maps-radial}). The break radius seen in star
counts using data and methods that do not discern stars by their
abundances is therefore most likely nothing more than the outermost of
the series of break (or peak) radii displayed by the MAPs; the series
in \figurename~\ref{fig:rpeak} ends around $13\kpc$. More metal-poor
MAPs fill-in the density beyond the peak of more metal-rich MAPs until
the most metal-poor MAPs are reached, and the total surface-density
starts to decline with the steep outer scale length of the most
metal-poor low-\afe\ MAPs. Because we simultaneously fit for
$\Sigma(R)$ and the flaring of the disk, we can be sure that the steep
decline in midplane density is truly because of a declining
$\Sigma(R)$ and not just a consequence of the flaring of the disk.

\subsection{Implications for disk formation and evolution}

Our new results on the stellar population dependent structure of the
MW disk provide stringent and qualitatively new constraints on models
of the formation and evolution of galactic disks. We discuss
qualitatively the implications of our results for some of the main
evolutionary mechanisms, but more detailed comparisons to models are
beyond the scope of this paper.

As displayed in \figurename s~\ref{fig:rpeak} and
\ref{fig:maps-radial}, we find that when the MW disk is dissected into
MAPs, it consists of a set of donut-like rings, with a peak radius
that is a declining function of \feh\ for
low-\afe\ MAPs. high-\afe\ MAPs have peak radii constrained to be
$<5\kpc$, and are therefore consistent with a more traditional disk
structure. Whether or not they actually display at peak at $0 < R
/\kpc< 5$ or are really a single exponential over the full radial
range is a question that requires more data at $R < 5\kpc$.

These present-day patterns in the abundance and spatial structure of
the disk must be the consequence of the radius-dependent chemical
evolution, convolved with the subsequent orbit evolution.  One
possibility is that what we are seeing at low \afe\ are the
(approximate) equilibrium points of chemical evolution, where a steady
state of metal consumption and gas dilution is maintained. Most stars
in simple chemical-evolution models form near the equilibrium \feh,
the value of which depends primarily on how much gas is lost to
outflows (\eg, \citealt{BinneyMerrifield}; D.~Weinberg, \etal, 2016,
in preparation). Outflows are likely more effective in the outer disk
than in the inner disk; this scenario would naturally explain the
well-defined radial range spanned by each low-\afe\ MAP and the
anti-correlation between \feh\ and \Rb. The radial migration that the
low-\afe\ populations have likely experienced (see below) will have
smoothed the radial profiles such that the initial radial profile of
each low-\afe\ MAP was even more sharply peaked around a single
radius.

How do the high-\afe\ MAPs with $\Rb < 5\kpc$ fit into this scenario?
The radial profiles of MAPs with the same \feh\ but different
\afe\ are strikingly different, especially at low \feh. This suggests
that they are not connected by an evolutionary track from high- to
low-\afe\ at a given \feh . The similarity in the radial profile of
high-\afe\ MAPs, combined with the narrow range of \afe\ spanned at
each \feh\ along the high-\afe\ sequence \citep{Nidever14a}, points
towards formation and evolution scenarios of all high-\afe\ MAPs that
were similar enough to be structurally indistinguishable by now.
Because high-\afe\ populations are likely to be the oldest populations
in the disk (see discussion in \citealt{BovyMAPstructure} and
\citealt{Haywood13a, Bergemann14, Martig14b}), this implies that
similar physical conditions existed throughout the disk at early
times, a conclusion also reached by \citet{Nidever14a}, because of the
constant locus of the high-\afe\ sequence in the $(\feh,\afe)$ plane
throughout the disk. What, if anything, caused this to change for the
low-\afe\ sequence remains to be sorted out
\citep[\eg,][]{Stinson13a,Bird13a}.

We have also discovered that the thickness of low-\afe\ MAPs is not
constant, but instead flares in an approximately exponential manner
(\figurename s~\ref{fig:mapfits-flare} and
\ref{fig:maps-radialflare}). Such flaring is commonly seen in
simulations of the outer disk, because of the dynamical heating due to
orbiting satellites and mergers \citep[\eg,][]{Quinn93a}. However, we
measure the same level of flaring for the outer-disk, low-\feh\ MAPs
as we do for the centrally-concentrated, high-\feh\ MAPs. It is
improbable that the latter have been affected much by outer-disk
satellite heating. Flaring with an exponential profile and at the
level that we detect is an important prediction of $J_z$-conserving
radial migration---that is, any redistribution of angular momentum
that approximately conserves the vertical action $J_z$
\citep[\eg,][]{Minchev12a,Solway12a,Roskar13a}. Therefore, we consider
the observed flaring of low-\afe\ MAPs as another important indication
that radial migration significantly affects the distribution of
low-\afe\ stars. This is in addition to the fact that one needs radial
migration to explain the observed lack of correlation between age and
metallicity for low-\afe\ stars
\citep[\eg,][]{Sellwood02a,Schoenrich09a} and to explain the reversal
in the skew of the metallicity distribution function when going from
the inner Galaxy to the outer Galaxy \citep{Hayden15a}.

We find at $>5\sigma$ confidence that the high-\afe\ MAPs do not flare
in the same manner as the low-\afe\ MAPs. Instead, we find that they
must have nearly constant thicknesses. At face value, this finding
implies that little radial migration has occurred in the
high-\afe\ populations, because effects such as those from mergers
that can undo the flaring by mixing in-situ and migrated populations
\citep{Minchev14a} do not apply to the mono-abundance populations
considered here. The fact that large-scale migration does happen for
the low-\afe\ populations entails that whatever causes migration
likely only strongly affects kinematically-cold populations. Spiral
structure whose strength rapidly declines with height is an obvious
candidate for such a migration mechanism. However, determining the
exact implications of the constant thickness of high-\afe\ populations
for the level at which radial migration affects the thicker,
high-\afe\ populations requires chemo-dynamical models that employ a
realistic model for the diffusion of stellar orbits due to
migration-inducing perturbers. Whatever the case may be, the lack of
flaring in the high-\afe\ populations makes it unlikely that their
thickness is due to migration or satellite heating.

While our measurements of the scale heights of MAPs are noisier than
those of \citet{BovyMAPstructure}, they are good enough to confirm the
existence and ubiquity of intermediate $h_Z$ components and the
smoothness of the transition between the traditional ``thin'' and
``thick'' disks: the vertical structure of the disk cannot be
described by only two structurally distinct components.  Overall, the
smoothness of the $h_Z$ transition, the fact that the high-\afe,
large-$h_Z$ MAPs are centrally concentrated, and the level and
homogeneity of the flaring observed in low-\afe\ MAPs all point toward
smooth, internal processes dominating the evolution of the MW disk.

Few cosmological simulations exist that can be directly compared to
our results. A dissection of the radial profile of galactic disks into
mono-abundance or mono-age-metallicity (a convenient proxy for MAPs in
simulations) populations has not been attempted in any simulation. The
analysis of \citet{Stinson13a} comes closest of any simulation,
because they plot the radial profile of MAPs in the MAGIC simulations
in their Figure 1 for a few MAPs. However, only a narrow radial range
around the solar circle is displayed, and larger-scale trends were not
investigated. In agreement with our measurements, they find that
high-\afe\ MAPs are well described by single exponentials in the
radial direction. For the low-\afe, low-metallicity MAPs they find
flat radial profiles, with some of them showing a shallow
peak. However, their low-\afe, high-metallicity MAPs do not display
the broken-exponential that we find here, but are instead consistent
with exponentials. Most other recent simulations, such as those of
\citet{Minchev13a}, \citet{Bird13a}, and \citet{Martig14a} only
dissect the disk in terms of mono-age populations and find
centrally-peaked radial profiles for all such populations. Using
\afe\ as an age proxy this does not come as a surprise, as discarding
the rich \feh\ structure in, \eg, \figurename~\ref{fig:rpeak}, will
lead to very different radial profiles from those observed for MAPs
here. The simplified one-dimensional simulations of a thin,
axisymmetric, gravitationally-unstable disk by \citet{Forbes12a}
similar to the turbulent disks found in cosmological simulations
\citep[\eg,][]{Bournaud09a} do lead to donut-like mono-age populations
by shutting off the gas supply of the inner regions of the disk over
time.

The amount of disk flaring for age- or abundance-selected populations
in these recent simulations varies, from little to no flaring for
undisturbed disks found by \citet{Martig14a} to the strong flaring in
the high-resolution simulation studied by \citet{Bird13a}. That the
high-\afe\ or old populations have a constant thickness while the
low-\afe\ or younger populations flare significantly has not been seen
in any simulation. If the flaring is due to migration in these
simulations, this may indicate that their resolution is not sufficient
to distinguish between the response of kinematically cold and warm
populations.

The overall thickness of the disk is likely to be constant due to the
mix of centrally-concentrated, thick (high-\afe) components and more
extended, thinner-but-flaring (low-\afe) components. This was also
recently found in the simulation of \citet{Minchev15a}.

\needspace{8ex}
\section{Conclusions}\label{sec:conclusion}

We summarize our main results as follows:\\ $\bullet$ The excellent
radial coverage of APOGEE and its APOGEE-RC subsample has enabled a
detailed investigation of the radial structure of abundance-selected
components of the disk (MAPs). Any analysis of the spatial density
distribution of Galactic low-latitude tracers requires explicit
accounting for the 3D dependence of interstellar extinction, even if
target selections and observations are in the NIR. We have applied a
new likelihood-based formalism \citep{BovySF} for determining the
radial profiles of stellar tracers in the presence of dust
extinction. We have been able to obtain good fits to the observed star
counts, even in regions of very high extinction, with simple models
for the spatial stellar distribution.\\$\bullet$ The radial profile of
high-\afe\ MAPs is consistent with a single exponential over the large
radial range over which they are observed ($4 \lesssim R/\kpc \lesssim
14\kpc$), with no sign of a steeper fall-off at large
$R$. Furthermore, all high-\afe\ MAPs are consistent with having
\emph{the same} exponential $\Sigma(R)$, with a scale length of $h_R =
2.2\pm0.2\kpc$. This agrees with the results of
\citet{BovyMAPstructure}, who find a common $h_R = 2.01\pm0.05\kpc$
for the high-\afe\ MAPs and those of \citet{Nidever14a}, who find that
the high-\afe\ sequence remains in the same place in the $(\feh,\afe)$
plane throughout the disk.\\ $\bullet$ We discovered that the radial
surface-density profiles $\Sigma(R)$ of low-\afe\ MAPs are complex:
they are not a single exponential and are not even monotonically
decreasing outward. Each MAP displays a peak radius \Rb\ with an
approximately exponential drop-off away from \Rb\ at smaller and
larger radii. Thus, the low-\afe\ stellar disk may be thought of as a
sequence of narrow, donut-like annuli of increasing \Rb\ for
decreasing \feh.\\$\bullet$ The peak radius of the low-\afe\ MAPs
depends strongly on metallicity, peaking far inside the solar radius
for the metal-rich low-\afe\ MAPs, and well outside the solar circle
for the metal-poor low-\afe\ MAPs. This is consistent with the known
radial metallicity gradient. The MAP with solar abundances peaks at
the solar radius, clearly demonstrating that the Sun is typical for
its Galactic location.\\$\bullet$ The thickness of the high-\afe\ MAPs
is constant with $R$ and does not display any flaring. The constraint
on the inverse flaring scale length of a model with exponential
flaring is strong: $\Rf = 0.00\pm0.02\kpc^{-1}$, when combining
constraints from multiple MAPs. This argues against the local vertical
structure of the thick disk components being set by outward radial
migration.\\$\bullet$ Low-\afe\ MAPs present clear evidence of
flaring, with an exponential $h_Z(R)$ profile and a common flaring
scale length of $8.5\pm0.7\kpc$. This flaring is present both for
low-\feh, outer-disk MAPs and for high-\feh, inner-disk
MAPs.\\$\bullet$ We confirm the result of \citet{BovyMAPstructure},
who found that all MAPs have a single vertical scale height, with a
continuous distribution of them from the thinnest to the thickest
components of the disk. The high precision abundances from APOGEE
($\sigma_{\feh} \approx 0.05\dex$ and $\sigma_{\afe} \approx
0.02\dex$) renders it unlikely that this smooth increase in $h_Z$ is
due to contamination between nearby abundance bins.

The measurements of the vertical profile of MAPs here are somewhat
noisy, because of a lack of data at intermediate and high
latitudes. To make progress on this front requires high-latitude data
with a known selection function. Such data will be provided by the
APOGEE-2 survey \citep{Sobeck14a}.

We have not attempted to determine updated total stellar surface
densities $\Sigma(R_0)$ or total disk masses associated with each MAP
here, as was done by \citet{BovyNoThickDisk} for the SEGUE G-dwarf
sample. Doing so for the RC tracers defined using the cuts of
\citet{BovyRC} has large uncertainties, because the manner in which
giants trace the underlying population depends on the star formation
history. Additionally, the severe cuts used to define the RC make the
sampling of the underlying stellar population highly non-trivial. We
plan to determine total stellar-population masses for the MAPs in the
future by using the full APOGEE giant sample and the SEGUE G-dwarf
sample in combination with the spatial densities measured here.

The current results provide greatly improved constraints on the global
abundance--spatial distribution of stars, which calls for rigorous and
global chemical-evolution modeling. As age constraints for stars
beyond the solar neighborhood ($\gtrsim1\kpc$) become available for
extensive samples, generalizing the current analysis to include age in
addition to abundances in defining MAPs will be an obvious step. This
opens up the prospect of a global map of the Galactic disk in
age--abundance--position space, even before the release of \emph{Gaia}
data.

\acknowledgements It is a pleasure to thank the anonymous referee,
Brett Andrews, Bob Benjamin, and Chris McKee for helpful comments and
discussions. Some of the results in this paper have been derived using
the HEALPix \citep{Gorski05a} and \texttt{healpy}
packages. J.B. received support from a John N. Bahcall Fellowship and
the W.M. Keck Foundation. H.W.R. received funding for this research
from the European Research Council under the European Union's Seventh
Framework Programme (FP 7) ERC Grant Agreement
n. [321035]. T.C.B. acknowledges partial support for this work from
grants PHY 08-22648; Physics Frontier Center/Joint Institute or
Nuclear Astrophysics (JINA), and PHY 14-30152; Physics Frontier
Center/JINA Center for the Evolution of the Elements (JINA-CEE),
awarded by the US National Science Foundation. J.B. and
H.W.R. acknowledge the generous support and hospitality of the Kavli
Institute for Theoretical Physics in Santa Barbara during the
`Galactic Archaeology and Precision Stellar Astrophysics' program,
where some of this research was performed.

Funding for SDSS-III has been provided by the Alfred P. Sloan
Foundation, the Participating Institutions, the National Science
Foundation, and the U.S. Department of Energy Office of Science. The
SDSS-III web site is http://www.sdss3.org/.

SDSS-III is managed by the Astrophysical Research Consortium for the
Participating Institutions of the SDSS-III Collaboration including the
University of Arizona, the Brazilian Participation Group, Brookhaven
National Laboratory, Carnegie Mellon University, University of
Florida, the French Participation Group, the German Participation
Group, Harvard University, the Instituto de Astrofisica de Canarias,
the Michigan State/Notre Dame/JINA Participation Group, Johns Hopkins
University, Lawrence Berkeley National Laboratory, Max Planck
Institute for Astrophysics, Max Planck Institute for Extraterrestrial
Physics, New Mexico State University, New York University, Ohio State
University, Pennsylvania State University, University of Portsmouth,
Princeton University, the Spanish Participation Group, University of
Tokyo, University of Utah, Vanderbilt University, University of
Virginia, University of Washington, and Yale University.

\end{document}